\documentclass[aps,prd,a4paper,superscriptaddress,nofootinbib,10pt, one column]{revtex4-1}
\usepackage{amsmath,amssymb,bm,latexsym,soul,nicefrac,graphicx,subfigure,multirow,dcolumn,aas_macros,xcolor,hyperref}
\usepackage{array, booktabs, cancel}
\usepackage{CJK}
\usepackage{geometry}
\usepackage{acronym}
\usepackage{graphicx,url,bm,makecell,verbatim}

\definecolor{Ggreen}{RGB}{255,0,0} 

\newcommand{\TRC}{MOE Key Laboratory of TianQin Mission, TianQin Research Center for Gravitational Physics, Frontiers Science Center for TianQin, CNSA Research Center for Gravitational Waves, Sun Yat-sen University (Zhuhai Campus), Zhuhai 519082, P.R. China}
\newcommand{\SPA}{School of Physics and Astronomy, Sun Yat-sen University (Zhuhai Campus), Zhuhai 519082, P.R. China}
\newcommand{\D}{\mathrm{d}}

\begin{document}

\title{Constraining the Hubble constant to a precision of about $1\%$ using multi-band dark standard siren detections}

\author{Liang-Gui Zhu}
\affiliation{\SPA}
\affiliation{\TRC}

\author{Ling-Hua Xie}
\affiliation{\SPA}

\author{Yi-Ming Hu}
\email{huyiming@mail.sysu.edu.cn}
\affiliation{\SPA}
\affiliation{\TRC}

\author{Shuai Liu}
\email{liushuai5@mail.sysu.edu.cn}
\affiliation{\SPA}
\affiliation{\TRC}

\author{En-Kun Li}
\email{lienk@mail.sysu.edu.cn}
\affiliation{\SPA}
\affiliation{\TRC}

\author{Nicola R. Napolitano}
\affiliation{\SPA}

\author{Bai-Tian Tang}
\affiliation{\SPA}

\author{Jian-dong Zhang}
\affiliation{\SPA}
\affiliation{\TRC}

\author{Jianwei Mei}
\affiliation{\SPA}
\affiliation{\TRC}

\date{\today}

\begin{abstract}

  Gravitational wave signal from the inspiral of stellar-mass binary black hole can be used as standard sirens to perform cosmological inference.
This inspiral covers a wide range of frequency bands, from the millihertz band to the audio-band, allowing for detections by both space-borne and ground-based gravitational wave detectors. 
In this work, we conduct a comprehensive study on the ability to constrain the Hubble constant using the dark standard sirens, or gravitational wave events that lack electromagnetic counterparts.
To acquire the redshift information, we weight the galaxies within the localization error box with photometric information from several bands and use them as a proxy for the binary black hole redshift.
We discover that TianQin is expected to constrain the Hubble constant to a precision of roughly $30\%$ through detections of $10$ gravitational wave events; in the most optimistic case, the Hubble constant can be constrained to a precision of $< 10 \%$, assuming TianQin I+II.
In the optimistic case, the multi-detector network of TianQin and LISA is capable of constraining the Hubble constant to within $5\%$ precision.
It is worth highlighting that the multi-band network of TianQin and Einstein Telescope is capable of constraining the Hubble constant to a precision of about $1\%$.
We conclude that inferring the Hubble constant without bias from photo-z galaxy catalog is achievable, and we also demonstrate self-consistency using the P$-$P plot.
On the other hand, high-quality spectroscopic redshift information is crucial for improving the estimation precision of Hubble constant.
\\
\\
{\bf keywords}: gravitational wave standard siren, Hubble constant, stellar-mass binary black hole, photometric luminosity, multi-band gravitational wave detection    

\end{abstract}

\maketitle

\acrodef{GW}{gravitational wave}
\acrodef{EM}{electromagnetic}
\acrodef{FLRW}{Friedmann-Lema\^itre-Robertson-Walker}
\acrodef{LCDM}[$\Lambda$CDM]{Lambda cold dark matter}
\acrodef{SNIa}[SN Ia]{type Ia supernova}
\acrodefplural{SNIa}[SNe Ia]{type Ia supernovae}
\acrodef{CMB}{cosmic microwave background}
\acrodef{StBH}{stellar-mass black hole}
\acrodef{SBBH}{stellar-mass binary black hole}
\acrodef{BNS}{binary neutron star}
\acrodef{NSBHB}{neutron star-black hole binary}
\acrodefplural{NSBHB}{neutron star-black hole binaries}
\acrodef{sGRB}{short $\gamma$-ray burst}
\acrodef{HCS}{heliocentric coordinate system}
\acrodef{SNR}{signal-to-noise ratio}
\acrodef{FIM}{Fisher information matrix}
\acrodef{SDSS}{Sloan Digital Sky Survey}
\acrodef{GWENS}{Gravitational Wave Events in Sloan}
\acrodef{DES}{Dark Energy Survey}
\acrodef{CDF}{cumulative distribution function}


\newpage
\setcounter{page}{1}

\section{Introduction}     \label{introduction}

The \ac{GW} observations of compact binary coalescences can be used as standard sirens thanks to the fact that the intrinsic \ac{GW} strength can be deduced from the phase evolution \cite{Schutz:1986gp}.
When combined with redshift data from \ac{EM} measurements, such standard sirens can be used to determine cosmological parameters.

The Hubble constant can be determined using data from the late Universe measurements, represented by the \ac{SNIa} observations, and from the early Universe measurements, represented by the \ac{CMB} anisotropies observations.
However, there is a significant inconsistency between these two measurements, and because the inconsistency has grown over $4 \sigma$ \cite{Planck:2015fie, Planck:2018vyg, Freedman:2017yms, Freedman:2019jwv, Freedman:2021ahq, Riess:2019cxk, Riess:2020sih, Riess:2020fzl}, this discrepancy, also known as the ``Hubble tension'', has become a hot topic.
Although luminosity distance measurements are sometimes subject to high statistical errors, they remain an important probe as the \ac{GW} observation can provide a direct measurement of the luminosity distance that is independent of the cosmic distance scale ladder.
Thus, standard sirens possess the potential to clarify the Hubble tension \cite{Chen:2017rfc, Feeney:2018mkj, Borhanian:2020vyr, Bian:2021ini}.

The direct detection of the \ac{GW} signals from compact binary coalescences by Advanced LIGO and Virgo \cite{2016PhRvL.116f1102A, 2016PhRvL.116x1103A, 2017PhRvL.118v1101A, 2017ApJ...851L..35A, 2017PhRvL.119n1101A, 2017PhRvL.119p1101A, LIGOScientific:2018mvr, LIGOScientific:2020stg, Abbott:2020uma, Abbott:2020tfl, Abbott:2020khf, Abbott:2020niy, LIGOScientific:2021usb, LIGOScientific:2021qlt} opened an era of \ac{GW} astronomy.
Among different types of \ac{GW} signals, the \acp{BNS} and \acp{NSBHB} mergers are ideal standard sirens since they have the potential to be detected through both the \ac{GW} and the \ac{EM} channels.
The current ground-based \ac{GW} detectors, including KAGRA \cite{Akutsu:2018axf} and LIGO-India \cite{Unnikrishnan:2013qwa}, is expected to detect dozens of \ac{GW} events of \acp{BNS} and \acp{NSBHB} during the course of the operation, and a few percent precision of the Hubble constant are expected to be reached from the \ac{GW} cosmology \cite{Nissanke:2013fka, Chen:2017rfc, Vitale:2018wlg, Mortlock:2018azx}.
The first multimessenger observations of a \ac{BNS} merger event GW170817 \cite{2017PhRvL.119p1101A, Monitor:2017mdv, Soares-Santos:2017lru, GBM:2017lvd} provided the first standard siren measurement of the Hubble constant, $H_0 = 69_{-8}^{+17} \ {\rm km/s/Mpc}$ \cite{Abbott:2019yzh} (also see \cite{Abbott:2017xzu, Fishbach:2018gjp, Guidorzi:2017ogy, Hotokezaka:2018dfi}).

The \ac{SBBH} mergers should be dark in the \ac{EM} channel \cite[cf.][]{McKernan:2019hqs, Graham:2020gwr}; therefore, the cosmological constraint from \acp{SBBH} can only be derived through the ``dark standard siren'' \cite{Schutz:1986gp, Soares-Santos:2019irc}.
In this scenario, the redshift information is provided by matching the \ac{GW} source sky localization and the galaxies catalogs.
It is expected that the current ground-based \ac{GW} detectors will continue to constrain the Hubble constant efficiently via \ac{SBBH} \ac{GW} events \cite{Soares-Santos:2019irc, Palmese:2020aof, Vasylyev:2020hgb, Abbott:2019yzh, Chen:2017rfc, Fishbach:2018gjp, DelPozzo:2012zz, Taylor:2011fs, Nair:2018ign, Farr:2019twy, Gray:2019ksv, Bera:2020jhx, Finke:2021aom, Gray:2021sew, LIGOScientific:2021aug}, and indeed constraint of the Hubble constant has already been obtained with the observation of GW170814 and GW190814, provided a measurement precision of about $57 \%$ \cite{Soares-Santos:2019irc, Palmese:2020aof, Vasylyev:2020hgb, Abbott:2019yzh}.

Future ground-based \ac{GW} detectors, such as the Einstein Telescope (ET) \cite{Punturo:2010zz, Sathyaprakash:2012jk} and Cosmic Explorer  \cite{Dwyer:2014fpa, Evans:2016mbw}, will be much more sensitive and capable of detecting \ac{GW} events at higher redshift.
This enables the potential of not only measuring the Hubble constant but also constraining other cosmological parameters \cite{Zhao:2010sz, Zhao:2017cbb, Taylor:2012db, Seikel:2012uu, Messenger:2011gi, Messenger:2013fya, DelPozzo:2015bna, Cai:2016sby, Du:2018tia, Zhang:2018byx, Mendonca:2019yfo, Zhang:2019loq, Jin:2020hmc, Yu:2020vyy, You:2020wju, Bonilla:2021dql}.
Space-borne \ac{GW} detectors operating in the millihertz band, such as TianQin \cite{Luo:2015ght} and LISA \cite{LISA:2017pwj}, can observe \ac{GW} signals at cosmological distances, including massive black hole binaries (MBHBs) \cite{Klein:2015hvg, Barausse:2020mdt, Wang:2019}, extreme mass ratio inspirals \cite{Babak:2017tow,Gair:2017ynp, Fan:2020zhy}, and \acp{SBBH} \cite{Kyutoku:2016ppx, Liu:2020eko, Liu:2021yoy}.
Space-borne \ac{GW} detectors are expected to have excellent capability for sky localization, which will also enable them to constrain the Hubble constant as well as other cosmological parameters \cite{Holz:2005df, Babak:2010ej, Petiteau:2011we, Tamanini:2016zlh, Caprini:2016qxs, Cai:2017yww, Wang:2019tto, Zhu:2021aat, Wang:2020dkc, Wang:2021srv, MacLeod:2007jd, Laghi:2021pqk,Kyutoku:2016zxn, DelPozzo:2017kme}.

\acp{SBBH} are very interesting \ac{GW} sources due to the vast frequency range of their \ac{GW} signals, which cover a wide frequency range from millihertz to kilohertz.
This feature enables the \acp{SBBH} to be detectable in multiple band \ac{GW} detectors \cite{Sesana:2016ljz, Sesana:2017vsj}. 
Space-borne \ac{GW} detectors can observe the early inspiral signal while ground-based \ac{GW} detectors can study the final merger.
Both low and high frequency can be complimentary,
with space-borne detectors capable of more precise phase information \cite{Kyutoku:2016ppx, Liu:2020eko} and ground-based detectors can accumulate higher \acp{SNR} \cite{Zhao:2017cbb}, 
thus improving the overall parameter estimation precision of the \ac{GW} source \cite{Vitale:2016rfr, Moore:2019pke, Ewing:2020brd, Grimm:2020ivq}, in order to facilitate the extraction of physical/astronomical information \cite{Barausse:2016eii, Vitale:2016rfr, Wong:2018uwb, Gerosa:2019dbe, Cutler:2019krq, Liu:2020nwz} and measurement of the expansion of the Universe \cite{Muttoni:2021veo}.

We study the potential of constraining the Hubble constant with TianQin using the \ac{SBBH} \ac{GW} sources.
Furthermore, the anticipated operation time of the various detectors allows simultaneous observation through a multi-detector network of TianQin \cite{Luo:2020bls, Mei:2020lrl} and LISA \cite{LISA:2017pwj}, as well as a multi-band network of TianQin and ET \cite{Maggiore:2019uih}.
Thus, we study how such networks might be used to better constrain the Hubble constant.

This paper is organized as follows.
In Section \ref{analysis-method}, we present the cosmological analytical framework and the methods needed to spatially localize \ac{GW} sources and weight candidate host galaxies.
In Section \ref{simulations}, we introduce the necessary astrophysical context for the simulations and present the method for simulating observational data.
In Section \ref{results}, we illustrate the constraint processes of the cosmological parameters and show the constraint results on the Hubble constant.
In Section \ref{discussion}, we discuss several critical concerns raised by the analyses and simulations.
In Section \ref{conclusion-outlook}, we summarize our results and discuss the need for additional research.

\section{Methodology}     \label{analysis-method}

Throughout the work, we adopt a spatially-flat \ac{LCDM} cosmology.
The Hubble parameter, which describes the expansion rate of the scale factor in the late Universe, can be expressed as
\begin{equation}  \label{H_z}
  H(z) = H_0 \sqrt{\Omega_M(1+z)^3 + \Omega_{\Lambda}},
\end{equation}
where $H_0 \equiv H(z=0)$ is the Hubble constant that describing the current rate of expansion, and $\Omega_M$, $\Omega_\Lambda = 1- \Omega_M$ are the fractional densities of total matter and dark energy with respect to the critical density $\rho_c= 3H_0^2/(8\pi G)$ (where $G$ is Newton's gravitational constant).
According to this cosmology predicts that the luminosity distance $D_L$ of a source with redshift $z$ is
\begin{equation} \label{DL_z}
  D_L = c(1+z)  \int_{0}^{z} \frac{1}{H(z')} \D z' ,
\end{equation}
where $c$ is the speed of light in a vacuum.
Throughout this work, we utilize the injected true numbers $H_0 = 67.8 $ km/s/Mpc and $\Omega_M = 0.307$ \cite{Planck:2015fie}.

\subsection{Standard siren and dark standard siren}    \label{standard-siren}

The two polarizations of \ac{GW} signal of an inspiralling binary with component masses $m_1$ and $m_2$ can be described as \cite{Colpi:2016fup}
\begin{subequations}
 \begin{align}
 h_+ (t) &=  \left(\frac{G \mathcal{M}_z}{c^2} \right)^{5/3} \left(\frac{\pi f(t)}{c} \right)^{2/3} \frac{2 (1+ \cos^2 \iota)}{D_L} \cos \big(\Psi (t, \mathcal{M}_z, \eta) \big),    \label{h_plus}    \\
 h_{\times} (t) &=  \left(\frac{G \mathcal{M}_z}{c^2} \right)^{5/3} \left(\frac{\pi f(t)}{c} \right)^{2/3} \frac{4 \cos \iota}{D_L} \sin \big(\Psi (t, \mathcal{M}_z, \eta) \big),    \label{h_cross}
 \end{align}
\end{subequations}
where $\mathcal{M}_z = (1+z)\mathcal{M} = (1+z)(m_1m_2)^{3/5}/(m_1+m_2)^{1/5}$ is the redshifted chirp mass (with a perfect degeneracy between the redshift and the physical mass), $\eta = m_1 m_2 / (m_1 + m_2)^2$ is the symmetric mass ratio, $\iota$ is the inclination angle of the binary orbital angular momentum relative in the line of sight, and $\Psi (t, \mathcal{M}_z, \eta)$ is the phase of the \ac{GW} signal.
Notably, the overall amplitude is determined solely by the redshifted chirp mass, the inclination, and the luminosity distance.
By observing the \ac{GW} phase evolution, the mass parameter $\mathcal{M}_z$ can be reliably estimated, and the inclination angle can be determined by observing the amplitude ratio of different polarizations.
Therefore, compact binary coalescences are referred to as ``standard sirens'', as it is possible to infer the luminosity distance $D_L$ directly from the \ac{GW} data.

To determine the cosmological parameters, one must still have redshift information, which the \ac{GW} data analysis can only supply infrequently.
The next section discusses several methods for obtaining redshift information \cite{2018SSPMA..48g9805Z}:
\begin{itemize}
  \item{The \ac{EM} counterpart. The coalescence of \ac{BNS} and \ac{NSBHB} are frequently accompanied by \ac{EM} radiation such as short Gamma ray bursts \cite{Nissanke:2009kt, Blanchard:2017csd, GBM:2017lvd} or kilonovae \cite{Li:1998bw,2010MNRAS.406.2650M,Tanvir:2017pws}, the redshift can be measured directly \cite{Guidorzi:2017ogy, Abbott:2017xzu, Vitale:2018wlg}. }
\item{The ``dark standard siren''. Galaxies are clustered on a small scale. Assuming the \ac{GW} sources are linked with a host galaxy, one can obtain a statistical understanding of the redshift from galaxy information even in the absence of an \ac{EM} counterpart \cite{Soares-Santos:2019irc, Palmese:2020aof, MacLeod:2007jd, Petiteau:2011we, DelPozzo:2012zz, DelPozzo:2017kme, Nair:2018ign, Gray:2019ksv, Laghi:2021pqk}. }
\item{The neutron star mass distribution. \ac{EM} observations, which deduced a relatively narrow distribution for neutron star masses in \acp{BNS}. 
  This intrinsic distribution can be exploited to overcome the mass-redshift degeneracy \cite{Kiziltan:2013oja, Taylor:2011fs, Taylor:2012db, DelPozzo:2015bna}. }
\item{The tidal deformation of the neutron stars. During coalescence, a compact object with finite size will experience tidal deformation, resulting in phase correction of the \ac{GW} waveform. Because the degree of tidal deformation is determined by both the intrinsic mass and the equation of the state of the neutron star, the correcting phase can overcome the mass-redshift degeneracy \cite{Messenger:2011gi, Messenger:2013fya}. }
\item{The cross-correlation method. After \ac{EM} observations have mapped the spatial distribution of the galaxies in redshift space, and \ac{GW} detections can also map out the spatial distribution of the \ac{GW} events in luminosity distance space, then the cross-correlation of spatial distributions between the galaxies and the \ac{GW} sources can be used to extract redshift information \cite{Oguri:2016dgk, Zhang:2018ekk, Mukherjee:2020hyn, Bera:2020jhx, Diaz:2021pem}. }
\item{Other methods. There are also efforts to use the mass distribution of the \ac{SBBH} population \cite{Farr:2019twy, You:2020wju}, the intrinsic redshift probability distribution of compact binary mergers \cite{Ding:2018zrk, Leandro:2021qlc}, and high-order correction of the \ac{GW} waveform phase caused by cosmic acceleration \cite{Seto:2001qf, Nishizawa:2010xx, Nishizawa:2011eq}, to break the degeneracy and obtain the redshift information. }
\end{itemize}
In this work, we consider the dark standard siren scenario with \acp{SBBH} to constrain the Hubble constant $H_0$.

\subsection{Bayesian framework}    \label{bayes-framework}

We adopt a Bayesian analytical framework to estimate cosmological parameters using the data from the dark standard siren and the catalogs of survey galaxies \cite{Chen:2017rfc, Abbott:2017xzu, Fishbach:2018gjp, Zhu:2021aat}.
Consider a set of \ac{GW} detection data composed of $N$ \ac{GW} events $\mathcal{D}_{\rm GW} \equiv \{ d_{\rm GW}^1, d_{\rm GW}^2, \ldots, d_{\rm GW}^i, \ldots, d_{\rm GW}^N \}$ as well as the corresponding \ac{EM} observation data set $\mathcal{D}_{\rm EM} \equiv \{ d_{\rm EM}^1, d_{\rm EM}^2, \ldots, d_{\rm EM}^i, \ldots, d_{\rm EM}^N \}$, the posterior probability distribution of the cosmological parameter set $\vec{H} \equiv \{H_0, \Omega_M \}$ is given by
\begin{equation} \label{bayes_formula}
p(\vec{H} |\mathcal{D}_{\rm GW}, \mathcal{D}_{\rm EM}, I) = \frac{p_0(\vec{H}|I) p(\mathcal{D}_{\rm GW}, \mathcal{D}_{\rm EM}|\vec{H}, I)}{p(\mathcal{D}_{\rm GW}, \mathcal{D}_{\rm EM}|I)} = \frac{p_0(\vec{H}|I) \prod_i p(d_{\rm GW}^i, d_{\rm EM}^i|\vec{H}, I)}{p(\mathcal{D}_{\rm GW}, \mathcal{D}_{\rm EM}|I)},
\end{equation}
where $p_0(\vec{H}|I)$ is the prior probability distribution for $\vec{H}$, $I$ indicates all the relevant background information.
The normalization factor $p(\mathcal{D}_{\rm GW}, \mathcal{D}_{\rm EM}|I)$ from which is independent of $\vec{H}$ is also known as Bayes evidence.
Therefore, we can derive from
\begin{equation} \label{posterior}
p(\vec{H} |\mathcal{D}_{\rm GW}, \mathcal{D}_{\rm EM}, I) \propto p_0(\vec{H}|I) \prod_i p(d_{\rm GW}^i, d_{\rm EM}^i|\vec{H}, I).
\end{equation}

For a \ac{GW} event, with the corresponding \ac{GW} data $d_{\rm GW}^i$ and the \ac{EM} data $d_{\rm EM}^i$, the likelihood can be expressed as
\begin{equation} \label{likeli1}
p(d_{\rm GW}^i, d_{\rm EM}^i| \vec{H}, I) = \frac{\int p(d_{\rm GW}^i, d_{\rm EM}^i, D_L, z, \alpha, \delta | \vec{H}, I) \D D_L \, \D z \, \D \alpha \, \D\delta }{\beta(\vec{H} | I)},
\end{equation}
where $\alpha$ and $\delta$ represent the longitude and latitude, respectively.
To eliminate systematic biases due to the selection effect, we introduced a correction term $\beta(\vec{H} | I)$ as the denominator \cite{Abbott:2017xzu, Chen:2017rfc, Mandel:2018mve}.
The integrand in the numerator of Eq. (\ref{likeli1}) can be factorized as
\begin{align} \label{likeli2}
& p(d_{\rm GW}^i, d_{\rm EM}^i, D_L, z, \alpha, \delta | \vec{H}, I)   \nonumber \\
 = &  p(d_{\rm GW}^i, d_{\rm EM}^i | D_L, z, \alpha, \delta, \vec{H}, I) p_0(D_L, z, \alpha, \delta | \vec{H}, I)    \nonumber \\
 = & p(d_{\rm GW}^i | D_L, \alpha, \delta, I) p(d_{\rm EM}^i | z, \alpha, \delta, I) p_0(D_L, z, \alpha, \delta | \vec{H}, I)    \nonumber \\
 = & p(d_{\rm GW}^i | D_L, \alpha, \delta, I) p(d_{\rm EM}^i | z, \alpha, \delta, I) p_0(D_L | z, \vec{H}, I) p_0(z, \alpha, \delta | \vec{H}, I) .
\end{align}
where $p_0$ represents the prior.

Assuming that the \ac{GW} noise is Gaussian and stationary, one has \cite{Finn:1992}
\begin{align}  \label{likeli_GW}
p(d_{\rm GW}^i | D_L, \alpha, \delta, I) &= \int p(d_{\rm GW}^i | D_L, \alpha, \delta, \vec{\theta}', I) \D \vec{\theta}'    \nonumber \\
&\varpropto  \int \exp \Big(-\frac{1}{2} \big \langle d_{\rm GW}^i - h(D_L, \alpha, \delta, \vec{\theta}') \big | d_{\rm GW}^i - h(D_L, \alpha, \delta, \vec{\theta}') \big \rangle  \Big) \D \vec{\theta}',
\end{align}
where $\langle \cdot|\cdot \rangle$ is the inner product as defined in Eq.(\ref{inner_product}), $h$ is the waveform of the \ac{GW} signal, and $\vec{\theta}'$ represents the parameters of the \ac{GW} source that is unrelated to the cosmological inference.
It should be noted that we marginalize over the parameters of the \ac{GW} source parameters that are not directly related to the constraining cosmological parameters, such as the inclination angle $\iota$.
For the dark standard siren scenario, since we assume no \ac{EM} signal associated with the \ac{GW} event, we set $p(d_{\rm EM}^i | z, \alpha, \delta, I) = \text{const.}$ \cite{Chen:2017rfc, Fishbach:2018gjp}.
We assume that $p_0(D_L | z, \vec{H}, I) \equiv \delta( D_L - \hat D_L(z, \vec{H}) )$ under the cosmology, where $\hat D_L(z, \vec{H})$ is defined in Eq. (\ref{DL_z}).

In the \ac{EM} observations, the sky localization of the galaxy is very accurate (relative to the sky localization of \ac{GW} source), and for a galaxies catalog from a photometric sky survey, the prior $p_0(z, \alpha, \delta | \vec{H}, I)$ in Eq. (\ref{likeli2}) can be expressed as
\begin{equation} \label{prior_EM}
  p_0(z, \alpha, \delta | \vec{H}, I) = \sum_{j=1}^{N_{\rm gal}}  W_j P(z|\bar z_j, \sigma_{z;j}) \delta(\alpha - \alpha_j) \delta(\delta - \delta_j) ,
\end{equation}
where $N_{\rm gal}$ is the total number of the galaxies catalog, and $P(x|\bar x, \sigma_x)$ is a Gaussian distribution on $x$, with expectation $\bar{x}$ and standard deviation $\sigma_x$, $x = \{z, L\}$, and $W_j$ is the weight of galaxy, reflect a priori confidence that the galaxy could host a compact binary.
While the metallicity, morphology, and rate of star formation could be different by a lot, one can assume that the potential for each galaxy to host a compact binary is the same, i.e., $W_j = 1/ N_{\rm gal}$.
Alternatively, one may expect the compact binary merger rate to be proportional to the galaxy stellar mass $M$, which is in turn related to its luminosities in various bands,
i.e., $W_j \propto M_{j} \simeq  \hat{M} \big( P(L|\bar L_j^1, \sigma_{L;j}^1), P(L|\bar L_j^2, \sigma_{L;j}^2), ..., P(L|\bar L_j^k, \sigma_{L;j}^k), ..., P(L|\bar L_j^{N_{\rm band}}, \sigma_{L;j}^{N_{\rm band}}) \big)$, where $\bar L_j^k$ and $\sigma_{L;j}^k$ are the mean value and standard deviation, respectively, of the luminosity of $j$-th galaxy in $k$-th band. 
The variable $N_{\rm band}$ represents the total number of bands in the photometric sky surveys, and more details about the function $\hat{M} (..., P(L|\bar L_j^k, \sigma_{L;j}^k), ...)$ is described in Section \ref{weighting}.

Taking the preceding analysis into account, substituting Eq. (\ref{likeli2}) into Eq. (\ref{likeli1}) and marginalizing over the parameter $D_L$, Eq. (\ref{likeli1}) becomes
\begin{equation} \label{likeli3}
p(d_{\rm GW}^i, d_{\rm EM}^i| \vec{H}, I) = \frac{\int p(d_{\rm GW}^i | \hat D_L(z, \vec{H}), \alpha, \delta, I) p_0(z, \alpha, \delta | \vec{H}, I) \D z \, \D \alpha \, \D\delta }{\beta(\vec{H} | I)}.
\end{equation}
Following the statistical method presented in \cite{Zhu:2021aat} for evaluating the survey galaxies catalog’s selection biases, we use a smooth prior distribution of the catalog redshift as
\begin{equation} \label{p_noLSS}
p_\textrm{c}(z|\vec{H}, I)  \varpropto  \frac{1}{2 \Delta z} \int_{(z-\Delta z)}^{(z+\Delta z)} \int \!\!\!\! \int_{4 \pi} \int  p_0(z',\alpha,\delta | \vec{H},I) \D \alpha \, \D \delta \, \D z',
\end{equation}
where $\Delta z$ is chosen to be much larger than the redshift interval of the galaxy clusters.
The correction term $\beta(\vec{H} | I)$ in Eq. (\ref{likeli3}) can be written approximately as
\begin{equation} \label{N_term}
\beta(\vec{H} | I) \approx \int p(d_{\rm GW}^i|\hat D_L(z, \vec{H}),I) p_\textrm{c}(z|\vec{H}, I) \D z.
\end{equation}
Notice that for catalogs that are composed of from multiple sources, like the GLADE catalog \cite{Dalya:2018cnd}, in order to maintain self-consistency, one needs to take extra care to deal with the selection effects $p_{\rm c}(z| \vec{H}, I)$ separately for different sources.

\subsection{Localization and distance of GW source}     \label{GWparas-estimate}

For regular triangular space-borne \ac{GW} detectors like TianQin and LISA, the recorded signal can be expressed as \cite{Klein:2015hvg, Liu:2020eko}
\begin{align} \label{GW_response}
h (t) &= \frac{\sqrt{3}}{2} \left( F^+(t) h_+(t + t_D) + F^{\times}(t) h_{\times}(t + t_D) \right),  \\
t_D &\approx -\frac{R_0}{c} \sin \theta'_S \cos \big( \phi_{\rm d}(t) - \phi'_S \big),
\end{align}
where $t$ is the \ac{HCS} time, $t_D$ is the time delay between the solar system barycenter to the detector, $R_0 = 1 \rm AU$, the primed angles $\theta'_S$ and $\phi'_S$ are the altitude and azimuth of the \ac{GW} source in the \ac{HCS} respectively, quantities related to the source are labeled with subscript ``S'', the ones related to the detector are labeled with subscript ``d''.
We have $\phi_{\rm d}(t) = 2 \pi t / T + \phi_0 $, where $T = 1 \rm yr$ is the orbital period of the detector around the Sun, $\phi_0$ is the initial orbital phase of the detector at $t=0$.
In the relatively low frequency region, $f \ll f_* = c /(2 \pi L)$ (where $L$ is the arm length of the interferometer), the antenna pattern functions $F^{+, \times}(t)$ can be approximately expressed as \cite{Thorne:1989lfp}
\begin{subequations}
 \begin{align}
 F^+ (t) &=  \frac{1}{2} \left(1 + \cos^2 \theta_S(t) \right) \cos 2\phi_S(t) \cos 2\psi_S(t) - \cos \theta_S(t) \sin 2\phi_S(t) \sin 2\psi_S(t) ,    \label{F_plus}    \\
 F^{\times} (t) &=  \frac{1}{2} \left(1 + \cos^2 \theta_S(t) \right) \cos 2\phi_S(t) \sin 2\psi_S(t) + \cos \theta_S(t) \sin 2\phi_S(t) \cos 2\psi_S(t),    \label{F_cross}
 \end{align}
\end{subequations}
where $\theta_S(t)$, $\phi_S(t)$, and $\psi_S(t)$ are the altitude, azimuth, and polarization angles, respectively, in the detector-based coordinate system at $t$.
In comparison to Eq. (\ref{F_plus}, \ref{F_cross}), the azimuth angle of the antenna pattern functions the second independent channel differ by $+\pi /4$ for TianQin and LISA; and by $+2\pi/3$ and $+4\pi/3$ for the second and third independent laser interferometer, respectively, for ET reference \cite{Freise:2008dk}.
This difference is due to the fact that ET unlike space-borne detectors, uses an independent interferometer.
The variation of $\theta_S(t)$, $\phi_S(t)$, and $\psi_S(t)$ with time dependent on the motion of the detector in space, the detailed description of the detector's response to the \ac{GW} signals can be found in \cite{Liu:2020eko, Hu:2018yqb} for TianQin, in \cite{Cutler:1997ta, Cornish:2002rt} for LISA, and in \cite{Jaranowski:1998qm, Sathyaprakash:2012jk} for ET.

Notice that although in the relatively high frequency range of $f \gtrsim f_*$, the low frequency limit is no longer valid for TianQin and LISA, and the sensitivity would drop in higher frequencies.
It will cause some complexity for data analysis. However, as long as we absorb the effect into the sensitivity curve, the majority of the conclusions discussed in our study remain intact \cite{Wang:2019, Liu:2020eko, Robson:2019}.
Moreover, a full analytical formula for the frequency response of a space-based detector is given in \cite{Zhang:2020khm}.

\begin{figure}[htbp]
\centering
\includegraphics[height=8.cm, width=12.cm]{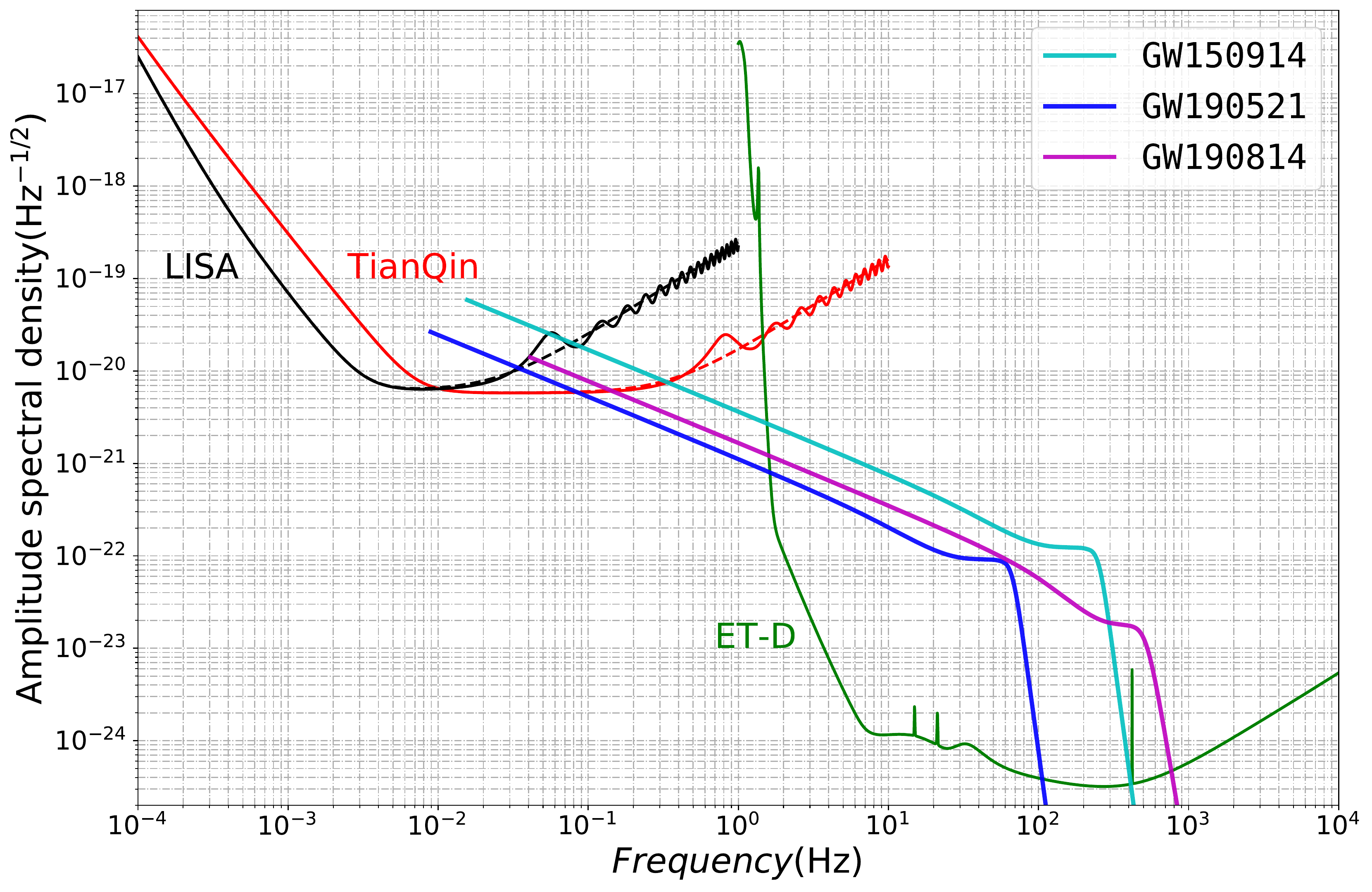}
\caption{Anticipated sensitivity curves for TianQin (red line), LISA (black line), and ET-D (green line).
  Additionally, the amplitude spectra densities $\sqrt{f} |\widetilde{h}(f)|$ of GW150914 \cite{2016PhRvL.116f1102A} (cyan line), GW190521 \cite{Abbott:2020tfl} (blue line), and GW190814 \cite{Abbott:2020khf} (magenta line) detected by Advanced LIGO and Virgo, assuming a 5 years merger time. }
\label{sensi_cur}
\end{figure}

With $N$ independent detectors observing the same \ac{GW} source, the \ac{GW} signals detected by different detectors can be collectively expressed as a vector,
\begin{equation} \label{h-vector}
\boldsymbol{h} = \Big[ h_1, h_2, \cdots, h_k, \cdots, h_N \Big]^{\rm T},
\end{equation}
where $h_k$ represents the \ac{GW} strain recorded by the $k$-th detector.
The total \ac{SNR} $\rho$ of a \ac{GW} source provided by multiple independent detectors is defined as
\begin{equation} \label{SNR}
\rho = \sqrt{ \left\langle \boldsymbol{h} | \boldsymbol{h} \right\rangle },
\end{equation}
where the inner product is defined as \cite{Finn:1992, Cutler:1994ys}
\begin{equation} \label{inner_product}
\left\langle \boldsymbol{h} | \boldsymbol{h} \right\rangle = \sum_k \left\langle h_k | h_k \right\rangle \equiv \sum_k 4 \mathfrak{Re} \int_{0}^{\infty} \frac{\widetilde{h}_k^{*} (f) \widetilde{h}_k(f) }{S_{n;k} (f)} \D f,
\end{equation}
where $\widetilde{h}_k (f)$ is the Fourier transform of $h_k(t)$, $*$ represents complex conjugate, and $S_{n;k} (f)$ is the sensitivity curve functions of the $k$-th detector.
The sensitivity curves of TianQin \cite{Wang:2019, Liu:2020eko}, LISA \cite{Robson:2019}, and ET (ET-D) \cite{Hild:2010id, Sathyaprakash:2012jk, Maggiore:2019uih} are shown in FIG. \ref{sensi_cur}.

For a \ac{GW} source characterized by a parameter set of $\vec{\theta} \equiv (\mathcal{M}_z, \eta, D_L, \alpha, \delta, \iota, t_c, \Psi_c, \psi)$ detected from multiple independent detectors, the \ac{FIM} can provide a Cram\'{e}r-Rao lower bound on the parameter estimation uncertainty \cite{Vallisneri_2008}.
The \ac{FIM} is defined as follows:

\begin{equation} \label{FIM}
\Gamma_{mn} \equiv \left\langle \frac{\partial \boldsymbol{h}}{\partial \theta_m} \bigg| \frac{\partial \boldsymbol{h}}{\partial \theta_n} \right\rangle = \sum_k \left\langle \frac{\partial h_k}{\partial \theta_m} \bigg| \frac{\partial h_k}{\partial \theta_n} \right\rangle,
\end{equation}
where $\theta_m$ indicates the $m$-th parameter of the \ac{GW} event.
The covariance matrix $\Sigma$ equal to the inverse of the \ac{FIM}, $\Sigma = \Gamma^{-1}$, we can adopt $\Delta \theta_m = \sqrt{\Sigma_{mm}} = \sqrt{(\Gamma^{-1})_{mm}}$ as the estimation error of the parameter, with the sky localization error being $\Delta \Omega = 2\pi |\sin \delta| \sqrt{\Sigma_{\alpha\alpha} \Sigma_{\delta\delta} - \Sigma_{\alpha\delta}^2}$.

\subsection{Galaxy weighting}     \label{weighting}

Assuming that the formation rate of the \ac{SBBH} per unit stellar mass is uniform across all galaxies, we can expect that the probability of a galaxy hosting the \ac{SBBH} is proportional to its total stellar mass.

The $K$-band luminosity is a commonly used parameter to account for the galaxy's mass.
The $K$-band is commonly accepted to trace the galaxy's old stellar population and, thus, is approximately proportional to the galaxy’s stellar mass, as well as being weakly correlated with the galaxy’s color \cite{Bell:2003cj, Lin:2004ak}.
Following this logic, the $K$-band luminosity weighting method was applied to the dark standard siren cosmological analysis using LIGO\&Virgo data \cite{Fishbach:2018gjp, Gray:2019ksv, Abbott:2019yzh} \footnote{The $B$-band luminosity information is also used, which reflects the star formation rate of the galaxy.}.

In this work, we aim to adopt a different approach to improve the galaxy weighting process in order to obtain more accurate cosmological parameter estimation.
This is achieved by deriving the galaxy’s stellar masses from the multi-band photometry of the galaxy samples.

\subsubsection{Galaxy sample and the completeness of the catalog}     \label{data:gwens}
We use the \ac{GWENS} galaxy catalog \footnote{The \ac{GWENS} catalog is available at: \url{https://astro.ru.nl/catalogs/sdss_gwgalcat} } \cite{Rahman:2019, Abbott:2019yzh}, which has been assembled on purpose for the \ac{EM}-\ac{GW} multimessenger observations.
It uses the \ac{SDSS} Data Release 14 (DR14) \cite{Abolfathi:2017vfu}, which maps over $45$ million galaxies, covering about a quarter of the whole sky. The \ac{GWENS} catalog provides photometric information in five bands of $ugriz$ for all galaxies, corresponding to {\it cModel} magnitudes (see \ac{SDSS} DR14 paper \cite{Abolfathi:2017vfu}) corrected for Galactic extinction from \cite{Schlegel:1997yv}.
Additionally, the \ac{GWENS} catalog contains photometric redshifts (photo-$z$) for the majority of the sample (about $98.5\%$) and spectroscopic redshifts for a small portion.

\begin{figure}[htbp]
\centering
\includegraphics[height=7.5cm, width=12.cm]{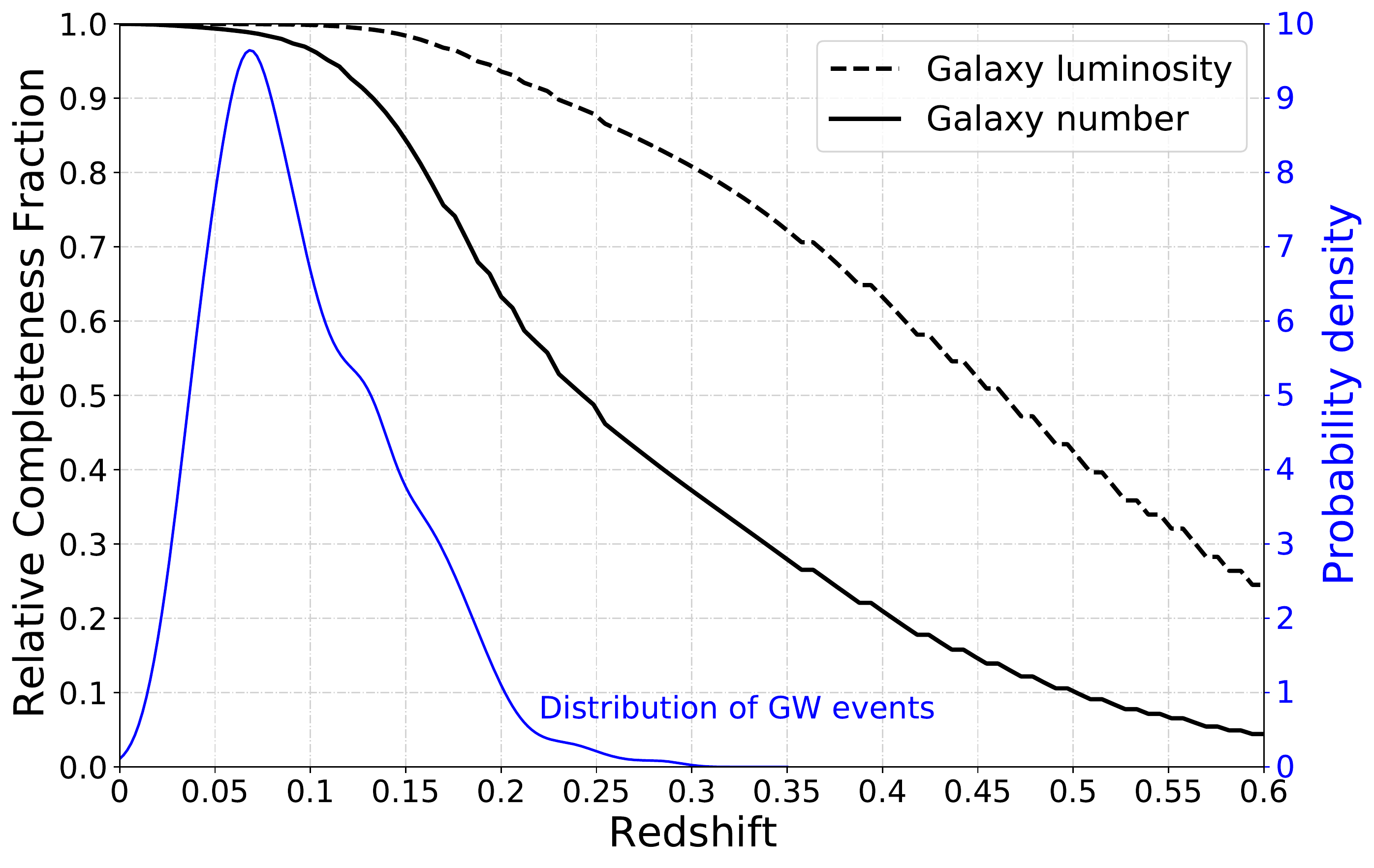}
  \caption{The solid and dashed black lines represent completeness fractions of the \ac{GWENS} catalog relative to local values, in terms of number and luminosity (in $z$-band), respectively.
  The blue line represents the probability distribution of simulated detectable \ac{GW} events, with the scales shown on the right side.}
\label{survey_completeness}
\end{figure}

The completeness of a galaxy catalog plays a vital role in the dark standard of siren study.
FIG. \ref{survey_completeness} shows the relative completeness fraction of the \ac{GWENS} catalog at different redshifts.
This is obtained by using the galaxy luminosity distribution within $z < 0.05$ as the fiducial distribution, assuming that galaxies are distributed uniformly in the co-moving volume, and neglecting the evolution of the galaxy luminosity distribution with redshift. 
According to previous GW forecast studies \citep{Liu:2020eko} conclude that the \ac{SBBH} inspirals detectable by TianQin are mainly concentrated in the $z < 0.2$, region with the highest redshift not exceeding $z = 0.3$.
The \ac{GWENS} catalog has a relative completeness fraction of around $0.63$ at $z = 0.2$ and $0.37$ at $z = 0.3$ in terms of the number of galaxies;
however, in terms of the total luminosity contributed by galaxies, the relative completeness fraction remains above $0.9$ at $z = 0.2$ and equal to or greater than $0.8$ at $z = 0.3$.
These levels of relative completeness ensure that the majority of host galaxies for \ac{GW} sources are included in the list, and the \ac{GWENS} catalog meets our requirements.
There exist other galaxy catalogs available, such as the GLADE catalog \cite{Dalya:2018cnd} and the DES catalog \cite{Abbott:2021hoj, Drlica-Wagner:2017tkk, Abbott:2018jhe}; however, we chose the \ac{GWENS} catalog due to its higher completeness and broader sky coverage.

\subsubsection{Stellar masses}     \label{data:masses}

\begin{figure}[htbp]
\centering
\includegraphics[height=8.cm, width=12.cm]{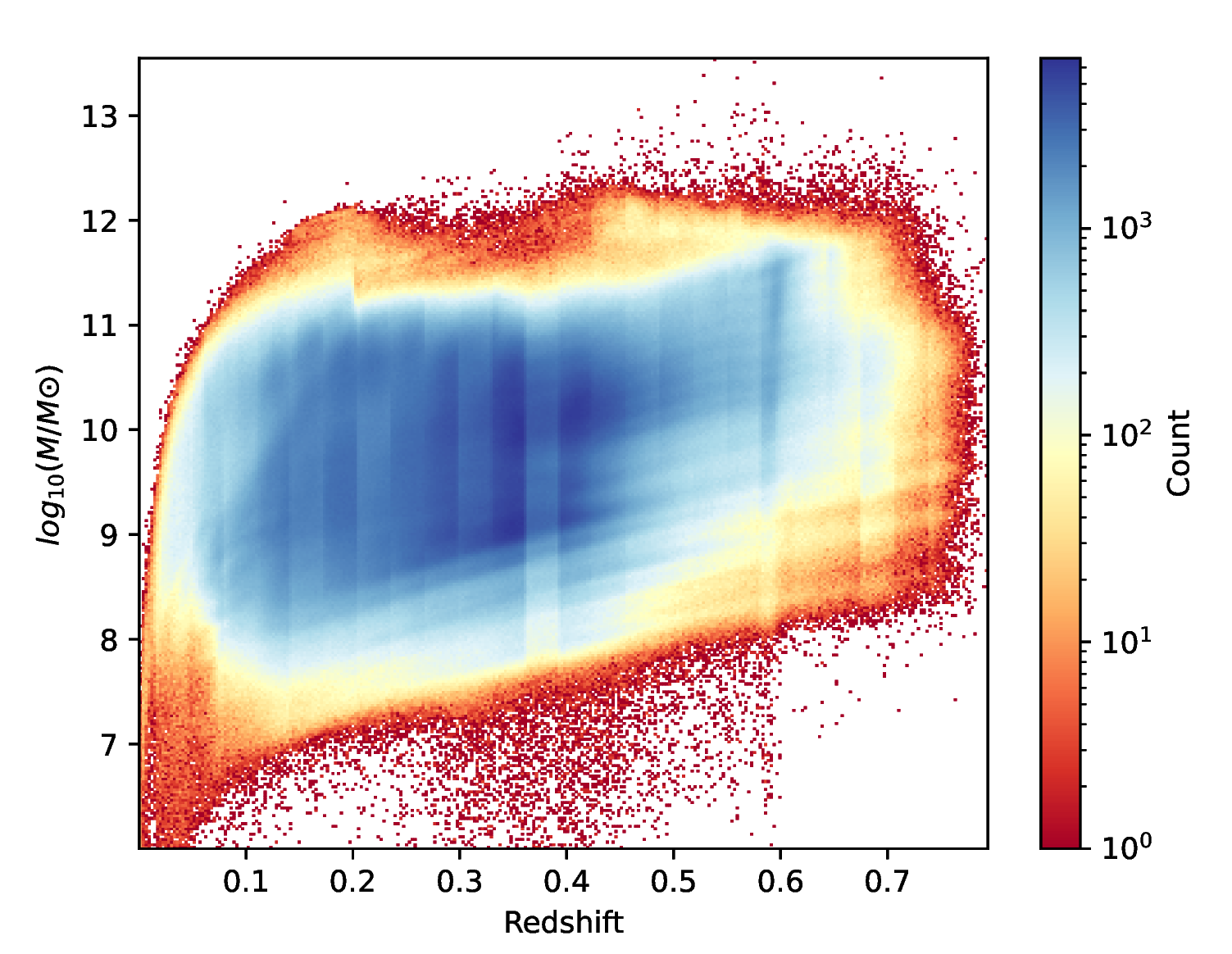}
\caption{Distribution of total stellar masses for \ac{GWENS} galaxies derived from the  {\tt Le Phare} software.
Note that the redshift in horizontal coordinates represents the median of the posterior probability distribution of the photo-$z$ of galaxies.  }
\label{galaxies_mass}
\end{figure}

Galaxy masses are calculated using {\tt Le Phare}, a widely used spectral energy distribution (SED) fitting software \cite{Arnouts:1999bb, Ilbert:2006dp}. {\tt Le Phare} matches observed galaxy colors to predicted colors using a set of theoretical SED libraries derived from simple stellar population models, characterized by a series of stellar parameters such as age $\tau$, metallicity $Z$, and a star formation history. The SED is convolved with the filter transmission curves (including instrument efficiency) adopted for the specific observation dataset in order to provide synthetic luminosities in the chosen bands. The best-fit parameters are obtained via minimizing the $\chi^{2}$ difference between the synthetic color SED and the photometric data.
The input data set is made up of multi-band photometric magnitudes from the \ac{GWENS} catalog \cite{Rahman:2019, Abbott:2019yzh}. The output data set consists of the best stellar population parameters, such as age, metallicity, rate of star formation, and stellar mass.
We adopt stellar templates from \cite{Bruzual:2003tq} in the simple stellar population models, together with an initial mass function from \cite{Chabrier:2003ki} and an exponentially decaying star formation history.
We use a diverse collection of models with three metallicities ($0.005Z_{\odot}$, $Z_{\odot}$, and $2.5Z_{\odot}$) and different ages (${\tau} \leq {\tau}_{\rm max}$), with the maximum age ${\tau}_{\rm max}$, set by the age of the Universe at the redshift of the galaxy, with a maximum value at $z=0$ of $13.5 \ \rm Gyr$.
We also consider internal extinction using the models from \cite{Calzetti:1994vw}.
Finally, in the {\tt Le Phare} run, we fix the galaxy redshift to the median of the probability density function of photo-$z$ reported in the \ac{GWENS} catalog to reduce degeneracies between redshift and galaxy colors. 
FIG. \ref{galaxies_mass} shows the final distribution of stellar masses for the $45\,141\,307$ \ac{GWENS} galaxies.
The stellar masses range, from dwarf-like systems ($\sim 10^7 M_{\odot}$) to massive galaxies ($\sim 10^{12}$), as do the redshifts (from $z = 0$ to $z \sim 0.8$).

To check the presence of biases in our mass estimates, we compare them to those obtained for \ac{SDSS} DR 12 (Portsmouth SED-fit Stellar Masses, which is a catalog of stellar masses of $1\,489\,670$ systems from the Portsmouth Group \footnote{Available at \url{https://www.sdss.org/dr12/spectro/galaxy_portsmouth/}}, see also \cite{Maraston:2012jf} ), for which we have discovered an overlap of $\sim 53\,700$ galaxies with \ac{GWENS}.
The mean deviation between our estimates and the \ac{SDSS} ``starburst model'' catalog is $\Delta_{M}=0.14$ and the scatter $\sigma_M=0.32$ in logarithmic space, meaning that the two stellar mass estimates are basically consistent with each other (note that in the case we compared against the Portsmouth passive model catalog, we obtained a $\Delta_{M}=-0.07$ and scatter $\sigma_M=0.18$).
The scatter obtained by these comparisons is quite small considering the two approaches’ rather different models (e.g., \cite{Maraston:2004em} and \cite{Maraston:2008nn} templates, no reddening, and a different magnitude definition).

\section{Simulations}    \label{simulations}

To perform a quantitative analysis of the Hubble constant constraint imposed by \ac{GW} signals using future space-borne and ground-based \ac{GW} detectors, we first simulate a \ac{GW} event catalog, then identify candidate host galaxies according to the spatial localization of the \ac{GW} source, and finally obtain their statistical redshift information.

\subsection{Stellar-mass black hole binary signals}     \label{SBBH-population}

We adopt the ``Power Law + Peak'' model to populate the \ac{SBBH} simulation \cite{LIGOScientific:2020kqk}.
Based on the 50 \ac{GW} events published by the first (GWTC-1) \cite{LIGOScientific:2018mvr} and the second Gravitational-Wave Transient Catalog (GWTC-2) \cite{Abbott:2020niy}, this mode has been associated with the highest Bayesian factor \cite{LIGOScientific:2020kqk}.
The model includes a power-law mass distribution for the primary component, with a smooth truncation at the lower mass limit and a Gaussian peak at the high mass end to account for the pile-up effect of pulsational pair-instability supernovae \cite{LIGOScientific:2020kqk, Talbot:2018cva}.
The mass ratio $q$, which describes the ratio between the secondary component mass and the primary component mass, is modeled by a power law \cite{LIGOScientific:2018jsj, Roulet:2018jbe, Fishbach:2019bbm}, $p(q | m_1) \propto q^{\beta_q}$, and $\beta_q = 1.3_{-1.6}^{+2.4}$ \cite{LIGOScientific:2020kqk}.
According to \cite{LIGOScientific:2020kqk}, we adopt the associated \ac{SBBH} merger rate as $R_{\rm SBBH} = 58_{-29}^{+54} \ {\rm Gpc}^{-3} {\rm yr}^{-1}$ \footnote{Notice that a revised version reports an updated value of $R_{\rm SBBH} = 52_{-26}^{+52} \ {\rm Gpc}^{-3} {\rm yr}^{-1}$.}, which assumes that $m_1 \ge 2 M_{\odot}$ and includes GW190814.

Moreover, we ignore the spins of inspiralling \acp{SBBH} because the black hole spin effect has a negligible effect on the \ac{GW} cosmology study \cite{Mangiagli:2018kpu, Nishizawa:2016jji}.
The eccentricity of the \ac{SBBH}, which is not well constrained by LIGO and Virgo observations, should have non-negligible effects on space-borne \ac{GW} detections.
Throughout this work, we set the eccentricity to $e_0 = 0.01$ at \ac{GW} frequency equal to $0.01$ Hz as a representative value \cite{Nishizawa:2016jji, Liu:2020eko}.

Furthermore, we assume that \acp{SBBH} are uniformly distributed in the co-moving volume.
Each \ac{GW} event is hosted randomly in a galaxy from the GWENS catalog, assuming that the probability of each galaxy being hosted by a \ac{GW} event is proportional to its total stellar mass.
The orientation parameter distribution is chosen to be isotropic, i.e., $\alpha \in \textrm{U}[0, 2\pi]$, $\cos \delta \in \textrm{U}[-1, 1]$, $\alpha_L \in \textrm{U}[0, 2\pi]$, and $\cos \delta_L \in \textrm{U}[-1, 1]$; the spins are fixed at $\chi_{1,2} \equiv 0$; the remaining parameters are assumed to obey uniform distribution, $t_c \in \textrm{U}[0, 20] \ \rm yr$, and $\Phi_c \in \textrm{U}[0, 2\pi]$.
Finally, we produce the simulated \ac{GW} signals with the IMRPhenomPv2 waveform \cite{Hannam:2014}.

\subsection{Detections with GW detectors}     \label{gw-catalog}

In this work, we investigate different detector configurations, including those using space-borne and ground-based \ac{GW} detectors, as described below, 
\begin{itemize}
  \item{\emph {TianQin}}: the default situation in which three satellites form a constellation and operate in a ``3 months on + 3 months off'' mode, with a mission life time of 5 years \cite{Luo:2015ght};
  \item{\emph {TianQin I+II}}: twin constellations of satellites with perpendicular orbital planes, that operates in a relay mode and can avoid the 3 months gap in data \cite{Wang:2019, Liu:2020eko, Liang:2021bde};
  \item{\emph {TianQin+LISA}}: a multi-detector \ac{GW} detector network of TianQin and LISA, we adopt LISA configuration according to \cite{LISA:2017pwj, Robson:2019}, and considering 4 years of overlap in operation time;
  \item{\emph {TianQin I+II+LISA}}: similar to above but with the TianQin I+II configuration considered;
  \item{\emph {TianQin+ET}}: a multi-band \ac{GW} detector network of TianQin and ET, with 5 years of overlap in operation time, we adopt ET configuration according to \cite{Punturo:2010zz, Hild:2010id, Maggiore:2019uih}, and assume they will continue operating for 15 years after the end of the TianQin mission;
  \item{\emph {TianQin I+II+ET}}: similar to TianQin+ET but with the TianQin I+II configuration considered.
\end{itemize}

In TABLE \ref{Detection_rate}, we list the anticipated detection rates with respect to different detection thresholds for \ac{SBBH} inspirals that merge within $20$ years from the start of TianQin detection.
Throughout this work, we adopt two \ac{SNR} thresholds for different detector configurations, one with $\rho_{\rm thre} = 8$ for space-borne \ac{GW} detectors \cite{Liu:2020eko, Kyutoku:2016ppx}, including TianQin, TianQin I+II, TianQin+LISA, and TianQin I+II+LISA; and the other with $\rho_{\rm thre} = 5$ for multi-band \ac{GW} detections \cite{Wong:2018uwb, Ewing:2020brd}, such as TianQin+ET and TianQin I+II+ET.
It is worth mentioning that, for sources with $\rho \ge 5$ detected with space-borne \ac{GW} detectors/networks, they can all be detected by ET with $\rho \gtrsim 20$, as long as ET is operating when they merge.
Therefore we do not list the detection rate of ET separately in TABLE \ref{Detection_rate}, and we do not consider \ac{GW} sources that can only be detected by ET in this work.

\begin{table}[]
  \caption{Total detection rate of the \acp{SBBH} with $t_c < 20 \rm yr$ over the entire observation time that based on the ``Power Law + Peak'' \ac{SBBH} population model for five detector configurations: TianQin, TianQin I+II, LISA, TianQin+LISA, and TianQin I+II+LISA.  }
    \vspace{12pt}
    \renewcommand\arraystretch{1.5}
    \centering
    \begin{tabular}{c|ccccc}
        \hline
        \hline
        \multirow{2}*{\ac{SNR}} & \multicolumn{5}{c}{Total detection rate} \\
        \cline{2-6}
        \thead[c]{~~Threshold~~} & \thead[c]{~~~~~~~~TianQin~~~~~~~~} & \thead[c]{~~~~TianQin I+II~~~~~} &\thead[c]{~~~~~~~~~~LISA~~~~~~~~~~} & \thead[c]{~~~TianQin+LISA~~~} & \thead[c]{TianQin I+II+LISA}  \\
        \hline
        $\rho \ge 5$  & $44.2_{-21.1}^{+43.1}$   & $111.6_{-53.3}^{+108.1}$   & $74.2_{-36.2}^{+70.3}$   & $155.9_{-75.7}^{+147.4}$   & $245.8_{-120.1}^{+234.9}$       \\
        \hline
        $\rho \ge 8$  & $10.8_{-5.5}^{+10.1}$   & $28.4_{-14.9}^{+26.0}$   & $16.8_{-8.2}^{+16.7}$   &  $38.3_{-19.4}^{+36.1}$ & $63.8_{-32.7}^{+58.7}$     \\
        \hline
        $\rho \ge 12$ \footnote{A conservative \ac{SNR} threshold \cite{Liu:2020eko}.} & $3.0_{-1.4}^{+3.1}$   & $7.8_{-3.6}^{+7.8}$   & $4.6_{-1.9}^{+4.6}$   &  $10.6_{-5.2}^{+9.9}$   & $17.5_{-8.5}^{+17.6}$      \\
        \hline
        \hline
    \end{tabular}
    \label{Detection_rate}
\end{table}

\begin{figure}[htbp]
\centering
\includegraphics[height=9.cm, width=15.cm]{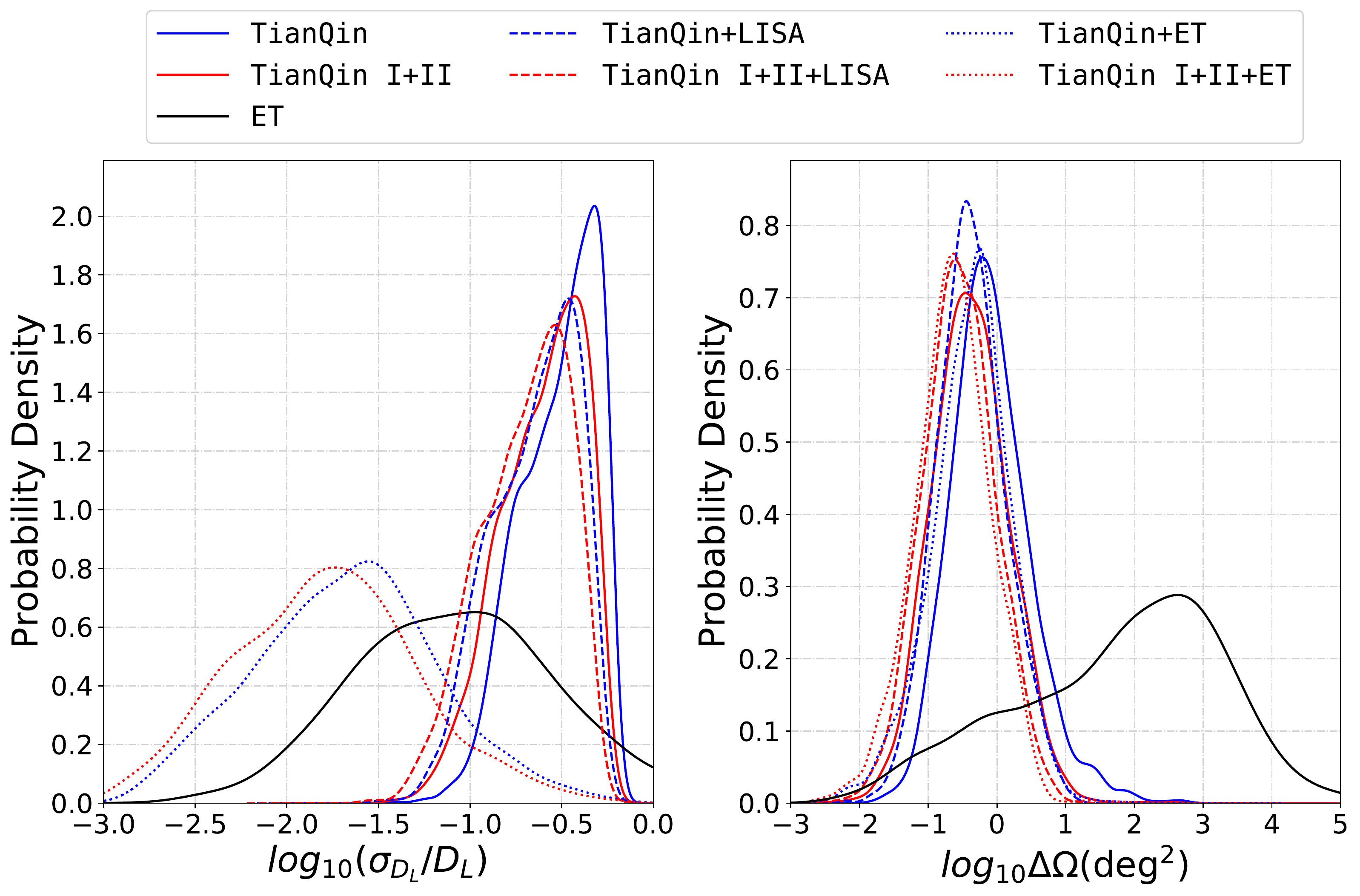}
\caption{Spatial localization error distribution of TianQin (solid blue line), TianQin I+II (solid red line), TianQin+LISA (dashed blue line), TianQin I+II+LISA (dashed red line), TianQin+ET (dotted blue line), TianQin I+II+ET (dotted red line), and ET (solid black line).  }
\label{GW_error}
\end{figure}

The precision of the Hubble constant constrained by \ac{GW} signals from \acp{SBBH} mainly depends on the spatial localization errors of the \ac{GW} sources.
In FIG. \ref{GW_error}, we illustrate the marginalized distribution on relative error of luminosity distance $\sigma_{D_L}/D_L$ and sky localization $\Delta \Omega$.
Due to the fact that \ac{GW} events are useful for reconstructing the `$D_L-z$ relation' only when the luminosity distance can be reasonably estimated, we only include the \ac{GW} sources with $\sigma_{D_L}/D_L < 0.6$ (approximately corresponding to $\Delta D_L/D_L < 1$ at the confidence level of $90\%$) for consideration in this work.

We observe that for events detected by TianQin or TianQin I+II, the relative error of the luminosity distance for most \ac{GW} sources is greater than $0.1$, but the sky position of most \ac{GW} sources can be localized to better than $1 \ \rm deg^2$.
The network of TianQin (or TianQin I+II) and LISA can marginally improve the spatial localization of the \ac{GW} sources.
This is because TianQin has better sensitivity than LISA at higher frequencies, which is where the \ac{SBBH} inspiral signals are concentrated.

On the other hand, for ET-detected events, due to the time-dependent modulation of antenna beam-pattern related to the Earth’s rotation, a typical sky localization error is at the level of $10^2 \sim 10^3 \ \rm deg^2$.
Noting the strong degeneracy between sky localization and luminosity distance, because the polarization angle of a \ac{GW} signal is dependent on the relative position of the \ac{GW} source and the detector; such a large sky localization error results in a large uncertainty on polarization, which translates into a large uncertainty on inclination angle and eventually on luminosity distance.
Although the majority of \ac{GW} signals detected by ET have a high \acp{SNR} (on the order of $10^2 \sim 10^3$), the typical relative error on the luminosity distance is about $0.1$.

Of course, the multi-band \ac{GW} cosmology is only meaningful if one can identify the common origin of a binary from both frequency bands.
Fortunately, this can be achieved thanks to the excellent parameter recovery ability in either frequency band.
Two binary black hole signals that share highly consistent merging time ($\Delta t_c \lesssim 1 \ {\rm s}$), location ($\Delta \Omega \lesssim 1 \ {\rm deg}^2$), mass ($\Delta \mathcal{M}_z /\mathcal{M}_z \sim 10^{-7}$), and distance ($\Delta D_L /D_L \sim 10^{-1}$) can be easily identified as the same binary \cite{Liu:2020eko}.
Even in the pessimistic scenario that the \ac{SNR} in TianQin band is too weak for independent detection, the archival search methods \cite{Wong:2018uwb, Moore:2019pke, Ewing:2020brd} can be used to find the signal.
Searching for archive data triggered by ET detections is a very practical way to achieve multi-band identification.

One remarkable conclusion we can draw from FIG. \ref{GW_error} is that the multi-band \ac{GW} detection can significantly improve the estimation precision of the luminosity distance.
Accurate \ac{GW} detection from space-borne detectors can provide very precise sky localization, breaking the degeneracy between sky localization and luminosity distance for ground-based detectors.
The relative error on luminosity distance can be improved by one order of magnitude compared with TianQin/LISA and by half an order of magnitude compared with ET.

\subsection{Statistical redshift}     \label{statistical-z}

The spatial localization information of the \ac{GW} source cannot be directly used to select candidate host galaxies of the \ac{GW} source.
The possible range of luminosity distance $[D_L^-, D_L^+] \equiv [(\bar D_L - 2\sigma_{D_L}), (\bar D_L + 2\sigma_{D_L})]$ (where $\bar D_L$ is the mean value) needs to be converted into the possible range of redshift $[z^-, z^+]$ first.
The candidate host galaxies are then selected from the survey galaxy catalog using the redshift-space error box.
This conversion depends on both a specific cosmological model and the prior of the corresponding cosmological parameters.
We convert the luminosity distance range into the redshift possible range under the standard \ac{LCDM} model, with the prior of $H_0 \in {\rm U}[30, 120] \ \rm km/s/Mpc$ and $\Omega_M \in {\rm U}[0.04, 0.6]$.
The conversion relation is given by
\begin{subequations}
 \begin{align}
 D_L^- &=  c(1+z^-) \int_{0}^{z^-} \frac{1}{H^-(z')} \D z',    \label{DL_min}    \\
 D_L^+ &=  c(1+z^+) \int_{0}^{z^+} \frac{1}{H^+(z')} \D z',    \label{DL_max}
 \end{align}
\end{subequations}
where $H^{-}(z)$ and $H^+(z)$ are the specific Hubble parameter realizations that minimize and maximize Eq. (\ref{H_z}) within the prior of both $H_0$ and $\Omega_M$.

In our simulation, we obtain the candidate host galaxies of the \ac{GW} source from the \ac{GWENS} catalog.
To properly account for the redshift uncertainty, we consider two factors:
(1) The redshift information of galaxy in the catalog is almost all photo-$z$, including non-negligible photo-$z$ error $\Delta z_{\rm photo}$;
(2) Due to the peculiar velocity of galaxies, the observed spectroscopic redshift $z_{\rm obs}$ is different from the cosmological redshift $z$, and we denote this redshift error $\Delta z_{\rm pv} \equiv |z_{\rm obs} - z|$.
In general, $\Delta z_{\rm photo} \gg \Delta z_{\rm pv}$, so the final boundary of the redshift of candidate host galaxies is defined as $[z_{\min}, z_{\max}] = [(z^- - \Delta z_{\rm photo}), (z^+ + \Delta z_{\rm photo})]$.
The final spatial localization error box of candidate host galaxies is defined by $4\Delta \Omega \times [z_{\min}, z_{\max}]$ (the factor of $4$ corresponds to the $2\sigma$ confidence level for the \ac{GW} sky localization).
For a small number of galaxies with spectroscopic redshift, we discard $\Delta z_{\rm photo}$ and approximate $\Delta z_{\rm pv} \approx (1 + z_{\rm obs}) \langle v_{\rm p} \rangle / c$, assuming $\langle v_{\rm p} \rangle = 500 \ \rm km/s$ \cite{He:2019dhl}.
The impact of peculiar velocities and their reconstruction on the estimation of $H_0$ has been studied in Refs. \cite{Howlett:2019mdh, Nicolaou:2019cip, He:2019dhl, Mukherjee:2019qmm}, and it is possible to eliminate the redshift error $\Delta z_{\rm pv}$ for nearby galaxies.

To obtain the statistical redshift distributions of \ac{GW} sources, for each galaxy in the localization error box, we adopt the following two methods:
\begin{itemize}
  \item{\emph{fiducial method}}: assigning equal weight regardless of its position and luminosity;
  \item{\emph{weighted method}}: assigning a weight that accounts for both its positional and luminosity-related information.
\end{itemize}
When the sky localization from \ac{GW} detection is described by a covariance matrix $\Sigma_{\alpha \delta}$, the positional weight of a galaxy at location $(\alpha_j,\delta_j)$ is defined as  $ W_{\rm posi} \propto \exp \left\{- \frac{1}{2} \left[ \big(\alpha_j - \bar \alpha,\delta_j - \bar \delta\big) \Sigma_{\alpha \delta}^{-1} \big(\alpha_j - \bar \alpha,\delta_j - \bar \delta\big)^{\rm T} \right] \right\}$, where $(\bar \alpha, \bar \delta)$ are the best measured values of the sky localization.
Meanwhile, we apply a luminosity-related weight corresponding to total stellar mass of the galaxy, which is derived using multi-band photometric information through the {\tt Le Phare} software (see Section \ref{weighting} for details).

\begin{figure}[htbp]
\centering
\includegraphics[height=10.cm, width=16.cm]{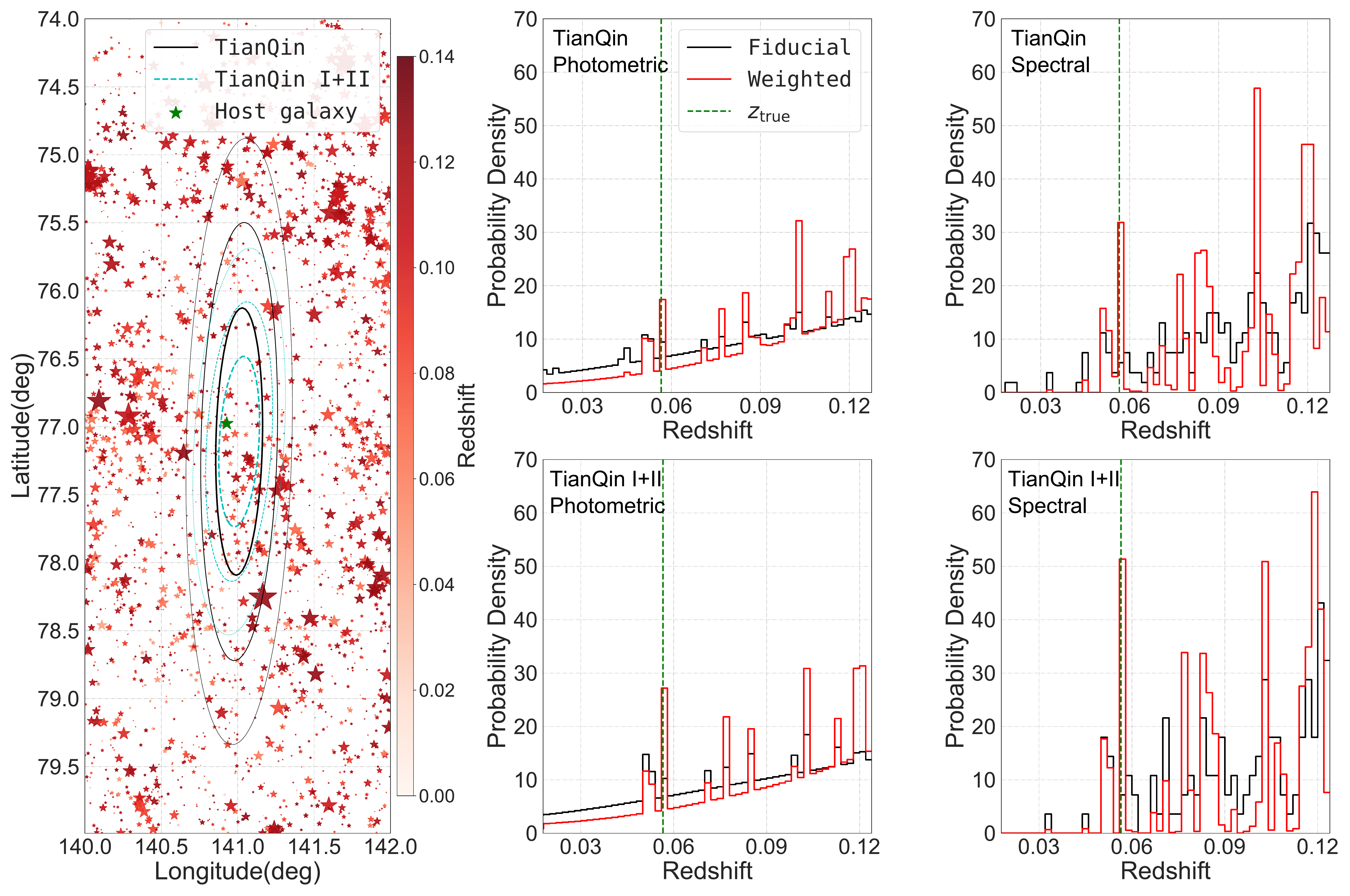}
\caption{Sky localization and statistical redshift distribution of an example \ac{SBBH} inspiral.
  The left panel shows sky localizations of TianQin (solid black contour lines) and TianQin I+II (dashed cyan contour lines), as well as a scatter plot of galaxies within the catalog. The green star labels the real host galaxy; the shade of the star represents its redshift, and the size of the star represents its luminosity-related weight.
  The center panels show the statistical photo-$z$ of TianQin (top) and TianQin I+II (bottom), respectively, using two different statistical redshift estimations with both the fiducial method (black histograms) and the weighted method (red histograms), and the vertical green dashed line represents the true redshift of the \ac{GW} source.
  The right panels show the same as the center panels but with spectroscopic galaxy redshifts.}
\label{z_distribution_12TQ}
\end{figure}

FIG. \ref{z_distribution_12TQ} illustrates an example localization of a \ac{SBBH} \ac{GW} signal and the impact of different weighting methods as well as redshift (photometric or spectroscopic) on the final redshift estimation of the \ac{SBBH}.
While the sky localization of \ac{GW} source is insufficiently precise to uniquely identify the host galaxy in the absence of an \ac{EM} counterpart; more accurate sky localization can eliminate the interference of many polluting galaxies.
The left panel shows the advantage of the more accurate sky localization of TianQin I+II over TianQin in selecting the galaxies; the middle panels show the different distributions of the statistical redshifts for the two detector configurations. One can conclude that with fewer candidate host galaxies, the true host galaxy’s redshift becomes more significant.
Additionally, the distribution of redshift determined by photometric data appears to be more smooth due to large photo-$z$ error, whereas the distribution determined by spectroscopic data shows larger fluctuations, implying a greater potential for constraining the Hubble constant.
Notably, because the quality of the statistical redshift distribution for the host galaxy is mainly dependent on the spatial localization precision provided by the \ac{GW} detection, we do not repeat the above illustration for other configurations where the sky localization is not significantly different.

It is slightly counter-intuitive that improved sky localization would result in a worse redshift estimate.
This is because the photo-$z$ estimate for any single galaxy is usually accompanied by a sizeable random error.
When the GW signal is localized to a larger area, it involves a greater number of galaxies, and the intrinsic clustering of galaxies effectively averages out the random error.
On the other hand, such averaging is less effective for more accurate sky localization.
However, we anticipate that future detections will trigger an interest in a comprehensive survey of galaxies within the GW localization error box.
Therefore, we expect that for events with a spatial resolution of less than $\Delta \Omega < 0.1 \ {\rm deg}^2$, spectroscopic redshift will be available for galaxies with an apparent magnitude $m_{\rm app} \leq +21 \ \rm mag$ \cite{Gong:2019yxt}.
In practice, we simply adopt the median value of the photo-$z$ as the ``true value'' of the spectroscopic redshift.

\section{Results}    \label{results}

\begin{figure}[t]
\centering
\includegraphics[height=8.cm, width=16.cm]{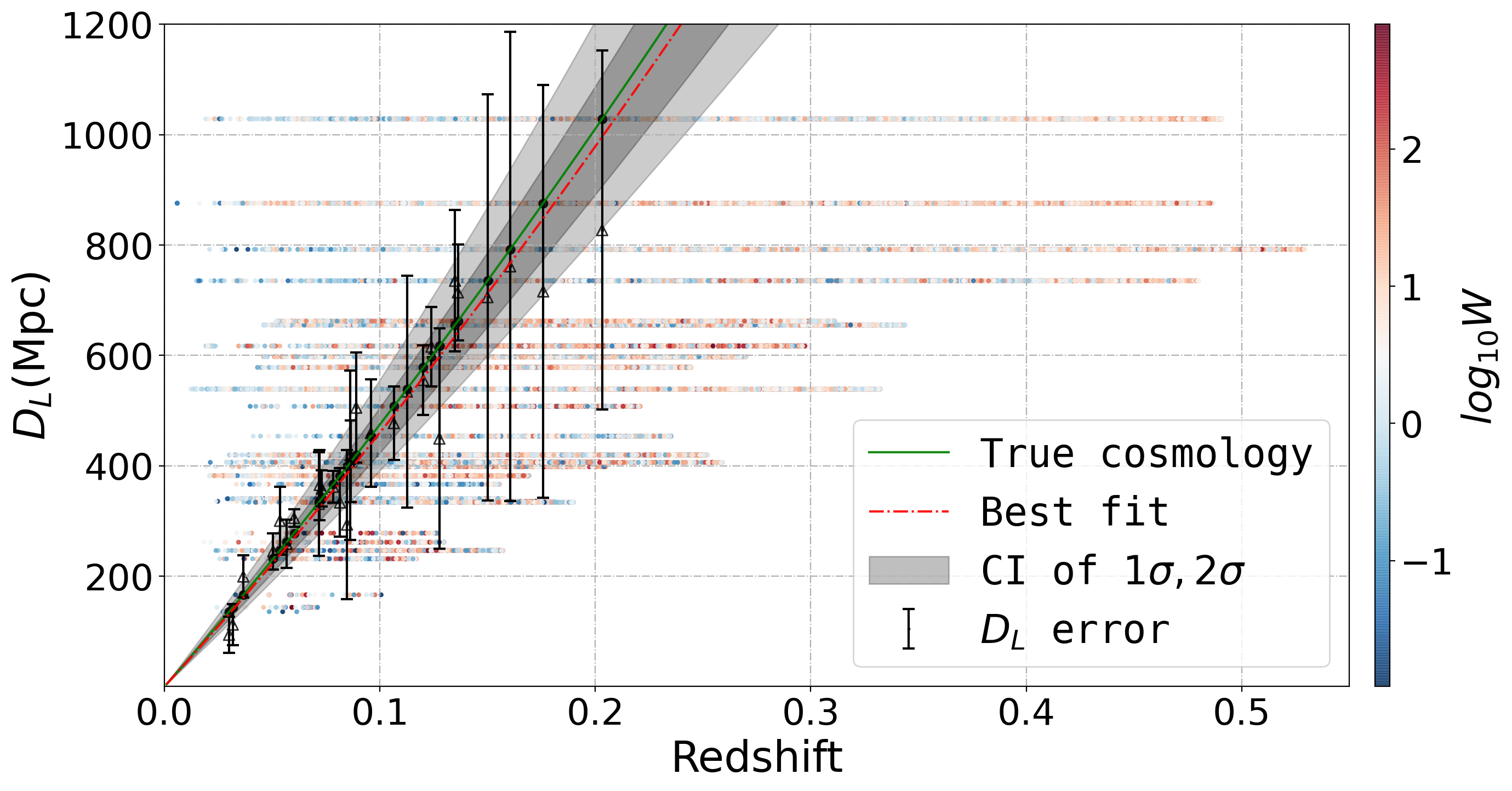}
\caption{Example of fitting the `$D_L - z$ relation' based on the dark standard siren observations of TianQin I+II.
The solid green line represents injected cosmological parameters $H_0$ and $\Omega_M$.
The dotted-dashed red line represents the most probable cosmology, while the two shaded areas represent confidence intervals of $68.27\%$ ($1 \sigma$) and $95.45\%$ ($2\sigma$), respectively.
The black error bars represent measurement errors in determining the luminosity distance to \ac{GW} sources.
  The median hollow black triangles in each error bar represent the mean of the $D_L$ measurement with random statistical deviation, while the solid black dot represents the true value of $D_L$.
  The horizontal colored dots represent the redshift of selected galaxies for that particular source. The color hues from blue to red show the logarithms of the total weights of galaxies, and we assigned a value to the $D_L$ of each candidate host galaxy equal to the true $D_L$ of the \ac{GW} source.  }
\label{final_costraint}
\end{figure}

To extract the Hubble constant from the dark standard sirens, we use a Markov chain Monte Carlo (MCMC) algorithm --- \textsf{emcee} package, which is a \textsf{Python} implementation of an affine-invariant MCMC ensemble sampler \cite{ForemanMackey:2012ig, ForemanMackey:2019ig}.
In FIG. \ref{final_costraint}, we illustrate how to estimate cosmological parameters using the \ac{SBBH} \ac{GW} detections of TianQin I+II.
We demonstrate that although a single event may suffer from large uncertainties due to its inability to identify the host galaxy, but a number of events can cancel out the random error and result in a more precise estimate of the Hubble constant.

Because of the uncertainty related to the luminosity distance, the redshift ($\Delta z_{\rm photo}$ and $\Delta z_{\rm pv}$), and the cosmological model, a single \ac{GW} event is frequently associated with a large number of galaxies with a wide range of redshifts (see Eq. (\ref{DL_min}, \ref{DL_max})), which are shown as the horizontally distributed dots in FIG. \ref{final_costraint}.
After constraining cosmological parameters, the redshift range of candidate host galaxies of the \ac{GW} source will be considerably reduced, which is referred to as posterior redshift.
The candidate host galaxies within the posterior redshift range ultimately determine the precision with which cosmological parameters can be estimated. 
The other candidate host galaxies outside the posterior redshift are just interference sources, and we cannot exclude them when the cosmological parameters are not constrained.

The results of constraints on $H_0$ using \ac{SBBH} \ac{GW} signals are shown in this section for different detector configurations, including TianQin, TianQin I+II, TianQin+LISA, TianQin I+II+LISA, TianQin+ET, and TianQin I+II+ET.
To alleviate the random effect of random realization, we repeat the calculation on 48 random realizations with different random seeds for each configuration and the two weighting methods. 
Furthermore, when reporting the constraining on $H_0$, the result is marginalized over $\Omega_M$ instead of being fixed at a specific $\Omega_M$ value.

\subsection{TianQin and TianQin I+II }    \label{results-1TQ}

\begin{figure}[htbp]
\centering
\includegraphics[height=8.cm, width=15.cm]{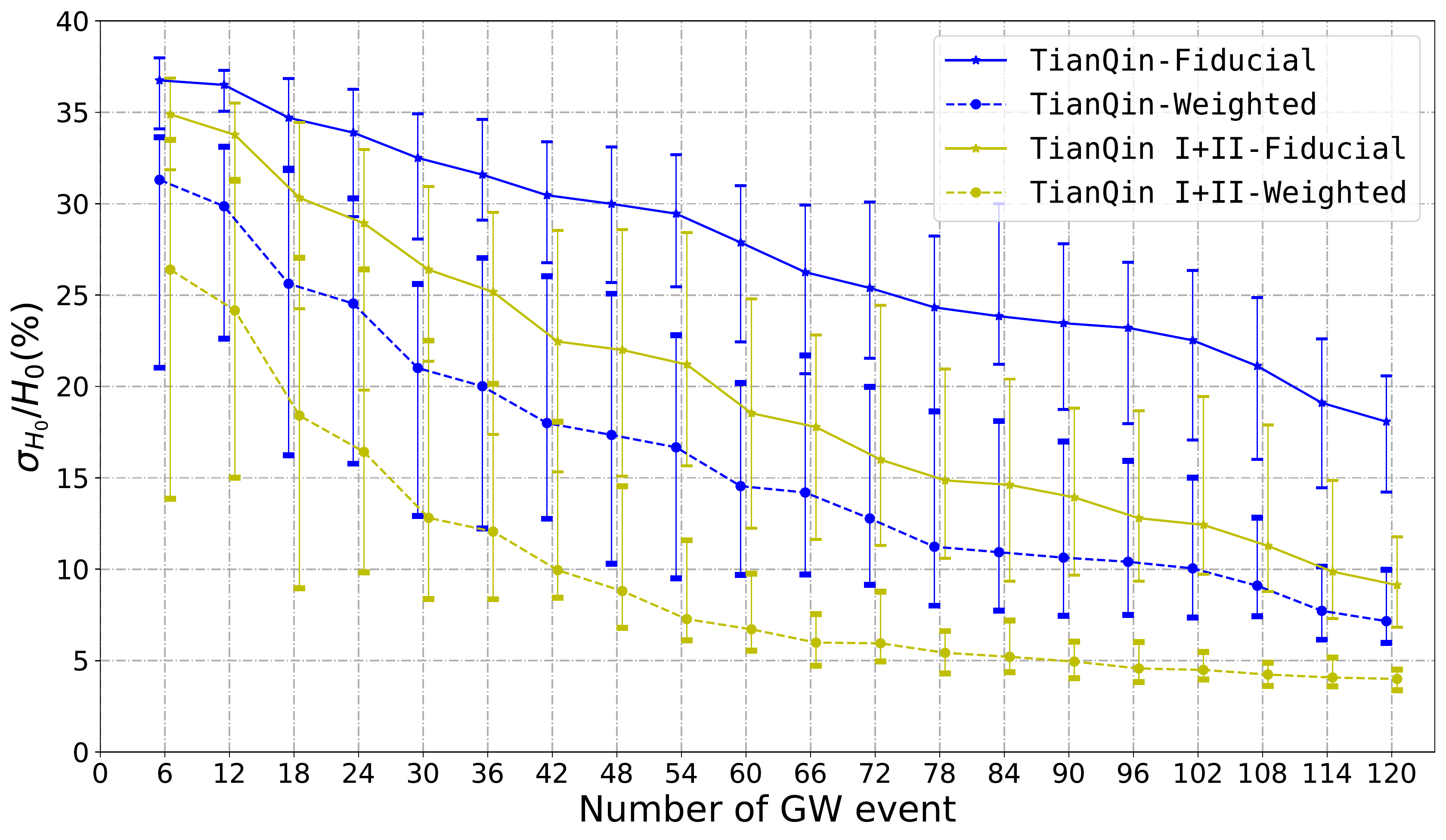}
\caption{Dependence of constraint precision of $H_0$ on the numbers of \ac{GW} events for TianQin (blue) and TianQin I+II (yellow).
The fiducial method and the weighted method are shown in solid and dashed lines, respectively.
  Each error bar represents a $68.27\%$ interval from 48 independent simulations. 
  The lines have been slightly shifted to improve the visual presentation. }
\label{H0_12TQ}
\end{figure}

The current studies of the population properties of \ac{SBBH} merger events observed by LIGO and Virgo have revealed a large uncertainty in the merger rate of \acp{SBBH} \cite{LIGOScientific:2018mvr, LIGOScientific:2018jsj, Abbott:2020niy, LIGOScientific:2020kqk}, which also results in large uncertainty in the prediction of the detection rate of \ac{GW} events derived from \acp{SBBH}.
We studied the variation of the constraint precision of $H_0$ with the number of \ac{GW} events in a range of $6$ to $120$ events to avoid the influence of the detection rate of \ac{GW} events on the constraint precision of $H_0$ and fully display the potential of detectors.
The constraints on $H_0$ for TianQin and TanQin I+II are shown in FIG. \ref{H0_12TQ}.
The uncertainty of $H_0$ shrinks as the number of detected \ac{GW} events increases.
In comparison to the fiducial method, the weighted method can improve the constraining precision of $H_0$ by a factor of about $2$.

\begin{figure}[htbp]
\centering
\includegraphics[height=7.5cm, width=7.5cm]{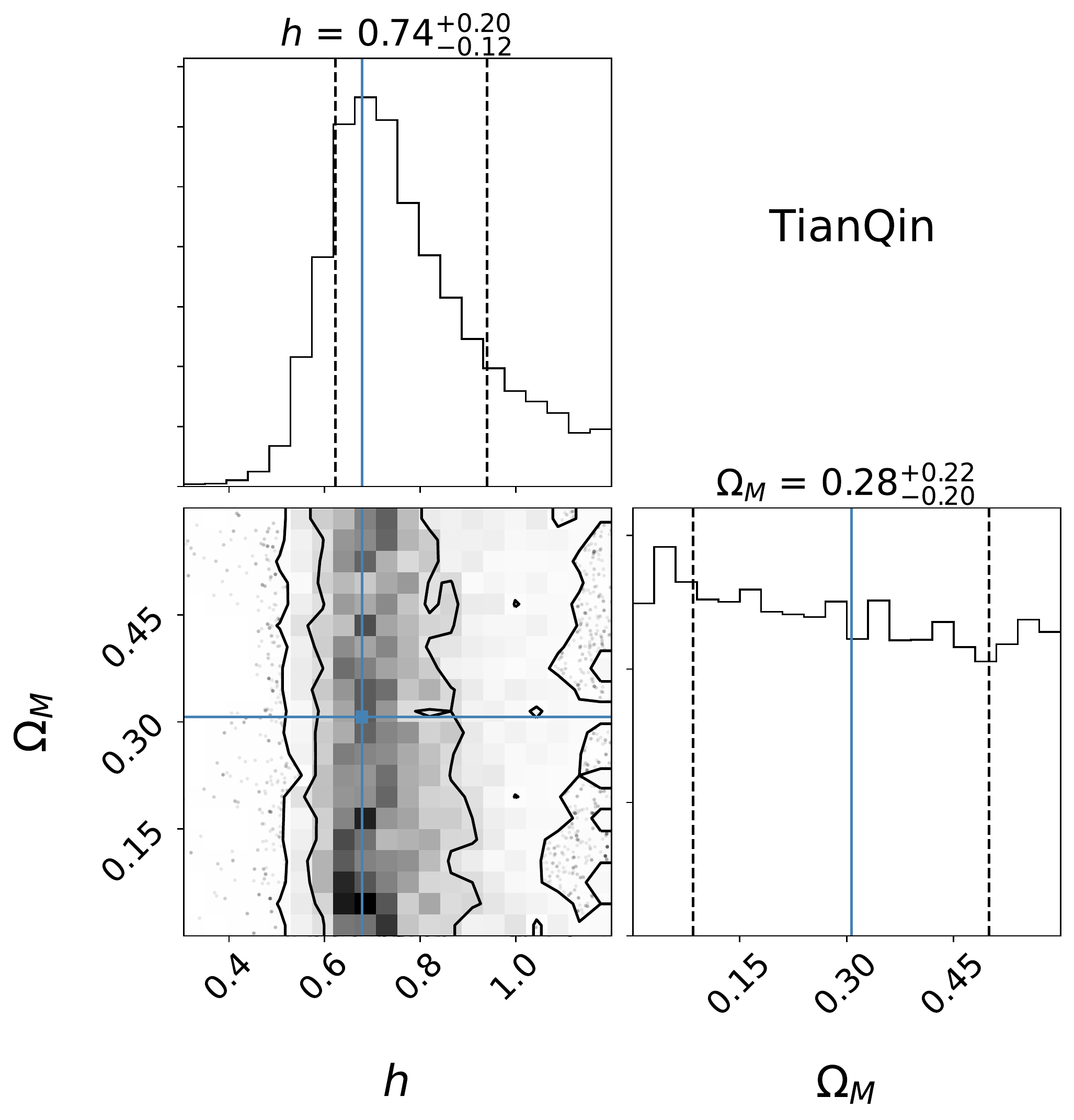} ~~~
\includegraphics[height=7.5cm, width=7.5cm]{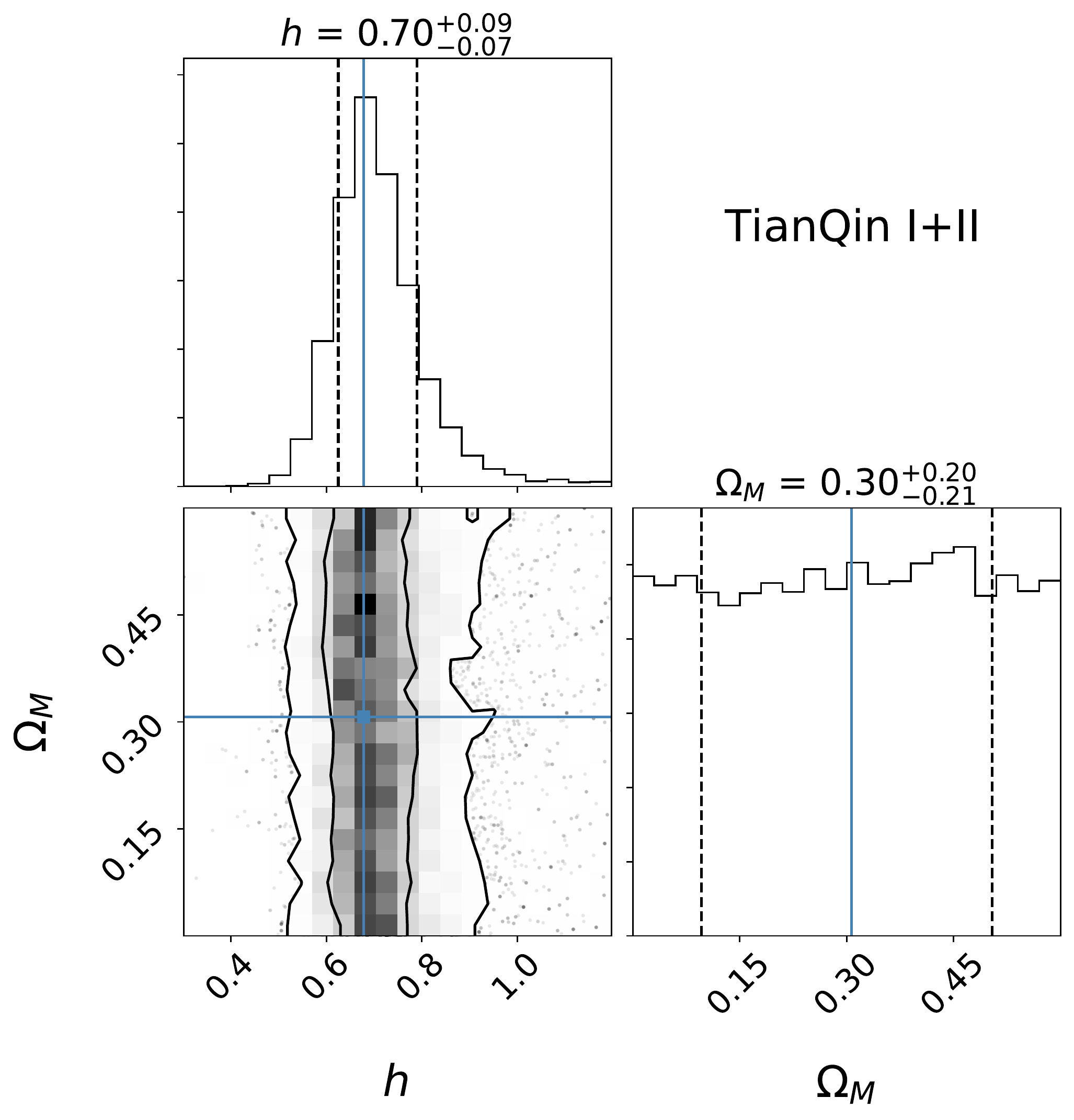}
    \caption{Example posterior probabilities of the parameters $h$ and $\Omega_M$ calculated using the weighted method for TianQin (left) and TianQin I+II (right).
In each plot, the lower left panel shows the joint posterior probability of $h$ and $\Omega_M$, with the contours representing confidence levels of $1 \sigma (68.27\%)$ and $2 \sigma (95.45\%)$, respectively; the upper and right panels show the marginalized posterior distribution of the same parameters, with the dashed lines indicating a $1 \sigma$ credible interval.
In each panel, the solid cyan lines mark the true values of the parameters.  }
\label{H0_example_12TQ}
\end{figure}

TianQin can detect approximately $11$ \ac{SBBH} \ac{GW} events over 5 years of operation (as illustrated in TABLE \ref{Detection_rate}), with which one can constrain $H_0$ to a precision of approximately $36.8\%$ using the fiducial method and approximately $30.9\%$ using the weighted method, respectively.
Due to the small number of \ac{GW} events, the constraints on cosmological parameters are very imprecise. 
A typical constraint result of the parameters $h$ ($h \equiv \frac{H_0}{100 \ {\rm km/s/Mpc}}$) and $\Omega_M$ from TianQin using the weighted method is shown in the left plot of FIG. \ref{H0_example_12TQ}.

The constraint on $H_0$ with TianQin I+II is tighter than that with TianQin, as shown in FIG. \ref{H0_12TQ}.
Given the same number of \ac{GW} events and the weighting method of candidate host galaxies, TianQin I+II can significantly improve the constraint of $H_0$.
This improvement is mainly due to the more accurate spatial localization provided by TianQin I+II, as shown in FIG. \ref{GW_error}.
Using the fiducial method and the weighted method, approximately $28$ \ac{GW} events are expected to be detected by TianQin I+II, and the precision of $H_0$ is expected to be approximately $26.3\%$ and $14.4\%$, respectively.
A typical cosmological parameter estimation using the weighted method for TianQin I+II is shown in the right plot of FIG. \ref{H0_example_12TQ}.
The non-Gaussian tail of the posterior probability distribution of $H_0$ becomes shorter as the rate of \ac{GW} detection increases and the spatial localization of \ac{GW} sources becomes more precise.
Additionally, when using \ac{SBBH} \ac{GW} events, either for TianQin or TianQin I+II, there are no effective constraints on $\Omega_M$ parameter.

It is worth noting that the precision of $H_0$ estimation with TianQin using the weighted method is higher than that obtained using the fiducial method with TianQin I+II, indicating that in comparison to improved spatial localization, the more accurate weighting method has a larger impact on $H_0$ estimation.

\subsection{Multi-detector network of TianQin and LISA}    \label{results-TQLISAnetwork}

\begin{figure}[htbp]
\centering
\includegraphics[height=7.2cm, width=12.cm]{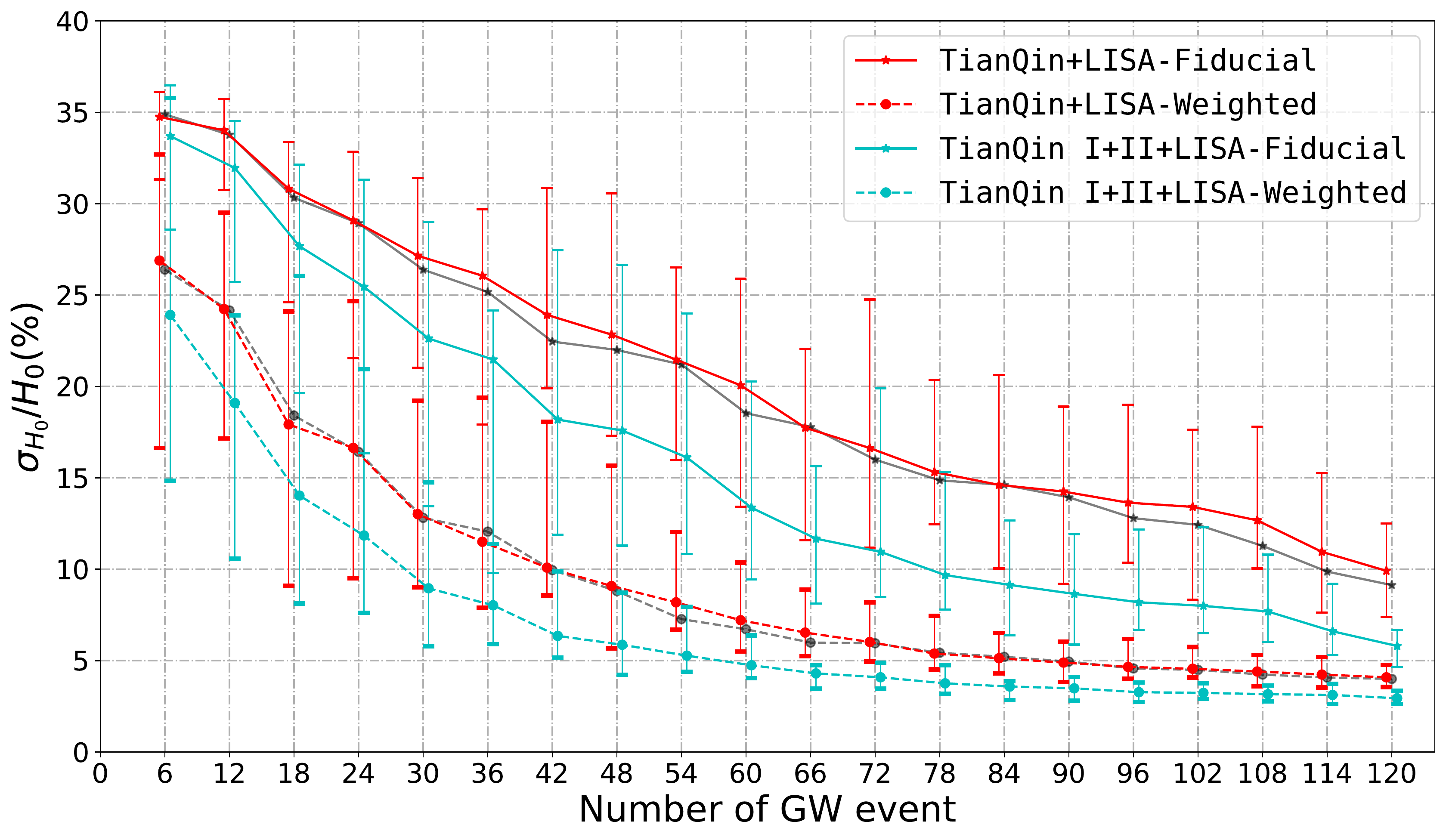}
\caption{Dependence of constraint precision of $H_0$ on the numbers of \ac{GW} events for TianQin+LISA (red) and TianQin I+II+LISA (cyan).
The fiducial method and the weighted method are shown in solid and dashed lines, respectively.
Each error bar represents a $68.27\%$ interval generated by 48 independent simulations.
For comparison, similar results for TianQin I+II (gray) are also shown. 
The lines have been slightly shifted to improve the visual presentation.  }
\label{H0_12TQLISA}
\end{figure}

\begin{figure}[htbp]
\centering
\includegraphics[height=7.5cm, width=7.5cm]{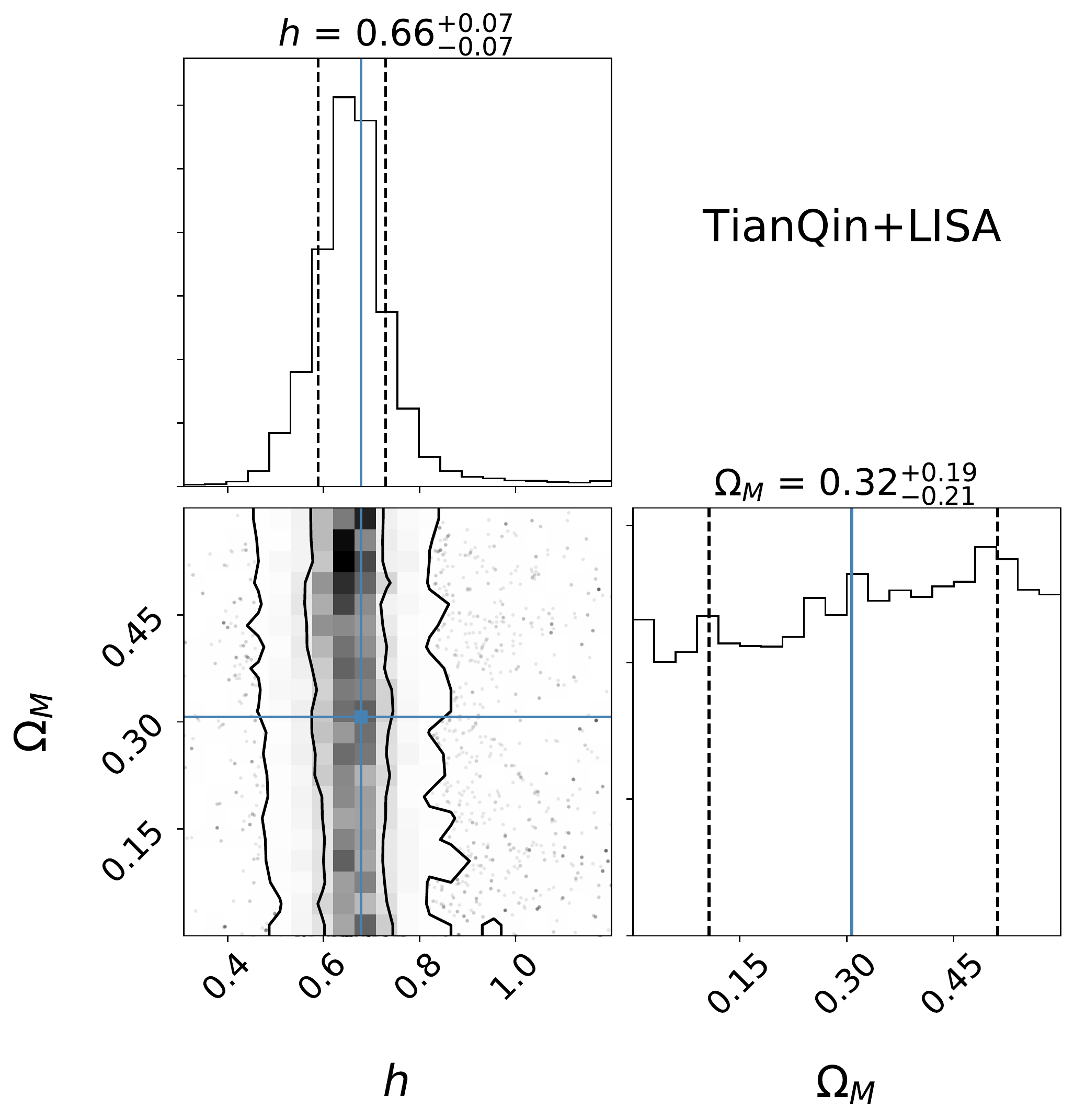} ~~~
\includegraphics[height=7.5cm, width=7.5cm]{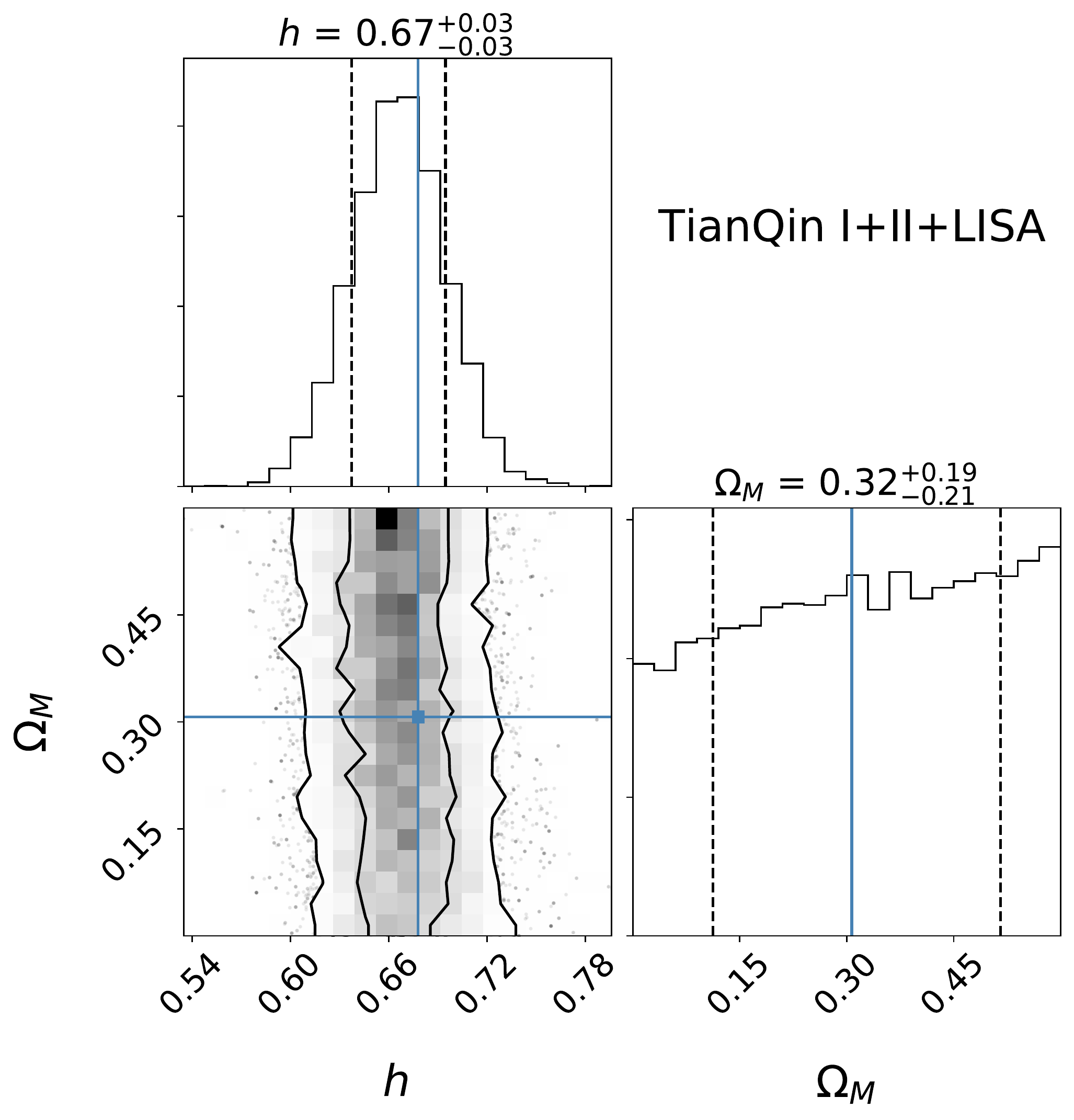}
\caption{Example posterior probabilities of the parameters $h$ and $\Omega_M$ calculated using the weighted method for TianQin+LISA (left) and TianQin I+II+LISA (right).
Each plot represents the joint posterior probability of $h$ and $\Omega_M$, with contours representing confidence levels of $1 \sigma (68.27\%)$ and $2 \sigma (95.45\%)$, respectively; the upper show the marginalized posterior distribution of the same parameters, with the dashed lines representing the $1 \sigma$ credible interval.
In each panel, the solid cyan lines mark the true values of the parameters.  }
\label{H0_example_12TQLISA}
\end{figure}

We then analyze scenarios in which LISA is added to the detector network.
In FIG. \ref{H0_12TQLISA}, we show the dependence of the precision of $H_0$ versus detection numbers, assuming TianQin+LISA and TianQin I+II+LISA.
Again we observe that increasing the number of detections results in a more precise $H_0$ measurement.
In FIG. \ref{H0_example_12TQLISA}, we give a representative joint posterior probability of $h$ and $\Omega_M$ using the weighted method under the expected total detection rates (see TABLE \ref{Detection_rate} for details).
We notice that TianQin+LISA has very similar constraining power to TianQin I+II because the two detection configurations have very similar localization precisions (see FIG. \ref{GW_error}).
Meanwhile, TianQin I+II+LISA can reduce uncertainty on $H_0$ by half as a result of improved localization.
On the other hand, neither of the two detector network configurations, imposes any meaningful constraint on $\Omega_M$.

For TianQin+LISA (TianQin I+II+LISA), about $38 ~(64)$ \ac{GW} events are expected to be detected, and the precision of $H_0$ can be constrained to approximately $25.2\% ~(12.2\%)$ and approximately $11.6\% ~(4.1\%)$ by using the fiducial method and the weighted method, respectively.
Similar to the case of TianQin and TianQin I+II, the constraint of $H_0$ by TianQin+LISA using the weighted method is stronger than that by TianQin I+II+LISA using the fiducial method. 

Space-borne \ac{GW} detectors/networks have excellent sky localizing capabilities, which enables the constraints on the Hubble constant through the observation of \ac{SBBH} inspirals.
However, such \ac{SBBH} inspiral \ac{GW} signals are relatively quieter; the relatively lower \ac{SNR} leads to a larger uncertainty on luminosity distance, $\sigma_{D_L}/D_L \gtrsim 0.1$.
This fact limits the expected precision of $H_0$ from the \ac{SBBH} inspiral observation with space-borne detectors.

\subsection{Multi-band detection with TianQin and ET}    \label{results-TQETmuntiband}

A multi-band \ac{GW} observation of \acp{SBBH} can significantly improve the constraining on the Hubble constant under the dark standard siren scenario since the multi-band observation combines advantages of both space-borne and ground-based detectors.
Space-borne \ac{GW} detectors can provide accurate sky localization, while ground-based detectors' high \ac{SNR} enables more precise luminosity distance estimation.
This combination results in a smaller localization error box that is less susceptible to contamination from neighbouring galaxies, resulting in a more precise estimation of the redshift.

\begin{figure}[htbp]
\centering
\includegraphics[height=7.2cm, width=12.cm]{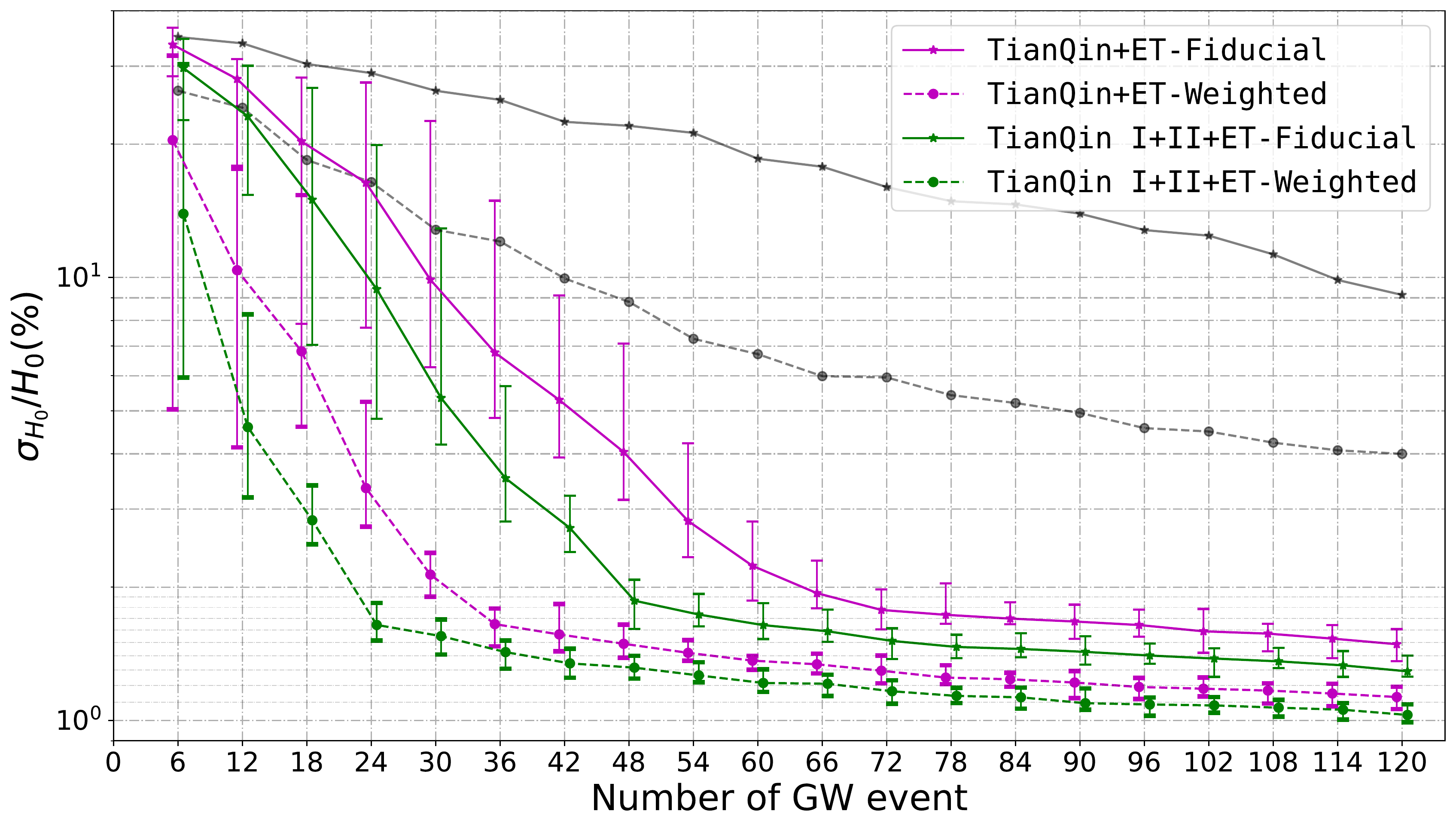}
\caption{Dependence of constraint precision of $H_0$ on the numbers of \ac{GW} events for TianQin+ET (magenta) and TianQin I+II+LISA (green).
The fiducial method and the weighted method are shown in solid and dashed lines, respectively.
  Each error bar represents a $68.27\%$ interval from 48 independent simulations.
  Comparable results for TianQin I+II (gray) are also provided. 
  The lines have been slightly shifted to improve the visual presentation. }
\label{H0_12TQET}
\end{figure}

\begin{figure}[htbp]
\centering
\includegraphics[height=7.5cm, width=7.5cm]{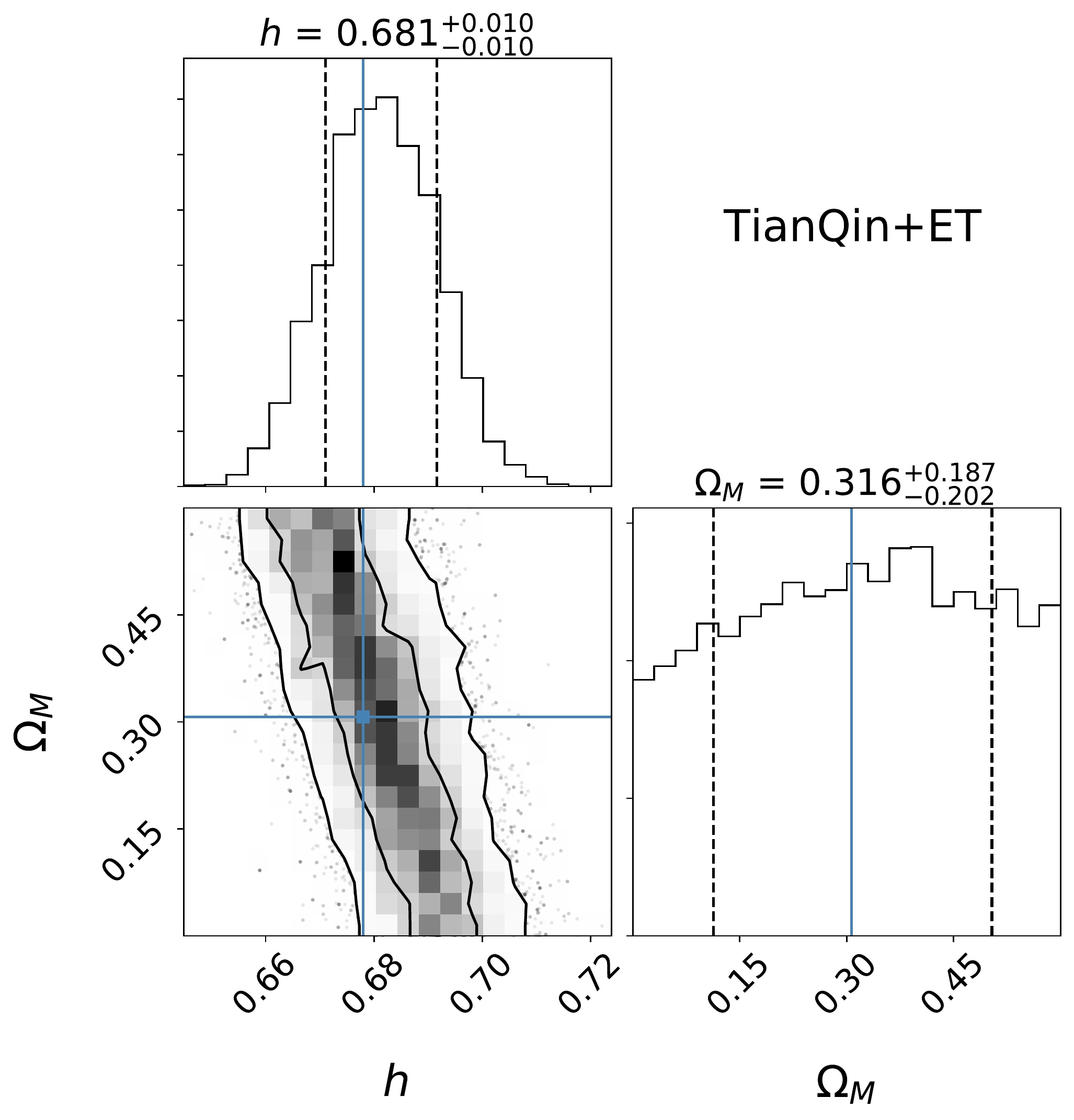} ~~~
\includegraphics[height=7.5cm, width=7.5cm]{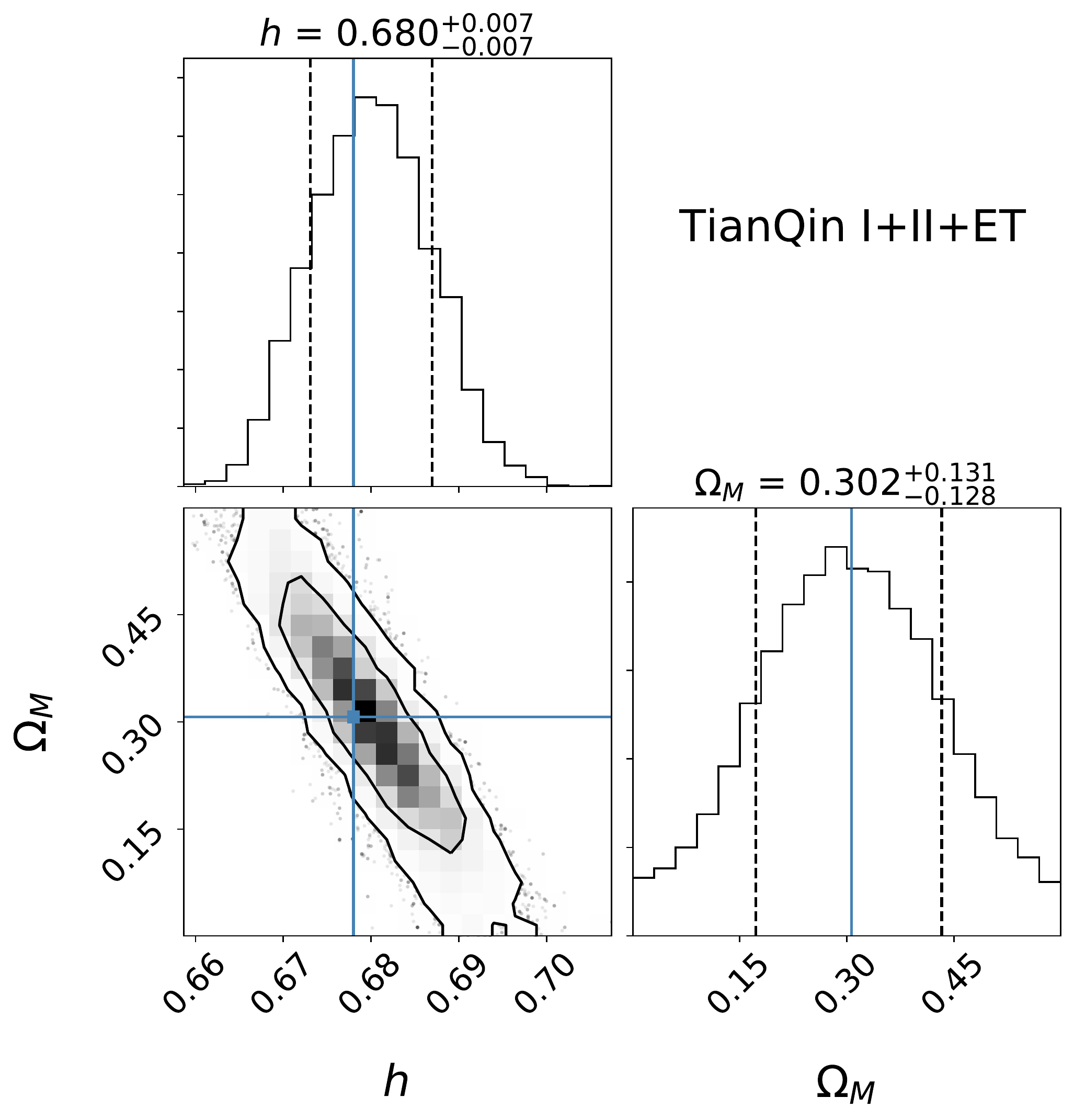}
    \caption{Example posterior probabilities of the parameters $h$ and $\Omega_M$ calculated using the weighted method for TianQin+ET (left) and TianQin I+II+ET (right).
In each plot, the lower left panel shows the joint posterior probability of $h$ and $\Omega_M$, with the contours representing confidence levels of $1 \sigma (68.27\%)$ and $2 \sigma (95.45\%)$, respectively; the upper and right panels show the marginalized posterior distribution of the same parameters, with the dashed lines indicating a $1 \sigma$ credible interval.
In each panel, the solid cyan lines mark the true values of the parameters.  }
\label{H0_example_12TQET}
\end{figure}

In FIG. \ref{H0_12TQET}, we present the dependence of constraint precision of $H_0$ versus different detection numbers, under the multi-band \ac{GW} detector networks TianQin+ET and TianQin I+II+ET.
Notably, the TianQin I+II lines (gray) approximate a power-law relationship, while the lines for multi-band networks indicate a saturation from relative uncertainties around $1.8\%$, which also corresponds to a turning point for the trend of the lines, respectively.
We observe that the multi-band network can quickly increase the precision of $H_0$ as more events are observed, but after the turning point, the precision improves only by $1/\sqrt{N}$, where $N$ is the number of \ac{GW} events.
Moreover, the precision is hard to reach the level of $1 \%$ even after 100 \ac{GW} events are detected.
Additionally, we notice that when adopting the weighted method, better precision leads to a quicker approach to saturation, which occurs in around 24 events for the TianQin I+II+ET network.

For TianQin+ET, about $44$ \ac{GW} events are expected to be detected via multiple band observation (as illustrated in TABLE \ref{Detection_rate}), and the precision of $H_0$ can be constrained to approximately $4.7\%$ and approximately $1.5\%$ by using the fiducial method and the weighted method, respectively.
Moreover, if TianQin I+II+ET is realized, approximately $112$ \ac{GW} events can be detected using multi-band observations in the optimistic case, and the precision of $H_0$ can be constrained to $\sim 1 \%$, using either the fiducial method (about $1.3 \%$) or the weighted method (about $1.1 \%$).
Constraining $H_0$ to a precision of $1 \%$ would be very exciting as it has the potential to shed light on the Hubble tension .

The typical constraints on $h$ and $\Omega_M$ using the weighted method for the multi-band networks are shown in FIG. \ref{H0_example_12TQET}.
The introduction of multi-band GW observation can greatly enhance the constraining power on the cosmological parameter $\Omega_M$.
While the fiducial method does not deliver significant and robust estimation, the weighted method can constrain $\Omega_M$ to a precision of $0.13$, or equivalently $42\%$ for relative uncertainty, with the TianQin I+II+ET network.

\section{Discussions}    \label{discussion}

\subsection{Important role of spectroscopic redshift}     \label{spectroscopic-redshift}

Considering that the sky localization error of almost all \ac{GW} sources is less than $10 \ {\rm deg}^2$, it may be possible to use current or future spectroscopic observation facilities, such as the Large Sky Area Multi-Object Fiber Spectroscopic Telescope \cite{1998SPIE.3352...76S, Cui_2012, 2012RAA....12..723Z}, the Dark Energy Spectroscopic Instrument \cite{Aghamousa:2016zmz}, the 4-meter Multi-Object Spectroscopic Telescope (4MOST) \cite{deJong:2012nj}, the TAIPAN \cite{daCunha:2017wwy}, the Chinese Space Station Telescope \cite{Gong:2019yxt}, the James Webb Space Telescope \cite{Gardner:2006ky, Kalirai:2018qfg}, and the Wide-Field Infrared Survey Telescope \cite{Green:2012mj, Chary:2020msh}, to perform galaxy survey and provide accurate estimations of redshifts for the galaxies locate in the error box.
In this section, we investigate to what extent a catalog of galaxies with spectroscopic redshifts can improve the constraint precision of $H_0$.
TABLE \ref{accuracies_summary} summarizes the constraint precision of $H_0$ under various detector configurations and under two assumptions that the survey galaxy catalog with photo-$z$ and with assumed spectroscopic redshift, respectively.

\begin{table}[]
  \caption{Expected constraint precision of $H_0$ under six detector configures, assuming the survey galaxy catalog with photo-$z$ and with spectroscopic redshift. The error represents $68.27\%$ confidence interval.  }
    \vspace{12pt}
    \renewcommand\arraystretch{1.5}
    \centering
    \begin{tabular}{|c|cc||cc|}
        \hline
        \multirow{3}*{Network }  & \multicolumn{4}{c|}{Constraint precision $\sigma_{H_0}/H_0$ (\%)}  \\
        \cline{2-5}
        ~  & \multicolumn{2}{c||}{Using photo-$z$ catalog} & \multicolumn{2}{c|}{Using spectroscopic redshift catalog}    \\
        \cline{2-5}
        \thead[c]{configuration}  &  \thead[c]{~~Fiducial method}  & \thead[c]{Weighted method}  &  \thead[c]{~~ Fiducial method}  & \thead[c]{Weighted method}  \\
        \hline
        TianQin            & ~~ $36.8_{-1.8}^{+1.7}$  & $30.9_{-5.1}^{+5.1}$    & ~~ $29.7_{-6.2}^{+5.4}$  & $22.2_{-9.8}^{+8.3}$         \\
        \hline
        TianQin I+II       & ~~ $26.3_{-6.9}^{+5.9}$  & $14.4_{-7.6}^{+9.4}$    & ~~ $15.1_{-6.9}^{+9.4}$  & $7.9_{-3.2}^{+2.9}$         \\
        \hline
        \hline
        TianQin+LISA       & ~~ $25.2_{-7.6}^{+6.4}$  & $11.6_{-6.4}^{+8.2}$    & ~~ $12.4_{-5.4}^{+6.2}$  & $6.4_{-2.4}^{+1.6}$         \\
        \hline
        TianQin I+II+LISA  & ~~ $12.2_{-5.3}^{+5.3}$  & $4.1_{-0.8}^{+0.8}$    & ~~ $5.7_{-2.0}^{+1.5}$  & $3.3_{-0.7}^{+0.7}$         \\
        \hline
        \hline
        TianQin+ET         & ~~ $4.67_{-1.60}^{+1.80}$  & $1.51_{-0.13}^{+0.13}$    & ~~ $2.10_{-0.63}^{+0.25}$  & $1.32_{-0.09}^{+0.10}$         \\
        \hline
        TianQin I+II+ET    & ~~ $1.32_{-0.09}^{+0.07}$  & $1.05_{-0.06}^{+0.05}$    & ~~ $1.08_{-0.09}^{+0.10}$  & $0.95_{-0.08}^{+0.08}$         \\
        \hline
    \end{tabular}
    \label{accuracies_summary}
\end{table}

We observe that the constraint precision of $H_0$ is quite low for TianQin.
This is because the expected detection rate is low; the detected events are insufficient to suppress random fluctuations, and so the effect of spectroscopic redshift is not significant.
However, for a network of detectors, one might anticipate a greater number of detected \ac{GW} events, which emphasizes the uncertainty associated with the galaxy redshift measurement. 
We observe that employing spectroscopic redshift can significantly improve the precision of $H_0$ by a factor of $2$.
However, for TianQin I+II+ET, the saturation described in Section \ref{results-TQETmuntiband} allows for extremely exact estimation of $H_0$ even without the spectroscopic redshift.
But still, by using the weighted method and the spectroscopic redshift, one can expect the $H_0$ to be constrained to a level better than $1 \%$.

\subsection{Information gained from multiple band photometry information}     \label{multiband-information}

\begin{figure}[htbp]
\centering
\includegraphics[width=0.90\textwidth]{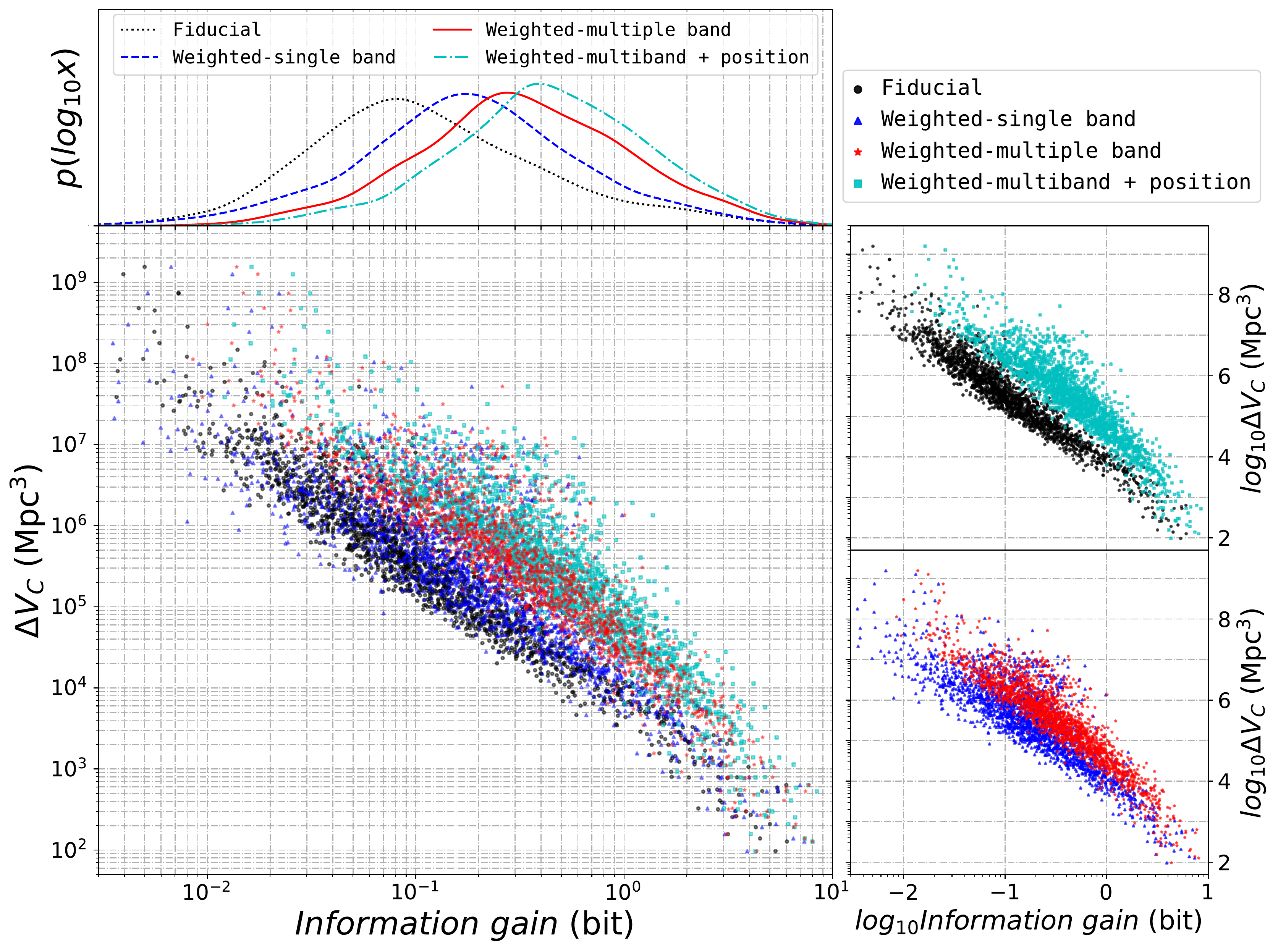}
\caption{Distribution of statistical redshift information gains for various weighting methods. The black, blue, red, and cyan dots or lines represent the results obtained by the fiducial method, single-band luminosity weighting, multiple band luminosities weighting, and the position plus multi-band luminosity weighting methods, respectively.
This figure contains 2000 mock GW events detected from TianQin. 
The top panel shows the distribution of information gains obtained using these four methods, the bottom left panel shows the scatter distribution of information gain $-$ localization co-moving volume of candidate host galaxies $(\Delta V_C)$, the right panels decompose the scatters for better distinction. }
\label{fig:informationGain}
\end{figure}

The use of multi-band luminosity can aid in our comprehension of the redshift information.
In this paper, we quantify the role of the multi-band luminosity information in this process.
For different weighting scenarios, we calculate the {\it information gain} of statistical redshift distributions on top of the prior distribution.
The information gain is defined as \cite{Sivia:2006book}
\begin{equation}   \label{information_gain}
\mathcal{H} = \int p(z|d_{\rm GW}, d_{\rm survey},\vec{H},I) \log_{2} \left[ \frac{ p(z|d_{\rm GW}, d_{\rm survey},\vec{H},I) }{ p_c(z| \vec{H}, I)}  \right] \D z ,
\end{equation}
where $p(z|d_{\rm GW}, d_{\rm survey},\vec{H},I)$ represents the posterior of redshift for candidate host galaxies of GW sources, $d_{\rm survey}$ represents data of galaxy survey, and $p_c(z| \vec{H}, I)$ is the prior distribution on redshift, which is defined by Eq. (\ref{p_noLSS}).

We illustrate the information gains on statistical redshift in FIG. \ref{fig:informationGain}, for four weighting methods: (1) the fiducial method, (2) single-band luminosity weighting method, (3) multiple band luminosities weighting method, and (4) position plus multi-band luminosity weighting method.
A total of 2000 mock GW events are used.
As can be shown, all methods yield a greater information gain when the localization error is smaller.
With more information (in terms of luminosity and position), more information about statistical redshift can be gained.
The peak of the marginalized distribution (shown on the top panel) demonstrates that for the four methods, each step provides an information gain increase of roughly 0.1 bits.
Moreover, the information gain from the single-band luminosity weighting method is approximately the same for any of the five bands of $ugriz$.
In most cases, the weighting method of using both location and multi-band luminosities can yield an information gain of as high as about 0.5 bits.

\subsection{Potential to address the Hubble tension}     \label{H0-tension}

Astronomers are puzzled by the Hubble tension, which describes the inconsistency between the two typical Hubble constant measurement methods, reported as $74.03 \pm 1.42 \ {\rm km/s/Mpc}$ from \acp{SNIa} observations \cite{Riess:2019cxk} (also see Ref. \cite{Riess:2020fzl}) and  $67.4 \pm 0.5 \ {\rm km/s/Mpc}$ from the $Planck$ \ac{CMB} anisotropies measurements \cite{Planck:2018vyg}.
Both methods have remarkable accuracies of approximately $1.9\%$ and $0.7\%$, respectively.
\ac{GW} cosmology with dark standard sirens provides a fascinating new means of potentially explaining or resolving the Hubble tension, as it is expected to be much less impacted by systematic errors. 
However, an equally accurate measurement would be required to achieve so.
With the TianQin detectors, even in the most optimal scenario where we assume TianQin I+II with the weighted method and a galaxy catalog with spectroscopic redshift, the estimated precision of $H_0$ can only reach $7.9 \%$.
When LISA is included, the precision can reach the level of $3.3\%$.

On the other hand, a multi-band \ac{GW} detector network has a very prominent potential to estimate $H_0$ accurately.
One can expect an $H_0$ precision of $1.51\%$ and $1.05\%$ for TianQin+ET and TianQin I+II+ET, respectively, which is very promising for addressing, or at the very least, shedding light on the nature of the Hubble tension.

According to some studies suggest that the multi-detector \ac{GW} detections of MBHB mergers can also achieve precision close to (or even better than) $1 \%$ for $H_0$ measurement.
This level of precision is typically facilitated by the fact that the MBHB mergers may be registered with a very high \ac{SNR} using space-borne \ac{GW} missions \cite{Klein:2015hvg, Holz:2005df, Wang:2019, Feng:2019wgq, Ruan:2020smc, Ruan:2019tje}.
Meanwhile, since MBHB mergers may be detected at a very long distance, they may also be used to recover additional cosmological parameters in addition to $H_0$.
However, the precision of $H_0$ with MBHB detections is highly sensitive to the detection of neighboring events, which usually have a very low detection rate; the high precision is not guaranteed \cite{Zhu:2021aat, Wang:2020dkc}.
On the other hand, the higher detection rate of \acp{SBBH} indicates a highly likely prospect of using their inspiral \ac{GW} signals to constrain $H_0$ with high precision.

\subsection{P$-$P plot check}     \label{pp-plot}

\begin{figure}[htbp]
\centering
\includegraphics[height=8.5cm, width=15.cm]{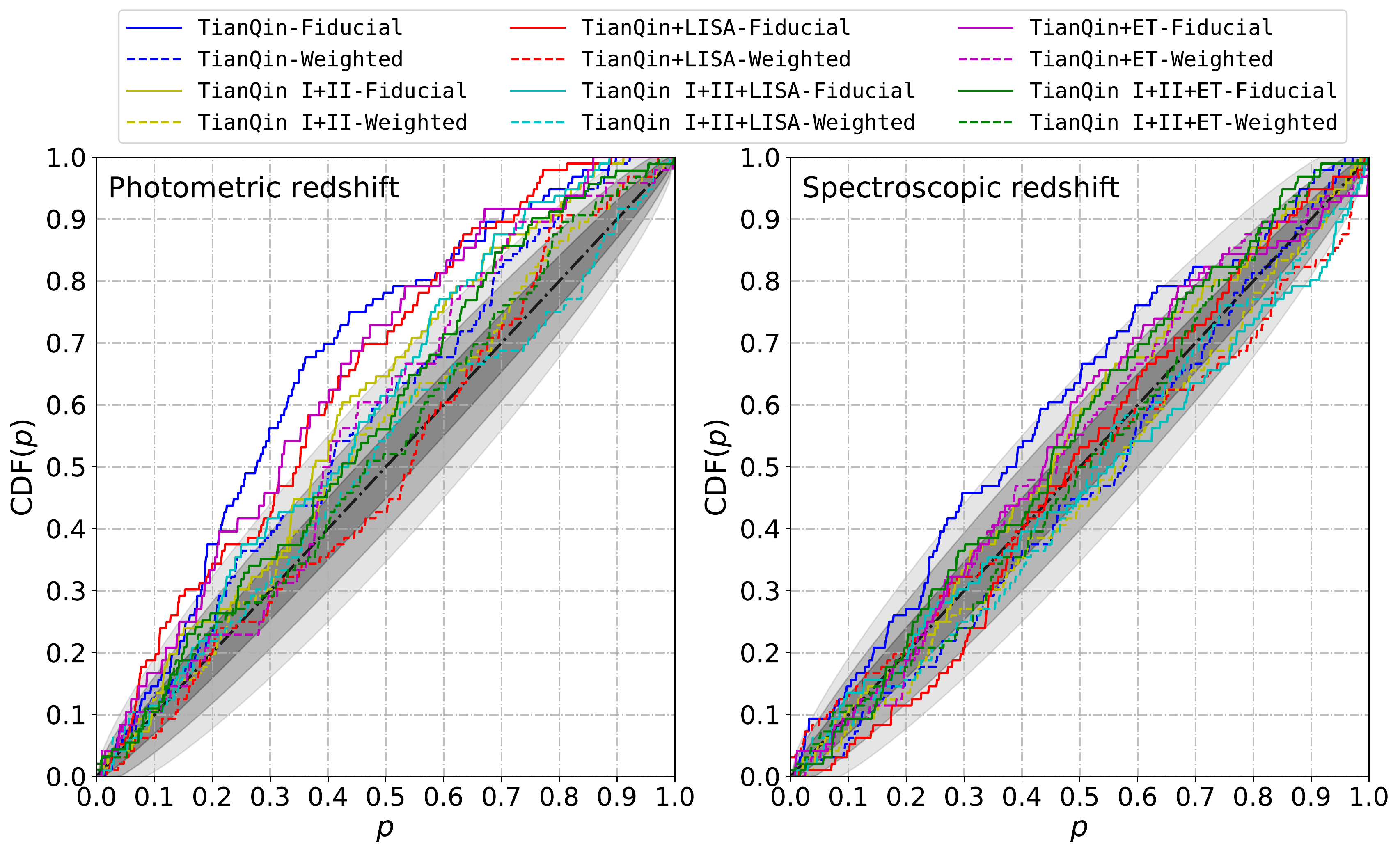}
\caption{P$-$P plot over $96$ independent simulations on the estimation of the Hubble constant for various detector configurations.
The $p$ on the $x$ axis is the credible level, and the CDF$(p)$ (cumulative distribution function, CDF) on the $y$ axis is the fraction of simulations with true parameters within the credible interval.
Results with different configurations are shown in different colors, with the fiducial method, the weighted method, and the diagonal line shown in solid, dashed, and dotted-dashed lines, respectively.
Different gray shadows represent $1 \sigma$, $2 \sigma$, and $3 \sigma$ confidence intervals, respectively.
The left panel makes the assumption that galaxies have photo-$z$ whereas the right panel makes the assumption that spectroscopic redshift is available. }
\label{PP_Plot}
\end{figure}

To compensate for the selection effect of galaxies observations, we adopt a correction factor in Eq. (\ref{N_term}) with the goal of obtaining an unbiased estimate of the Hubble constant.
We want to examine whether the conclusion is indeed unbiased, using both the weighted method and the fiducial method. 
To do this, we perform a series of P$-$P plots to ensure consistency, as shown in FIG. \ref{PP_Plot}.
The horizontal axis shows the credible level, while the vertical axis shows the fraction of simulations in which the true parameters are located within a certain credible interval.
For a fully consistent method, we expect the P$-$P plot to be alongside the diagonal line, with some degree of random fluctuations. 
For a more straightforward comparison, we perform Kolmogorov-Smirnov tests for different lines versus the diagonal line and present the calculated $p$-value in TABLE \ref{P_value}, with a larger $p$-value representing better consistency.

In the left panel, assuming that just photo-$z$ is available for the galaxies, the majority of the solid lines (or the fiducial method) deviate significantly from the diagonal, with a general pessimistic tendency. 
This is because the posterior probability of the cosmological parameters using the fiducial method does not gain sufficient redshift information from a large number of candidate host galaxies, resulting in a large credible interval, but the true value has most likely already been enclosed.
We observe, on the other hand, that the result from the weighted method (dashed lines) is more consistent with the diagonal line, fluctuating well within the $3\sigma$ confidence interval, which indicates a successful implementation of an unbiased estimator for the Hubble constant.

\begin{table}[htbp]
  \caption{Meta $p$-value of the distribution of $H_0$ estimations obtained from multiple independent simulations under various detector configurations, in comparison to an expected uniform distribution over $[0, 1]$. The null hypothesis is set so that the two distributions are consistent; therefore, a $p$-value less than $0.05$ indicates that the simulation results deviate significantly from the ideal result.   }
    \vspace{12pt}
    \renewcommand\arraystretch{1.2}
    \centering
    \begin{tabular}{|c|cc|cc|}
        \hline
        \multirow{3}*{Detector}  & \multicolumn{4}{c|}{ $p$-value }  \\
        \cline{2-5}
        ~  & \multicolumn{2}{c|}{Using photo-$z$ catalog} & \multicolumn{2}{c|}{Using spectroscopic redshift catalog}    \\
        \cline{2-5}
        \thead[c]{configuration}  &  \thead[c]{~~Fiducial method}  & \thead[c]{Weighted method}  &  \thead[c]{~~ Fiducial method}  & \thead[c]{Weighted method}  \\
        \hline
        TianQin      & ~~ $5.6 \times 10^{-9}$  & $0.0756$    & ~~ $9.8 \times 10^{-3}$  & $0.3756$         \\
        \hline
        TianQin I+II       & ~~ $3.8 \times 10^{-3}$  & $0.2511$    & ~~ $0.1774$  & $0.5081$         \\
        \hline
        TianQin+LISA       & ~~ $2.3 \times 10^{-5}$  & $0.2513$    & ~~ $0.2026$  & $0.2419$         \\
        \hline
        TianQin I+II+LISA  & ~~ $1.9 \times 10^{-3}$  & $0.6767$    & ~~ $0.1549$  & $0.4871$         \\
        \hline
        TianQin+ET         & ~~ $2.6 \times 10^{-3}$  & $0.1085$    & ~~ $0.1030$  & $0.2690$         \\
        \hline
        TianQin I+II+ET    & ~~ $0.0209$              & $0.4074$    & ~~ $0.2386$  & $0.5599$         \\
        \hline
    \end{tabular}
    \label{P_value}
\end{table}

If galaxies have spectroscopic redshifts, both methods’ P$-$P plots are statistically consistent with the diagonal for all configurations.
This finding emphasizes the importance of accurate measurement of galaxy redshifts in estimating the Hubble constant.

\subsection{Effect of eccentricity on multi-band GW detection}     \label{eccentricity-effect}

In this work, we focus on \ac{GW} sources with very small eccentricities ($e_0 = 0.01$ at $0.01$Hz).
This is a reasonable assumption for binary black holes generated via an isolated binary channel, as the GW radiation is expected to circularize the orbit.
However, if the \acp{SBBH} have a dynamical origin, they may retain higher eccentricities in the millihertz range \cite{OLeary:2005vqo, Kremer:2018tzm}, which may affect the multi-band \ac{GW} detection \cite{Barack:2003fp, Chen:2017gfm, Randall:2019znp}.
Indeed, the observations of GW190521 by LIGO and Virgo may suggest such a possibility \cite{Abbott:2020tfl, Gayathri:2020coq, Gayathri:2020fbl}.
For extreme eccentricities, the GW signals in the millihertz band may fall below the sensitivity of space-borne GW detectors \cite{Chen:2017gfm}.
Fortunately, the majority of \acp{SBBH} are not expected to be associated with extreme eccentricities \cite{Nishizawa:2016jji, Samsing:2018isx}, and the general conclusion of our study should not be altered by the inclusion of eccentricities.

\subsection{Other potential sources of constraint bias on the Hubble constant}     \label{potential-bias}

Numerous factors can introduce bias into the estimation of the Hubble constant. We describe the potential sources of the bias in this work as follows.
\begin{itemize}

  \item{The detectable \ac{SBBH} sources are primarily distributed in the $z \lesssim 0.2$, where Hubble's law can be used to approximate the expansion of the Universe.
Therefore, in the majority of cases, the model-dependent $\Omega_M$ has no effect on the outcomes of our $H_0$ analysis results. 
However, a small number of \ac{GW} events with $z \gtrsim 0.2$ show a certain constraint ability on $\Omega_M$, such as the constraint results of TianQin+ET and TianQin I+II+ET.
This implies that the cosmological model will introduce a bias when using higher redshift \ac{SBBH} events to constrain $H_0$ with high precision. }

  \item{In the \ac{GW} event catalog simulation, the \ac{GW} sources are randomly assigned to one of the galaxies in the survey galaxy catalog.
However, the survey galaxy catalog may not contain the true host galaxy of the \ac{GW} source in the actual $H 0$ estimation due to the selection effect such as the Malmquist bias.
The absence of the true redshift of the \ac{GW} source in the statistical redshift may also introduce a bias into the estimation of $H_0$. }

  \item{In the simulation setup, we use the median of photo-$z$ estimation as the true cosmological redshift of the galaxy.
However, the actual photo-$z$ measurements have an intrinsic scatter and deviation in relation to the spectroscopic redshifts \cite{Abazajian:2008wr, Abolfathi:2017vfu, Drlica-Wagner:2017tkk, DeVicente:2015kyp}. }

  \item{If some of the detected \acp{SBBH} have a primordial black hole genesis, \cite{DeLuca:2020qqa, Clesse:2020ghq, DeLuca:2020sae, DeLuca:2021wjr, Deng:2021ezy}, the concept of a host galaxy may lose relevance. 
Furthermore, drawing conclusions based on the assumption that all \acp{SBBH} are produced astronomically may be deceptive. }
\end{itemize}

Of course, there are methods to alleviate the bias.
For example, the bias of $H_0$ caused by the cosmological model can be avoided by fitting model-independent high-order expansion of the `$D_L - z$ relation' \cite{Visser:2003vq, Gong:2004sd}, or using the model-independent parameter estimation method such as Gaussian processes \cite{Seikel:2012uu, Cai:2016sby, Li:2019nux}.
The selection effect in the catalog of the galaxy survey can be corrected by conducting a follow-up deeper field galaxy survey triggered by the \ac{GW} detection \cite{Bartos:2014spa, Chen:2015nlv, Klingler:2019fbl}, utilizing both ground-based \cite{DES:2005dhi, Sevilla-Noarbe:2020jpu} and space telescopes \cite{Gong:2019yxt, Green:2012mj}.
By performing a follow-up spectroscopic measurement, the bias introduced by photo-$z$ can be eliminated.
Although the possibility of contamination by primordial black holes requires additional investigation, an internal consistency check as introduced in  \cite{Petiteau:2011we, Zhu:2021aat} may help in identifying and removing the anomaly.

\section{Conclusions and Outlook}    \label{conclusion-outlook}

In this work, we investigate the potential of using \ac{SBBH} \ac{GW} events (assuming the ``Power Law + Peak'' model) observed by space-borne \ac{GW} detectors as dark standard sirens to constrain the Hubble constant.
Several different detection scenarios are considered, for instance, single detectors, such as TianQin and TianQin I+II; multiple detector networks, such as TianQin+LISA and TianQin I+II+LISA; and multi-band \ac{GW} detector network, such as TianQin+ET and TianQin I+II+ET.
The redshift information is obtained statistically from the photometric survey galaxy catalog by matching the sky localization and possible redshift range of the \ac{SBBH} sources with the position and redshift of galaxies (we adopt the \ac{GWENS} catalog in this work).
Two methods are used to assign the weight of galaxies, including the fiducial method, in which each galaxy within the error box has a uniform weight, and the weighted method, in which the weight of a galaxy is proportional to the total stellar mass of the galaxy derived from the multi-band luminosity through the \hyperref[data:masses]{\tt Le Phare} method.

In comparison to single detector detection, both multi-detector networks and multi-band \ac{GW} observations can significantly improve detection rates and spatial localization.
As compared to the fiducial method, the weighted method can significantly improve the constraint precision of the Hubble constant.
Using the fiducial method and the weighted method, in the TianQin scenario, the constraint precision of $H_0$ is expected to be approximately $36.8 \%$ and approximately $30.9 \%$, respectively; in the TianQin I+II detection scenario, the precision of $H_0$ is expected to be approximately $26.3 \%$ and approximately $14.4 \%$, respectively.
Using the weighted method, the constraint precision of $H_0$ in the TianQin+LISA and TianQin I+II+LISA detection scenarios is expected to be approximately $11.6 \%$ and approximately $4.1 \%$ respectively.
In the multi-band \ac{GW} detection scenarios, the precision of the $H_0$ constrain by using the weighted method can approach around $1.5 \%$ and $1.1 \%$ under TianQin+ET configuration and TianQin I+II+ET configuration, respectively.
It should be emphasized that both the weighted method and the multi-band \ac{GW} detection significantly improve the constraint of $H_0$.

Apart from $H_0$, the space-borne \ac{GW} detector can hardly constrain any other cosmological parameter, such as $\Omega_M$, through the detections of \ac{SBBH} inspiral \ac{GW} signals.
However, other cosmological parameters can be constrained by other types of \ac{GW} sources, such as MBHB mergers  \cite{Petiteau:2011we, Tamanini:2016zlh, Zhu:2021aat}.
MBHB sources generally have a very high \acp{SNR}, and their event horizons can extend to high redshifts ($z \gg 1$). As a result, combining \ac{SBBH} and MBHB \ac{GW} observations enables a more thorough analysis of \ac{GW} cosmology study.

Additionally, we study the extent to which spectroscopic redshift information improves the constraining power of the Hubble constant, which increases not just the precision of the Hubble constant but also the accuracy of the constraint.
Finally, we evaluated the reliability of our Bayesian analysis framework and galaxy weighting method using the P$-$P plot method by performing a consistency test.

Certain concerns remain unresolved for future studies, such as incorporating a complete frequency response \cite{Zhang:2020khm} or the time-delay interferometry response \cite{Zhang:2019oet}.
MCMC can also be used in place of the \ac{FIM} method to obtain more realistic \ac{GW} parameter estimations \cite{Toubiana:2020cqv}.
The calibration uncertainty of the laser interferometer on the ground-based \ac{GW} detector might result in systematic errors in the luminosity distance measurement \cite{Karki:2016pht, Sun:2020wke, Estevez:2020pvj}, such as for ET.
We reserve such issues for future investigations.

\section*{Acknowledgements}

This work has been supported by the Guangdong Major Project of Basic and Applied Basic Research (Grant No. 2019B030302001),  the Natural Science Foundation of China (Grants  No. 12173104, 11805286, and 11690022), and the National Key Research and Development Program of China (No. 2020YFC2201400).
We acknowledge the use of the {\it Kunlun} cluster, a supercomputer owned by the School of Physics and Astronomy, Sun Yat-Sen University, and of the {\it Tianhe-2}, a supercomputer owned by the National Supercomputing Center in GuangZhou. 
The authors acknowledge the uses of the calculating utilities of \textsf{emcee} \cite{ForemanMackey:2012ig, ForemanMackey:2019ig}, \textsf{numpy} \cite{vanderWalt:2011bqk}, \textsf{scipy} \cite{Virtanen:2019joe}, and \textsf{LALSuite} \cite{lalsuite}, and the plotting utilities of \textsf{matplotlib} \cite{Hunter:2007ouj}, and \textsf{corner} \cite{corner}. 
The authors also thank Xiao-Dong Li, Martin Hendry, and Han Wang for helpful discussions.

\renewcommand{\refname}{Reference}
\bibliographystyle{unsrt}   
\bibliography{bibfile}   

\begin{thebibliography}{100}

\bibitem{Schutz:1986gp}
Bernard~F. Schutz.
\newblock {Determining the Hubble Constant from Gravitational Wave
  Observations}.
\newblock {\em Nature}, 323:310--311, 1986.

\bibitem{Planck:2015fie}
P.~A.~R. Ade et~al.
\newblock {Planck 2015 results. XIII. Cosmological parameters}.
\newblock {\em Astron. Astrophys.}, 594:A13, 2016.

\bibitem{Planck:2018vyg}
N.~Aghanim et~al.
\newblock {Planck 2018 results. VI. Cosmological parameters}.
\newblock {\em Astron. Astrophys.}, 641:A6, 2020.
\newblock [Erratum: Astron.Astrophys. 652, C4 (2021)].

\bibitem{Freedman:2017yms}
Wendy~L. Freedman.
\newblock {Cosmology at a Crossroads}.
\newblock {\em Nature Astron.}, 1:0121, 2017.

\bibitem{Freedman:2019jwv}
Wendy~L. Freedman, Barry~F. Madore, Dylan Hatt, Taylor~J. Hoyt, In~Sung Jang,
  Rachael~L. Beaton, Christopher~R. Burns, Myung~Gyoon Lee, Andrew~J. Monson,
  Jillian~R. Neeley, M.~M. Phillips, Jeffrey~A. Rich, and Mark Seibert.
\newblock The carnegie-chicago hubble program. {VIII}. an independent
  determination of the hubble constant based on the tip of the red giant
  branch.
\newblock {\em The Astrophysical Journal}, 882(1):34, aug 2019.

\bibitem{Freedman:2021ahq}
Wendy~L. Freedman.
\newblock {Measurements of the Hubble Constant: Tensions in Perspective}.
\newblock 6 2021.

\bibitem{Riess:2019cxk}
Adam~G. Riess, Stefano Casertano, Wenlong Yuan, Lucas~M. Macri, and Dan
  Scolnic.
\newblock {Large Magellanic Cloud Cepheid Standards Provide a 1\% Foundation
  for the Determination of the Hubble Constant and Stronger Evidence for
  Physics beyond $\Lambda$CDM}.
\newblock {\em Astrophys. J.}, 876(1):85, 2019.

\bibitem{Riess:2020sih}
Adam~G. Riess.
\newblock {The Expansion of the Universe is Faster than Expected}.
\newblock {\em Nature Rev. Phys.}, 2(1):10--12, 2019.

\bibitem{Riess:2020fzl}
Adam~G. Riess, Stefano Casertano, Wenlong Yuan, J.~Bradley Bowers, Lucas Macri,
  Joel~C. Zinn, and Dan Scolnic.
\newblock {Cosmic Distances Calibrated to 1\% Precision with Gaia EDR3
  Parallaxes and Hubble Space Telescope Photometry of 75 Milky Way Cepheids
  Confirm Tension with $\Lambda$CDM}.
\newblock {\em Astrophys. J. Lett.}, 908(1):L6, 2021.

\bibitem{Chen:2017rfc}
Hsin-Yu Chen, Maya Fishbach, and Daniel~E. Holz.
\newblock {A two per cent Hubble constant measurement from standard sirens
  within five years}.
\newblock {\em Nature}, 562(7728):545--547, 2018.

\bibitem{Feeney:2018mkj}
Stephen~M. Feeney, Hiranya~V. Peiris, Andrew~R. Williamson, Samaya~M. Nissanke,
  Daniel~J. Mortlock, Justin Alsing, and Dan Scolnic.
\newblock {Prospects for resolving the Hubble constant tension with standard
  sirens}.
\newblock {\em Phys. Rev. Lett.}, 122(6):061105, 2019.

\bibitem{Borhanian:2020vyr}
Ssohrab Borhanian, Arnab Dhani, Anuradha Gupta, K.~G. Arun, and B.~S.
  Sathyaprakash.
\newblock {Dark Sirens to Resolve the Hubble\textendash{}Lema\^\i{}tre
  Tension}.
\newblock {\em Astrophys. J. Lett.}, 905(2):L28, 2020.

\bibitem{Bian:2021ini}
Ligong Bian et~al.
\newblock {The Gravitational-Wave Physics II: Progress}.
\newblock {\em Sci. China Phys. Mech. Astron.}, 64:120401, 2021.

\bibitem{2016PhRvL.116f1102A}
B.~P. {Abbott}, R.~{Abbott}, T.~D. {Abbott}, M.~R. {Abernathy}, F.~{Acernese},
  K.~{Ackley}, C.~{Adams}, T.~{Adams}, P.~{Addesso}, R.~X. {Adhikari}, and
  et~al.
\newblock {Observation of Gravitational Waves from a Binary Black Hole Merger}.
\newblock {\em \prl}, 116(6):061102, February 2016.

\bibitem{2016PhRvL.116x1103A}
B.~P. {Abbott}, R.~{Abbott}, T.~D. {Abbott}, M.~R. {Abernathy}, F.~{Acernese},
  K.~{Ackley}, C.~{Adams}, T.~{Adams}, P.~{Addesso}, R.~X. {Adhikari}, and
  et~al.
\newblock {GW151226: Observation of Gravitational Waves from a 22-Solar-Mass
  Binary Black Hole Coalescence}.
\newblock {\em \prl}, 116(24):241103, June 2016.

\bibitem{2017PhRvL.118v1101A}
B.~P. {Abbott}, R.~{Abbott}, T.~D. {Abbott}, F.~{Acernese}, K.~{Ackley},
  C.~{Adams}, T.~{Adams}, P.~{Addesso}, R.~X. {Adhikari}, V.~B. {Adya}, and
  et~al.
\newblock {GW170104: Observation of a 50-Solar-Mass Binary Black Hole
  Coalescence at Redshift 0.2}.
\newblock {\em \prl}, 118(22):221101, June 2017.

\bibitem{2017ApJ...851L..35A}
B.~P. {Abbott}, R.~{Abbott}, T.~D. {Abbott}, F.~{Acernese}, K.~{Ackley},
  C.~{Adams}, T.~{Adams}, P.~{Addesso}, R.~X. {Adhikari}, V.~B. {Adya}, and
  et~al.
\newblock {GW170608: Observation of a 19 Solar-mass Binary Black Hole
  Coalescence}.
\newblock {\em \apjl}, 851(2):L35, December 2017.

\bibitem{2017PhRvL.119n1101A}
B.~P. {Abbott}, R.~{Abbott}, T.~D. {Abbott}, F.~{Acernese}, K.~{Ackley},
  C.~{Adams}, T.~{Adams}, P.~{Addesso}, R.~X. {Adhikari}, V.~B. {Adya}, and
  et~al.
\newblock {GW170814: A Three-Detector Observation of Gravitational Waves from a
  Binary Black Hole Coalescence}.
\newblock {\em \prl}, 119(14):141101, October 2017.

\bibitem{2017PhRvL.119p1101A}
B.~P. {Abbott}, R.~{Abbott}, T.~D. {Abbott}, F.~{Acernese}, K.~{Ackley},
  C.~{Adams}, T.~{Adams}, P.~{Addesso}, R.~X. {Adhikari}, V.~B. {Adya}, and
  et~al.
\newblock {GW170817: Observation of Gravitational Waves from a Binary Neutron
  Star Inspiral}.
\newblock {\em \prl}, 119(16):161101, October 2017.

\bibitem{LIGOScientific:2018mvr}
B.P. Abbott et~al.
\newblock {GWTC-1: A Gravitational-Wave Transient Catalog of Compact Binary
  Mergers Observed by LIGO and Virgo during the First and Second Observing
  Runs}.
\newblock {\em Phys. Rev. X}, 9(3):031040, 2019.

\bibitem{LIGOScientific:2020stg}
R.~Abbott et~al.
\newblock {GW190412: Observation of a Binary-Black-Hole Coalescence with
  Asymmetric Masses}.
\newblock {\em Phys. Rev. D}, 102(4):043015, 2020.

\bibitem{Abbott:2020uma}
B.~P. Abbott et~al.
\newblock {GW190425: Observation of a Compact Binary Coalescence with Total
  Mass $\sim 3.4 M_{\odot}$}.
\newblock {\em Astrophys. J. Lett.}, 892(1):L3, 2020.

\bibitem{Abbott:2020tfl}
R.~Abbott et~al.
\newblock {GW190521: A Binary Black Hole Merger with a Total Mass of $150
  M_{\odot}$}.
\newblock {\em Phys. Rev. Lett.}, 125(10):101102, 2020.

\bibitem{Abbott:2020khf}
R.~Abbott et~al.
\newblock {GW190814: Gravitational Waves from the Coalescence of a 23 Solar
  Mass Black Hole with a 2.6 Solar Mass Compact Object}.
\newblock {\em Astrophys. J. Lett.}, 896(2):L44, 2020.

\bibitem{Abbott:2020niy}
R.~Abbott et~al.
\newblock {GWTC-2: Compact Binary Coalescences Observed by LIGO and Virgo
  During the First Half of the Third Observing Run}.
\newblock 10 2020.

\bibitem{LIGOScientific:2021usb}
R.~Abbott et~al.
\newblock {GWTC-2.1: Deep Extended Catalog of Compact Binary Coalescences
  Observed by LIGO and Virgo During the First Half of the Third Observing Run}.
\newblock 8 2021.

\bibitem{LIGOScientific:2021qlt}
R.~Abbott et~al.
\newblock {Observation of Gravitational Waves from Two Neutron
  Star\textendash{}Black Hole Coalescences}.
\newblock {\em Astrophys. J. Lett.}, 915(1):L5, 2021.

\bibitem{Akutsu:2018axf}
T.~Akutsu et~al.
\newblock {KAGRA: 2.5 Generation Interferometric Gravitational Wave Detector}.
\newblock {\em Nature Astron.}, 3(1):35--40, 2019.

\bibitem{Unnikrishnan:2013qwa}
C.~S. Unnikrishnan.
\newblock {IndIGO and LIGO-India: Scope and plans for gravitational wave
  research and precision metrology in India}.
\newblock {\em Int. J. Mod. Phys. D}, 22:1341010, 2013.

\bibitem{Nissanke:2013fka}
Samaya Nissanke, Daniel~E. Holz, Neal Dalal, Scott~A. Hughes, Jonathan~L.
  Sievers, and Christopher~M. Hirata.
\newblock {Determining the Hubble constant from gravitational wave observations
  of merging compact binaries}.
\newblock 7 2013.

\bibitem{Vitale:2018wlg}
Salvatore Vitale and Hsin-Yu Chen.
\newblock {Measuring the Hubble constant with neutron star black hole mergers}.
\newblock {\em Phys. Rev. Lett.}, 121(2):021303, 2018.

\bibitem{Mortlock:2018azx}
Daniel~J. Mortlock, Stephen~M. Feeney, Hiranya~V. Peiris, Andrew~R. Williamson,
  and Samaya~M. Nissanke.
\newblock {Unbiased Hubble constant estimation from binary neutron star
  mergers}.
\newblock {\em Phys. Rev. D}, 100(10):103523, 2019.

\bibitem{Monitor:2017mdv}
B.P. Abbott et~al.
\newblock {Gravitational Waves and Gamma-rays from a Binary Neutron Star
  Merger: GW170817 and GRB 170817A}.
\newblock {\em Astrophys. J. Lett.}, 848(2):L13, 2017.

\bibitem{Soares-Santos:2017lru}
M.~Soares-Santos et~al.
\newblock {The Electromagnetic Counterpart of the Binary Neutron Star Merger
  LIGO/Virgo GW170817. I. Discovery of the Optical Counterpart Using the Dark
  Energy Camera}.
\newblock {\em Astrophys. J. Lett.}, 848(2):L16, 2017.

\bibitem{GBM:2017lvd}
B.~P. Abbott et~al.
\newblock {Multi-messenger Observations of a Binary Neutron Star Merger}.
\newblock {\em Astrophys. J. Lett.}, 848(2):L12, 2017.

\bibitem{Abbott:2019yzh}
B.~P. Abbott et~al.
\newblock {A Gravitational-wave Measurement of the Hubble Constant Following
  the Second Observing Run of Advanced LIGO and Virgo}.
\newblock {\em Astrophys. J.}, 909(2):218, 2021.

\bibitem{Abbott:2017xzu}
B.~P. Abbott et~al.
\newblock {A gravitational-wave standard siren measurement of the Hubble
  constant}.
\newblock {\em Nature}, 551(7678):85--88, 2017.

\bibitem{Fishbach:2018gjp}
M.~Fishbach et~al.
\newblock {A Standard Siren Measurement of the Hubble Constant from GW170817
  without the Electromagnetic Counterpart}.
\newblock {\em Astrophys. J. Lett.}, 871(1):L13, 2019.

\bibitem{Guidorzi:2017ogy}
C.~Guidorzi et~al.
\newblock {Improved Constraints on $H_0$ from a Combined Analysis of
  Gravitational-wave and Electromagnetic Emission from GW170817}.
\newblock {\em Astrophys. J. Lett.}, 851(2):L36, 2017.

\bibitem{Hotokezaka:2018dfi}
Kenta Hotokezaka, Ehud Nakar, Ore Gottlieb, Samaya Nissanke, Kento Masuda,
  Gregg Hallinan, Kunal~P. Mooley, and Adam.~T. Deller.
\newblock {A Hubble constant measurement from superluminal motion of the jet in
  GW170817}.
\newblock {\em Nature Astron.}, 3(10):940--944, 2019.

\bibitem{McKernan:2019hqs}
B.~McKernan, K.~E.~S. Ford, I.~Bartos, M.~J. Graham, W.~Lyra, S.~Marka,
  Z.~Marka, N.~P. Ross, D.~Stern, and Y.~Yang.
\newblock {Ram-pressure stripping of a kicked Hill sphere: Prompt
  electromagnetic emission from the merger of stellar mass black holes in an
  AGN accretion disk}.
\newblock {\em Astrophys. J. Lett.}, 884(2):L50, 2019.

\bibitem{Graham:2020gwr}
M.~J. Graham et~al.
\newblock {Candidate Electromagnetic Counterpart to the Binary Black Hole
  Merger Gravitational Wave Event S190521g}.
\newblock {\em Phys. Rev. Lett.}, 124(25):251102, 2020.

\bibitem{Soares-Santos:2019irc}
M.~Soares-Santos et~al.
\newblock {First Measurement of the Hubble Constant from a Dark Standard Siren
  using the Dark Energy Survey Galaxies and the LIGO/Virgo
  Binary\textendash{}Black-hole Merger GW170814}.
\newblock {\em Astrophys. J. Lett.}, 876(1):L7, 2019.

\bibitem{Palmese:2020aof}
A.~Palmese et~al.
\newblock {A statistical standard siren measurement of the Hubble constant from
  the LIGO/Virgo gravitational wave compact object merger GW190814 and Dark
  Energy Survey galaxies}.
\newblock {\em Astrophys. J. Lett.}, 900(2):L33, 2020.

\bibitem{Vasylyev:2020hgb}
Sergiy Vasylyev and Alex Filippenko.
\newblock {A Measurement of the Hubble Constant using Gravitational Waves from
  the Binary Merger GW190814}.
\newblock {\em Astrophys. J.}, 902(2):149, 2020.

\bibitem{DelPozzo:2012zz}
Walter Del~Pozzo.
\newblock {Inference of the cosmological parameters from gravitational waves:
  application to second generation interferometers}.
\newblock {\em Phys. Rev. D}, 86:043011, 2012.

\bibitem{Taylor:2011fs}
Stephen~R. Taylor, Jonathan~R. Gair, and Ilya Mandel.
\newblock {Hubble without the Hubble: Cosmology using advanced
  gravitational-wave detectors alone}.
\newblock {\em Phys. Rev. D}, 85:023535, 2012.

\bibitem{Nair:2018ign}
Remya Nair, Sukanta Bose, and Tarun~Deep Saini.
\newblock {Measuring the Hubble constant: Gravitational wave observations meet
  galaxy clustering}.
\newblock {\em Phys. Rev. D}, 98(2):023502, 2018.

\bibitem{Farr:2019twy}
Will~M. Farr, Maya Fishbach, Jiani Ye, and Daniel Holz.
\newblock {A Future Percent-Level Measurement of the Hubble Expansion at
  Redshift 0.8 With Advanced LIGO}.
\newblock {\em Astrophys. J. Lett.}, 883(2):L42, 2019.

\bibitem{Gray:2019ksv}
Rachel Gray et~al.
\newblock {Cosmological inference using gravitational wave standard sirens: A
  mock data analysis}.
\newblock {\em Phys. Rev. D}, 101(12):122001, 2020.

\bibitem{Bera:2020jhx}
Sayantani Bera, Divya Rana, Surhud More, and Sukanta Bose.
\newblock {Incompleteness Matters Not: Inference of $H_0$ from Binary Black
  Hole\textendash{}Galaxy Cross-correlations}.
\newblock {\em Astrophys. J.}, 902(1):79, 2020.

\bibitem{Finke:2021aom}
Andreas Finke, Stefano Foffa, Francesco Iacovelli, Michele Maggiore, and
  Michele Mancarella.
\newblock {Cosmology with LIGO/Virgo dark sirens: Hubble parameter and modified
  gravitational wave propagation}.
\newblock 1 2021.

\bibitem{Gray:2021sew}
Rachel Gray, Chris Messenger, and John Veitch.
\newblock {A Pixelated Approach to Galaxy Catalogue Incompleteness: Improving
  the Dark Siren Measurement of the Hubble Constant}.
\newblock 11 2021.

\bibitem{LIGOScientific:2021aug}
R.~Abbott et~al.
\newblock {Constraints on the cosmic expansion history from GWTC-3}.
\newblock 11 2021.

\bibitem{Punturo:2010zz}
M.~Punturo et~al.
\newblock {The Einstein Telescope: A third-generation gravitational wave
  observatory}.
\newblock {\em Class. Quant. Grav.}, 27:194002, 2010.

\bibitem{Sathyaprakash:2012jk}
B.~Sathyaprakash et~al.
\newblock {Scientific Objectives of Einstein Telescope}.
\newblock {\em Class. Quant. Grav.}, 29:124013, 2012.
\newblock [Erratum: Class.Quant.Grav. 30, 079501 (2013)].

\bibitem{Dwyer:2014fpa}
Sheila Dwyer, Daniel Sigg, Stefan~W. Ballmer, Lisa Barsotti, Nergis Mavalvala,
  and Matthew Evans.
\newblock {Gravitational wave detector with cosmological reach}.
\newblock {\em Phys. Rev. D}, 91(8):082001, 2015.

\bibitem{Evans:2016mbw}
Benjamin~P Abbott et~al.
\newblock {Exploring the Sensitivity of Next Generation Gravitational Wave
  Detectors}.
\newblock {\em Class. Quant. Grav.}, 34(4):044001, 2017.

\bibitem{Zhao:2010sz}
W.~Zhao, C.~Van Den~Broeck, D.~Baskaran, and T.~G.~F. Li.
\newblock {Determination of Dark Energy by the Einstein Telescope: Comparing
  with CMB, BAO and SNIa Observations}.
\newblock {\em Phys. Rev. D}, 83:023005, 2011.

\bibitem{Zhao:2017cbb}
Wen Zhao and Linqing Wen.
\newblock {Localization accuracy of compact binary coalescences detected by the
  third-generation gravitational-wave detectors and implication for cosmology}.
\newblock {\em Phys. Rev. D}, 97(6):064031, 2018.

\bibitem{Taylor:2012db}
Stephen~R. Taylor and Jonathan~R. Gair.
\newblock {Cosmology with the lights off: standard sirens in the Einstein
  Telescope era}.
\newblock {\em Phys. Rev. D}, 86:023502, 2012.

\bibitem{Seikel:2012uu}
Marina Seikel, Chris Clarkson, and Mathew Smith.
\newblock {Reconstruction of dark energy and expansion dynamics using Gaussian
  processes}.
\newblock {\em JCAP}, 06:036, 2012.

\bibitem{Messenger:2011gi}
C.~Messenger and J.~Read.
\newblock {Measuring a cosmological distance-redshift relationship using only
  gravitational wave observations of binary neutron star coalescences}.
\newblock {\em Phys. Rev. Lett.}, 108:091101, 2012.

\bibitem{Messenger:2013fya}
C.~Messenger, Kentaro Takami, Sarah Gossan, Luciano Rezzolla, and B.~S.
  Sathyaprakash.
\newblock {Source Redshifts from Gravitational-Wave Observations of Binary
  Neutron Star Mergers}.
\newblock {\em Phys. Rev. X}, 4(4):041004, 2014.

\bibitem{DelPozzo:2015bna}
Walter Del~Pozzo, Tjonnie G.~F. Li, and Chris Messenger.
\newblock {Cosmological inference using only gravitational wave observations of
  binary neutron stars}.
\newblock {\em Phys. Rev. D}, 95(4):043502, 2017.

\bibitem{Cai:2016sby}
Rong-Gen Cai and Tao Yang.
\newblock {Estimating cosmological parameters by the simulated data of
  gravitational waves from the Einstein Telescope}.
\newblock {\em Phys. Rev. D}, 95(4):044024, 2017.

\bibitem{Du:2018tia}
Minghui Du, Weiqiang Yang, Lixin Xu, Supriya Pan, and David~F. Mota.
\newblock {Future constraints on dynamical dark-energy using gravitational-wave
  standard sirens}.
\newblock {\em Phys. Rev. D}, 100(4):043535, 2019.

\bibitem{Zhang:2018byx}
Xuan-Neng Zhang, Ling-Feng Wang, Jing-Fei Zhang, and Xin Zhang.
\newblock {Improving cosmological parameter estimation with the future
  gravitational-wave standard siren observation from the Einstein Telescope}.
\newblock {\em Phys. Rev. D}, 99(6):063510, 2019.

\bibitem{Mendonca:2019yfo}
Josiel Mendon\c{c}a~Soares de~Souza and Riccardo Sturani.
\newblock {Cosmological model selection from standard siren detections by
  third-generation gravitational wave observatories}.
\newblock {\em Phys. Dark Univ.}, 32:100830, 2021.

\bibitem{Zhang:2019loq}
Jing-Fei Zhang, Ming Zhang, Shang-Jie Jin, Jing-Zhao Qi, and Xin Zhang.
\newblock {Cosmological parameter estimation with future gravitational wave
  standard siren observation from the Einstein Telescope}.
\newblock {\em JCAP}, 09:068, 2019.

\bibitem{Yu:2020vyy}
Jiming Yu, Yu~Wang, Wen Zhao, and Youjun Lu.
\newblock {Hunting for the host galaxy groups of binary black holes and the
  application in constraining Hubble constant}.
\newblock {\em Mon. Not. Roy. Astron. Soc.}, 498(2):1786--1800, 2020.

\bibitem{You:2020wju}
Zhi-Qiang You, Xing-Jiang Zhu, Gregory Ashton, Eric Thrane, and Zong-Hong Zhu.
\newblock {Standard-siren cosmology using gravitational waves from binary black
  holes}.
\newblock {\em Astrophys. J.}, 908(2):215, 2021.

\bibitem{Bonilla:2021dql}
Alexander Bonilla, Suresh Kumar, Rafael~C. Nunes, and Supriya Pan.
\newblock {Reconstruction of the dark sectors' interaction: A model-independent
  inference and forecast from GW standard sirens}.
\newblock 2 2021.

\bibitem{Luo:2015ght}
Jun Luo et~al.
\newblock {TianQin: a space-borne gravitational wave detector}.
\newblock {\em Class. Quant. Grav.}, 33(3):035010, 2016.

\bibitem{LISA:2017pwj}
Pau Amaro-Seoane et~al.
\newblock {Laser Interferometer Space Antenna}.
\newblock 2 2017.

\bibitem{Klein:2015hvg}
Antoine Klein et~al.
\newblock {Science with the space-based interferometer eLISA: Supermassive
  black hole binaries}.
\newblock {\em Phys. Rev. D}, 93(2):024003, 2016.

\bibitem{Barausse:2020mdt}
Enrico Barausse, Irina Dvorkin, Michael Tremmel, Marta Volonteri, and Matteo
  Bonetti.
\newblock {Massive Black Hole Merger Rates: The Effect of Kiloparsec Separation
  Wandering and Supernova Feedback}.
\newblock {\em Astrophys. J.}, 904(1):16, 2020.

\bibitem{Wang:2019}
Hai-Tian {Wang}, Zhen {Jiang}, Alberto {Sesana}, Enrico {Barausse}, Shun-Jia
  {Huang}, Yi-Fan {Wang}, Wen-Fan {Feng}, Yan {Wang}, Yi-Ming {Hu}, Jianwei
  {Mei}, and Jun {Luo}.
\newblock {Science with the TianQin observatory: Preliminary results on massive
  black hole binaries}.
\newblock {\em \prd}, 100(4):043003, August 2019.

\bibitem{Babak:2017tow}
Stanislav Babak, Jonathan Gair, Alberto Sesana, Enrico Barausse, Carlos~F.
  Sopuerta, Christopher P.~L. Berry, Emanuele Berti, Pau Amaro-Seoane, Antoine
  Petiteau, and Antoine Klein.
\newblock {Science with the space-based interferometer LISA. V: Extreme
  mass-ratio inspirals}.
\newblock {\em Phys. Rev. D}, 95(10):103012, 2017.

\bibitem{Gair:2017ynp}
Jonathan~R. Gair, Stanislav Babak, Alberto Sesana, Pau Amaro-Seoane, Enrico
  Barausse, Christopher P.~L. Berry, Emanuele Berti, and Carlos Sopuerta.
\newblock {Prospects for observing extreme-mass-ratio inspirals with LISA}.
\newblock {\em J. Phys. Conf. Ser.}, 840(1):012021, 2017.

\bibitem{Fan:2020zhy}
Hui-Min Fan, Yi-Ming Hu, Enrico Barausse, Alberto Sesana, Jian-dong Zhang,
  Xuefeng Zhang, Tie-Guang Zi, and Jianwei Mei.
\newblock {Science with the TianQin observatory: Preliminary result on
  extreme-mass-ratio inspirals}.
\newblock {\em Phys. Rev. D}, 102(6):063016, 2020.

\bibitem{Kyutoku:2016ppx}
Koutarou Kyutoku and Naoki Seto.
\newblock {Concise estimate of the expected number of detections for
  stellar-mass binary black holes by eLISA}.
\newblock {\em Mon. Not. Roy. Astron. Soc.}, 462(2):2177--2183, 2016.

\bibitem{Liu:2020eko}
Shuai Liu, Yi-Ming Hu, Jian-dong Zhang, and Jianwei Mei.
\newblock {Science with the TianQin observatory: Preliminary results on
  stellar-mass binary black holes}.
\newblock {\em Phys. Rev. D}, 101(10):103027, 2020.

\bibitem{Liu:2021yoy}
Shuai Liu, Liang-Gui Zhu, Yi-Ming Hu, Jian-dong Zhang, and Mu-Jie Ji.
\newblock {The capability for detection of GW190521-like binary black holes
  with TianQin}.
\newblock 10 2021.

\bibitem{Holz:2005df}
Daniel~E. Holz and Scott~A. Hughes.
\newblock {Using gravitational-wave standard sirens}.
\newblock {\em Astrophys. J.}, 629:15--22, 2005.

\bibitem{Babak:2010ej}
Stanislav Babak, Jonathan~R. Gair, Antoine Petiteau, and Alberto Sesana.
\newblock {Fundamental physics and cosmology with LISA}.
\newblock {\em Class. Quant. Grav.}, 28:114001, 2011.

\bibitem{Petiteau:2011we}
Antoine Petiteau, Stanislav Babak, and Alberto Sesana.
\newblock {Constraining the dark energy equation of state using LISA
  observations of spinning Massive Black Hole binaries}.
\newblock {\em Astrophys. J.}, 732:82, 2011.

\bibitem{Tamanini:2016zlh}
Nicola Tamanini, Chiara Caprini, Enrico Barausse, Alberto Sesana, Antoine
  Klein, and Antoine Petiteau.
\newblock {Science with the space-based interferometer eLISA. III: Probing the
  expansion of the Universe using gravitational wave standard sirens}.
\newblock {\em JCAP}, 04:002, 2016.

\bibitem{Caprini:2016qxs}
Chiara Caprini and Nicola Tamanini.
\newblock {Constraining early and interacting dark energy with gravitational
  wave standard sirens: the potential of the eLISA mission}.
\newblock {\em JCAP}, 10:006, 2016.

\bibitem{Cai:2017yww}
Rong-Gen Cai, Nicola Tamanini, and Tao Yang.
\newblock {Reconstructing the dark sector interaction with LISA}.
\newblock {\em JCAP}, 05:031, 2017.

\bibitem{Wang:2019tto}
Ling-Feng Wang, Ze-Wei Zhao, Jing-Fei Zhang, and Xin Zhang.
\newblock {A preliminary forecast for cosmological parameter estimation with
  gravitational-wave standard sirens from TianQin}.
\newblock {\em JCAP}, 11:012, 2020.

\bibitem{Zhu:2021aat}
Liang-Gui Zhu, Yi-Ming Hu, Hai-Tian Wang, Jian-Dong Zhang, Xiao-Dong Li, Martin
  Hendry, and Jianwei Mei.
\newblock {Constraining the cosmological parameters using gravitational wave
  observations of massive black hole binaries and statistical redshift
  information}.
\newblock 4 2021.

\bibitem{Wang:2020dkc}
Renjie Wang, Wen-Hong Ruan, Qing Yang, Zong-Kuan Guo, Rong-Gen Cai, and Bin Hu.
\newblock {Hubble parameter estimation via dark sirens with the LISA-Taiji
  network}.
\newblock 10 2020.

\bibitem{Wang:2021srv}
Ling-Feng Wang, Shang-Jie Jin, Jing-Fei Zhang, and Xin Zhang.
\newblock {Forecast for cosmological parameter estimation with
  gravitational-wave standard sirens from the LISA-Taiji network}.
\newblock {\em Sci. China Phys. Mech. Astron.}, 65:210411, 2022.

\bibitem{MacLeod:2007jd}
Chelsea~L. MacLeod and Craig~J. Hogan.
\newblock {Precision of Hubble constant derived using black hole binary
  absolute distances and statistical redshift information}.
\newblock {\em Phys. Rev. D}, 77:043512, 2008.

\bibitem{Laghi:2021pqk}
Danny Laghi, Nicola Tamanini, Walter Del~Pozzo, Alberto Sesana, Jonathan Gair,
  and Stanislav Babak.
\newblock {Gravitational wave cosmology with extreme mass-ratio inspirals}.
\newblock 2 2021.

\bibitem{Kyutoku:2016zxn}
Koutarou Kyutoku and Naoki Seto.
\newblock {Gravitational-wave cosmography with LISA and the Hubble tension}.
\newblock {\em Phys. Rev. D}, 95(8):083525, 2017.

\bibitem{DelPozzo:2017kme}
Walter Del~Pozzo, Alberto Sesana, and Antoine Klein.
\newblock {Stellar binary black holes in the LISA band: a new class of standard
  sirens}.
\newblock {\em Mon. Not. Roy. Astron. Soc.}, 475(3):3485--3492, 2018.

\bibitem{Sesana:2016ljz}
Alberto Sesana.
\newblock {Prospects for Multiband Gravitational-Wave Astronomy after
  GW150914}.
\newblock {\em Phys. Rev. Lett.}, 116(23):231102, 2016.

\bibitem{Sesana:2017vsj}
Alberto Sesana.
\newblock {Multi-band gravitational wave astronomy: science with joint space-
  and ground-based observations of black hole binaries}.
\newblock {\em J. Phys. Conf. Ser.}, 840(1):012018, 2017.

\bibitem{Vitale:2016rfr}
Salvatore Vitale.
\newblock {Multiband Gravitational-Wave Astronomy: Parameter Estimation and
  Tests of General Relativity with Space- and Ground-Based Detectors}.
\newblock {\em Phys. Rev. Lett.}, 117(5):051102, 2016.

\bibitem{Moore:2019pke}
Christopher~J. Moore, Davide Gerosa, and Antoine Klein.
\newblock {Are stellar-mass black-hole binaries too quiet for LISA?}
\newblock {\em Mon. Not. Roy. Astron. Soc.}, 488(1):L94--L98, 2019.

\bibitem{Ewing:2020brd}
Becca Ewing, Surabhi Sachdev, Ssohrab Borhanian, and B.~S. Sathyaprakash.
\newblock {Archival searches for stellar-mass binary black holes in LISA data}.
\newblock {\em Phys. Rev. D}, 103(2):023025, 2021.

\bibitem{Grimm:2020ivq}
Stefan Grimm and Jan Harms.
\newblock {Multiband gravitational-wave parameter estimation: A study of future
  detectors}.
\newblock {\em Phys. Rev. D}, 102(2):022007, 2020.

\bibitem{Barausse:2016eii}
Enrico Barausse, Nicol\'as Yunes, and Katie Chamberlain.
\newblock {Theory-Agnostic Constraints on Black-Hole Dipole Radiation with
  Multiband Gravitational-Wave Astrophysics}.
\newblock {\em Phys. Rev. Lett.}, 116(24):241104, 2016.

\bibitem{Wong:2018uwb}
Kaze W.~K. Wong, Ely~D. Kovetz, Curt Cutler, and Emanuele Berti.
\newblock {Expanding the LISA Horizon from the Ground}.
\newblock {\em Phys. Rev. Lett.}, 121(25):251102, 2018.

\bibitem{Gerosa:2019dbe}
Davide Gerosa, Sizheng Ma, Kaze W.~K. Wong, Emanuele Berti, Richard
  O'Shaughnessy, Yanbei Chen, and Krzysztof Belczynski.
\newblock {Multiband gravitational-wave event rates and stellar physics}.
\newblock {\em Phys. Rev. D}, 99(10):103004, 2019.

\bibitem{Cutler:2019krq}
Curt Cutler et~al.
\newblock {What we can learn from multi-band observations of black hole
  binaries}.
\newblock 3 2019.

\bibitem{Liu:2020nwz}
Chang Liu, Lijing Shao, Junjie Zhao, and Yong Gao.
\newblock {Multiband Observation of LIGO/Virgo Binary Black Hole Mergers in the
  Gravitational-wave Transient Catalog GWTC-1}.
\newblock {\em Mon. Not. Roy. Astron. Soc.}, 496(1):182--196, 2020.

\bibitem{Muttoni:2021veo}
Niccol\`o Muttoni, Alberto Mangiagli, Alberto Sesana, Danny Laghi, Walter
  Del~Pozzo, and David Izquierdo-Villalba.
\newblock {Multi-band gravitational wave cosmology with stellar origin black
  hole binaries}.
\newblock 9 2021.

\bibitem{Luo:2020bls}
Jun Luo et~al.
\newblock {The first round result from the TianQin-1 satellite}.
\newblock {\em Class. Quant. Grav.}, 37(18):185013, 2020.

\bibitem{Mei:2020lrl}
Jianwei Mei et~al.
\newblock {The TianQin project: current progress on science and technology}.
\newblock 8 2020.

\bibitem{Maggiore:2019uih}
Michele Maggiore et~al.
\newblock {Science Case for the Einstein Telescope}.
\newblock {\em JCAP}, 03:050, 2020.

\bibitem{Colpi:2016fup}
Monica Colpi and Alberto Sesana.
\newblock {\em {Gravitational Wave Sources in the Era of Multi-Band
  Gravitational Wave Astronomy}}, pages 43--140.
\newblock 2017.

\bibitem{2018SSPMA..48g9805Z}
Wen {Zhao}.
\newblock {Gravitational-wave standard siren and cosmology}.
\newblock {\em Scientia Sinica Physica, Mechanica \& Astronomica},
  48(7):079805, July 2018.

\bibitem{Nissanke:2009kt}
Samaya Nissanke, Daniel~E. Holz, Scott~A. Hughes, Neal Dalal, and Jonathan~L.
  Sievers.
\newblock {Exploring short gamma-ray bursts as gravitational-wave standard
  sirens}.
\newblock {\em Astrophys. J.}, 725:496--514, 2010.

\bibitem{Blanchard:2017csd}
P.~K. Blanchard et~al.
\newblock {The Electromagnetic Counterpart of the Binary Neutron Star Merger
  LIGO/VIRGO GW170817. VII. Properties of the Host Galaxy and Constraints on
  the Merger Timescale}.
\newblock {\em Astrophys. J. Lett.}, 848(2):L22, 2017.

\bibitem{Li:1998bw}
Li-Xin Li and Bohdan Paczynski.
\newblock {Transient events from neutron star mergers}.
\newblock {\em Astrophys. J. Lett.}, 507:L59, 1998.

\bibitem{2010MNRAS.406.2650M}
B.~D. {Metzger}, G.~{Mart{\'\i}nez-Pinedo}, S.~{Darbha}, E.~{Quataert},
  A.~{Arcones}, D.~{Kasen}, R.~{Thomas}, P.~{Nugent}, I.~V. {Panov}, and N.~T.
  {Zinner}.
\newblock {Electromagnetic counterparts of compact object mergers powered by
  the radioactive decay of r-process nuclei}.
\newblock {\em \mnras}, 406(4):2650--2662, August 2010.

\bibitem{Tanvir:2017pws}
N.~R. Tanvir et~al.
\newblock {The Emergence of a Lanthanide-Rich Kilonova Following the Merger of
  Two Neutron Stars}.
\newblock {\em Astrophys. J. Lett.}, 848(2):L27, 2017.

\bibitem{Kiziltan:2013oja}
B\"ulent Kiziltan, Athanasios Kottas, Maria De~Yoreo, and Stephen~E. Thorsett.
\newblock {The Neutron Star Mass Distribution}.
\newblock {\em Astrophys. J.}, 778:66, 2013.

\bibitem{Oguri:2016dgk}
Masamune Oguri.
\newblock {Measuring the distance-redshift relation with the cross-correlation
  of gravitational wave standard sirens and galaxies}.
\newblock {\em Phys. Rev. D}, 93(8):083511, 2016.

\bibitem{Zhang:2018ekk}
Pengjie Zhang.
\newblock {Accurate redshift determination of standard sirens by the luminosity
  distance space-redshift space large scale structure cross correlation}.
\newblock 11 2018.

\bibitem{Mukherjee:2020hyn}
Suvodip Mukherjee, Benjamin~D. Wandelt, Samaya~M. Nissanke, and Alessandra
  Silvestri.
\newblock {Accurate precision Cosmology with redshift unknown gravitational
  wave sources}.
\newblock {\em Phys. Rev. D}, 103(4):043520, 2021.

\bibitem{Diaz:2021pem}
Cristina~Cigarran Diaz and Suvodip Mukherjee.
\newblock {Mapping the cosmic expansion history from LIGO-Virgo-KAGRA in
  synergy with DESI and SPHEREx}.
\newblock 7 2021.

\bibitem{Ding:2018zrk}
Xuheng Ding, Marek Biesiada, Xiaogang Zheng, Kai Liao, Zhengxiang Li, and
  Zong-Hong Zhu.
\newblock {Cosmological inference from standard sirens without redshift
  measurements}.
\newblock {\em JCAP}, 04:033, 2019.

\bibitem{Leandro:2021qlc}
Hebertt Leandro, Valerio Marra, and Riccardo Sturani.
\newblock {Measuring the Hubble constant with black sirens}.
\newblock 9 2021.

\bibitem{Seto:2001qf}
Naoki Seto, Seiji Kawamura, and Takashi Nakamura.
\newblock {Possibility of direct measurement of the acceleration of the
  universe using 0.1-Hz band laser interferometer gravitational wave antenna in
  space}.
\newblock {\em Phys. Rev. Lett.}, 87:221103, 2001.

\bibitem{Nishizawa:2010xx}
Atsushi Nishizawa, Atsushi Taruya, and Shun Saito.
\newblock {Tracing the redshift evolution of Hubble parameter with
  gravitational-wave standard sirens}.
\newblock {\em Phys. Rev. D}, 83:084045, 2011.

\bibitem{Nishizawa:2011eq}
Atsushi Nishizawa, Kent Yagi, Atsushi Taruya, and Takahiro Tanaka.
\newblock {Cosmology with space-based gravitational-wave detectors --- dark
  energy and primordial gravitational waves ---}.
\newblock {\em Phys. Rev. D}, 85:044047, 2012.

\bibitem{Mandel:2018mve}
Ilya Mandel, Will~M. Farr, and Jonathan~R. Gair.
\newblock {Extracting distribution parameters from multiple uncertain
  observations with selection biases}.
\newblock {\em Mon. Not. Roy. Astron. Soc.}, 486(1):1086--1093, 2019.

\bibitem{Finn:1992}
Lee~S. Finn.
\newblock Detection, measurement, and gravitational radiation.
\newblock {\em Phys. Rev. D}, 46:5236--5249, Dec 1992.

\bibitem{Dalya:2018cnd}
Gergely D\'alya, G\'abor Galg\'oczi, L\'aszl\'o Dobos, Zsolt Frei, Ik~Siong
  Heng, Ronaldas Macas, Christopher Messenger, P\'eter Raffai, and Rafael~S.
  de~Souza.
\newblock {GLADE: A galaxy catalogue for multimessenger searches in the
  advanced gravitational-wave detector era}.
\newblock {\em Mon. Not. Roy. Astron. Soc.}, 479(2):2374--2381, 2018.

\bibitem{Thorne:1989lfp}
Kip~S. Thorne.
\newblock In S.~W. Hawking and W.~Israel, editors, {\em {Three hundred years of
  gravitation}}, pages 330--458. Cambridge University Press, Cambridge,
  England, 1987.

\bibitem{Freise:2008dk}
A.~Freise, S.~Chelkowski, S.~Hild, W.~Del~Pozzo, A.~Perreca, and A.~Vecchio.
\newblock {Triple Michelson Interferometer for a Third-Generation Gravitational
  Wave Detector}.
\newblock {\em Class. Quant. Grav.}, 26:085012, 2009.

\bibitem{Hu:2018yqb}
Xin-Chun Hu, Xiao-Hong Li, Yan Wang, Wen-Fan Feng, Ming-Yue Zhou, Yi-Ming Hu,
  Shou-Cun Hu, Jian-Wei Mei, and Cheng-Gang Shao.
\newblock {Fundamentals of the orbit and response for TianQin}.
\newblock {\em Class. Quant. Grav.}, 35(9):095008, 2018.

\bibitem{Cutler:1997ta}
Curt Cutler.
\newblock {Angular resolution of the LISA gravitational wave detector}.
\newblock {\em Phys. Rev. D}, 57:7089--7102, 1998.

\bibitem{Cornish:2002rt}
Neil~J. Cornish and Louis~J. Rubbo.
\newblock {The LISA response function}.
\newblock {\em Phys. Rev. D}, 67:022001, 2003.
\newblock [Erratum: Phys.Rev.D 67, 029905 (2003)].

\bibitem{Jaranowski:1998qm}
Piotr Jaranowski, Andrzej Krolak, and Bernard~F. Schutz.
\newblock {Data analysis of gravitational - wave signals from spinning neutron
  stars. 1. The Signal and its detection}.
\newblock {\em Phys. Rev. D}, 58:063001, 1998.

\bibitem{Robson:2019}
Travis {Robson}, Neil~J. {Cornish}, and Chang {Liu}.
\newblock {The construction and use of LISA sensitivity curves}.
\newblock {\em Classical and Quantum Gravity}, 36(10):105011, May 2019.

\bibitem{Zhang:2020khm}
Chunyu Zhang, Qing Gao, Yungui Gong, Bin Wang, Alan~J. Weinstein, and Chao
  Zhang.
\newblock {Full analytical formulas for frequency response of space-based
  gravitational wave detectors}.
\newblock {\em Phys. Rev. D}, 101(12):124027, 2020.

\bibitem{Cutler:1994ys}
Curt Cutler and Eanna~E. Flanagan.
\newblock {Gravitational waves from merging compact binaries: How accurately
  can one extract the binary's parameters from the inspiral wave form?}
\newblock {\em Phys. Rev. D}, 49:2658--2697, 1994.

\bibitem{Hild:2010id}
S.~Hild et~al.
\newblock {Sensitivity Studies for Third-Generation Gravitational Wave
  Observatories}.
\newblock {\em Class. Quant. Grav.}, 28:094013, 2011.

\bibitem{Vallisneri_2008}
Michele Vallisneri.
\newblock Use and abuse of the fisher information matrix in the assessment of
  gravitational-wave parameter-estimation prospects.
\newblock {\em Physical Review D}, 77(4), Feb 2008.

\bibitem{Bell:2003cj}
Eric~F. Bell, Daniel~H. McIntosh, Neal Katz, and Martin~D. Weinberg.
\newblock {The optical and near-infrared properties of galaxies. 1. Luminosity
  and stellar mass functions}.
\newblock {\em Astrophys. J. Suppl.}, 149:289, 2003.

\bibitem{Lin:2004ak}
Yen-Ting Lin, Joseph~J. Mohr, and S.~Adam Stanford.
\newblock {K-band properties of galaxy clusters and groups: Luminosity
  function, radial distribution and halo occupation number}.
\newblock {\em Astrophys. J.}, 610:745--761, 2004.

\bibitem{Rahman:2019}
M.~Rahman, S.~Nissanke, A.~Williamson, et~al.
\newblock 2019.
\newblock (in preparation).

\bibitem{Abolfathi:2017vfu}
Bela Abolfathi et~al.
\newblock {The Fourteenth Data Release of the Sloan Digital Sky Survey: First
  Spectroscopic Data from the Extended Baryon Oscillation Spectroscopic Survey
  and from the Second Phase of the Apache Point Observatory Galactic Evolution
  Experiment}.
\newblock {\em Astrophys. J. Suppl.}, 235(2):42, 2018.

\bibitem{Schlegel:1997yv}
David~J. Schlegel, Douglas~P. Finkbeiner, and Marc Davis.
\newblock {Maps of dust IR emission for use in estimation of reddening and CMBR
  foregrounds}.
\newblock {\em Astrophys. J.}, 500:525, 1998.

\bibitem{Abbott:2021hoj}
T.~M.~C. Abbott et~al.
\newblock {The Dark Energy Survey Data Release 2}.
\newblock 1 2021.

\bibitem{Drlica-Wagner:2017tkk}
A.~Drlica-Wagner et~al.
\newblock {Dark Energy Survey Year 1 Results: Photometric Data Set for
  Cosmology}.
\newblock {\em Astrophys. J. Suppl.}, 235(2):33, 2018.

\bibitem{Abbott:2018jhe}
T.~M.~C. Abbott et~al.
\newblock {The Dark Energy Survey Data Release 1}.
\newblock {\em Astrophys. J. Suppl.}, 239(2):18, 2018.

\bibitem{Arnouts:1999bb}
Stephane Arnouts, Stefano Cristiani, Lauro Moscardini, Sabino Matarrese,
  Francesco Lucchin, Adriano Fontana, and Emanuele Giallongo.
\newblock {Measuring and modeling the redshift evolution of clustering: The
  Hubble Deep Field North}.
\newblock {\em Mon. Not. Roy. Astron. Soc.}, 310:540--556, 1999.

\bibitem{Ilbert:2006dp}
Olivier Ilbert et~al.
\newblock {Accurate photometric redshifts for the cfht legacy survey calibrated
  using the vimos vlt deep survey}.
\newblock {\em Astron. Astrophys.}, 457:841--856, 2006.

\bibitem{Bruzual:2003tq}
G.~Bruzual and Stephane Charlot.
\newblock {Stellar population synthesis at the resolution of 2003}.
\newblock {\em Mon. Not. Roy. Astron. Soc.}, 344:1000, 2003.

\bibitem{Chabrier:2003ki}
Gilles Chabrier.
\newblock {Galactic stellar and substellar initial mass function}.
\newblock {\em Publ. Astron. Soc. Pac.}, 115:763--796, 2003.

\bibitem{Calzetti:1994vw}
Daniela Calzetti, Anne~L. Kinney, and Thaisa Storchi-Bergmann.
\newblock {Dust extinction of the stellar continua in starburst galaxies: The
  Ultraviolet and optical extinction law}.
\newblock {\em Astrophys. J.}, 429:582, 1994.

\bibitem{Maraston:2012jf}
Claudia Maraston et~al.
\newblock {Stellar masses of SDSS-III BOSS galaxies at z\textasciitilde{}0.5
  and constraints to galaxy formation models}.
\newblock {\em Mon. Not. Roy. Astron. Soc.}, 435:2764, 2013.

\bibitem{Maraston:2004em}
Claudia Maraston.
\newblock {Evolutionary population synthesis: Models, analysis of the
  ingredients and application to high-z galaxies}.
\newblock {\em Mon. Not. Roy. Astron. Soc.}, 362:799--825, 2005.

\bibitem{Maraston:2008nn}
Claudia Maraston, Gustav Stromback, Daniel Thomas, David~A. Wake, and Robert~C.
  Nichol.
\newblock {Modeling the color evolution of luminous red galaxies - improvements
  with empirical stellar spectra}.
\newblock {\em Mon. Not. Roy. Astron. Soc.}, 394:107, 2009.

\bibitem{LIGOScientific:2020kqk}
R.~Abbott et~al.
\newblock {Population Properties of Compact Objects from the Second LIGO-Virgo
  Gravitational-Wave Transient Catalog}.
\newblock {\em Astrophys. J. Lett.}, 913(1):L7, 2021.

\bibitem{Talbot:2018cva}
Colm Talbot and Eric Thrane.
\newblock {Measuring the binary black hole mass spectrum with an
  astrophysically motivated parameterization}.
\newblock {\em Astrophys. J.}, 856(2):173, 2018.

\bibitem{LIGOScientific:2018jsj}
B.~P. Abbott et~al.
\newblock {Binary Black Hole Population Properties Inferred from the First and
  Second Observing Runs of Advanced LIGO and Advanced Virgo}.
\newblock {\em Astrophys. J. Lett.}, 882(2):L24, 2019.

\bibitem{Roulet:2018jbe}
Javier Roulet and Matias Zaldarriaga.
\newblock {Constraints on binary black hole populations from
  LIGO\textendash{}Virgo detections}.
\newblock {\em Mon. Not. Roy. Astron. Soc.}, 484(3):4216--4229, 2019.

\bibitem{Fishbach:2019bbm}
Maya Fishbach and Daniel~E. Holz.
\newblock {Picky Partners: The Pairing of Component Masses in Binary Black Hole
  Mergers}.
\newblock {\em Astrophys. J. Lett.}, 891(1):L27, 2020.

\bibitem{Mangiagli:2018kpu}
Alberto Mangiagli, Antoine Klein, Alberto Sesana, Enrico Barausse, and Monica
  Colpi.
\newblock {Post-Newtonian phase accuracy requirements for stellar black hole
  binaries with LISA}.
\newblock {\em Phys. Rev. D}, 99(6):064056, 2019.

\bibitem{Nishizawa:2016jji}
Atsushi Nishizawa, Emanuele Berti, Antoine Klein, and Alberto Sesana.
\newblock {eLISA eccentricity measurements as tracers of binary black hole
  formation}.
\newblock {\em Phys. Rev. D}, 94(6):064020, 2016.

\bibitem{Hannam:2014}
Mark {Hannam}, Patricia {Schmidt}, Alejandro {Boh{\'e}}, Le{\"\i}la {Haegel},
  Sascha {Husa}, Frank {Ohme}, Geraint {Pratten}, and Michael {P{\"u}rrer}.
\newblock {Simple Model of Complete Precessing Black-Hole-Binary Gravitational
  Waveforms}.
\newblock {\em \prl}, 113(15):151101, October 2014.

\bibitem{Liang:2021bde}
Zheng-Cheng Liang, Yi-Ming Hu, Yun Jiang, Jun Cheng, Jian-dong Zhang, and
  Jianwei Mei.
\newblock {Science with the TianQin Observatory: Preliminary Results on
  Stochastic Gravitational-Wave Background}.
\newblock 7 2021.

\bibitem{He:2019dhl}
Jian-hua He.
\newblock {Accurate method to determine the systematics due to the peculiar
  velocities of galaxies in measuring the Hubble constant from
  gravitational-wave standard sirens}.
\newblock {\em Phys. Rev. D}, 100(2):023527, 2019.

\bibitem{Howlett:2019mdh}
Cullan Howlett and Tamara~M. Davis.
\newblock {Standard siren speeds: improving velocities in gravitational-wave
  measurements of $H_0$}.
\newblock {\em Mon. Not. Roy. Astron. Soc.}, 492(3):3803--3815, 2020.

\bibitem{Nicolaou:2019cip}
Constantina Nicolaou, Ofer Lahav, Pablo Lemos, William Hartley, and Jonathan
  Braden.
\newblock {The Impact of Peculiar Velocities on the Estimation of the Hubble
  Constant from Gravitational Wave Standard Sirens}.
\newblock {\em Mon. Not. Roy. Astron. Soc.}, 495(1):90--97, 2020.

\bibitem{Mukherjee:2019qmm}
Suvodip Mukherjee, Guilhem Lavaux, Fran\c{c}ois~R. Bouchet, Jens Jasche,
  Benjamin~D. Wandelt, Samaya~M. Nissanke, Florent Leclercq, and Kenta
  Hotokezaka.
\newblock {Velocity correction for Hubble constant measurements from standard
  sirens}.
\newblock {\em Astron. Astrophys.}, 646:A65, 2021.

\bibitem{Gong:2019yxt}
Yan Gong, Xiangkun Liu, Ye~Cao, Xuelei Chen, Zuhui Fan, Ran Li, Xiao-Dong Li,
  Zhigang Li, Xin Zhang, and Hu~Zhan.
\newblock {Cosmology from the Chinese Space Station Optical Survey (CSS-OS)}.
\newblock {\em Astrophys. J.}, 883:203, 2019.

\bibitem{ForemanMackey:2012ig}
Daniel Foreman-Mackey, David~W. Hogg, Dustin Lang, and Jonathan Goodman.
\newblock {emcee: The MCMC Hammer}.
\newblock {\em Publ. Astron. Soc. Pac.}, 125:306--312, 2013.

\bibitem{ForemanMackey:2019ig}
Foreman-Mackey et~al.
\newblock {emcee v3: A Python ensemble sampling toolkit for affine-invariant
  MCMC}.
\newblock {\em Journal of Open Source Software}, 4:1--2, 2019.

\bibitem{1998SPIE.3352...76S}
Ding~Qiang {Su}, Xiangqun {Cui}, Yanan {Wang}, and Zhengqiu {Yao}.
\newblock {Large-sky-area multiobject fiber spectroscopic telescope (LAMOST)
  and its key technology}.
\newblock In Larry~M. {Stepp}, editor, {\em Advanced Technology Optical/IR
  Telescopes VI}, volume 3352 of {\em Society of Photo-Optical Instrumentation
  Engineers (SPIE) Conference Series}, pages 76--90, August 1998.

\bibitem{Cui_2012}
Xiang-Qun Cui et~al.
\newblock The large sky area multi-object fiber spectroscopic telescope
  ({LAMOST}).
\newblock {\em Research in Astronomy and Astrophysics}, 12(9):1197--1242, aug
  2012.

\bibitem{2012RAA....12..723Z}
Gang {Zhao}, Yong-Heng {Zhao}, Yao-Quan {Chu}, Yi-Peng {Jing}, and Li-Cai
  {Deng}.
\newblock {LAMOST spectral survey {\textemdash} An overview}.
\newblock {\em Research in Astronomy and Astrophysics}, 12(7):723--734, July
  2012.

\bibitem{Aghamousa:2016zmz}
Amir Aghamousa et~al.
\newblock {The DESI Experiment Part I: Science,Targeting, and Survey Design}.
\newblock 10 2016.

\bibitem{deJong:2012nj}
Roelof~S. de~Jong et~al.
\newblock {4MOST - 4-metre Multi-Object Spectroscopic Telescope}.
\newblock {\em Proc. SPIE Int. Soc. Opt. Eng.}, 8446:84460T, 2012.

\bibitem{daCunha:2017wwy}
Elisabete da~Cunha et~al.
\newblock {The Taipan Galaxy Survey: Scientific Goals and Observing Strategy}.
\newblock {\em Publ. Astron. Soc. Austral.}, 34:47, 2017.

\bibitem{Gardner:2006ky}
Jonathan~P. Gardner et~al.
\newblock {The James Webb Space Telescope}.
\newblock {\em Space Sci. Rev.}, 123:485, 2006.

\bibitem{Kalirai:2018qfg}
Jason Kalirai.
\newblock {Scientific Discovery with the James Webb Space Telescope}.
\newblock {\em Contemp. Phys.}, 59(3):251--290, 2018.

\bibitem{Green:2012mj}
J.~Green et~al.
\newblock {Wide-Field InfraRed Survey Telescope (WFIRST) Final Report}.
\newblock 8 2012.

\bibitem{Chary:2020msh}
R.~Chary et~al.
\newblock {Joint Survey Processing of Euclid, Rubin and Roman: Final Report}.
\newblock 8 2020.

\bibitem{Sivia:2006book}
{D. S. Sivia with J. Skilling}.
\newblock {\em {Information gain: quantifying the worth of an experiment}},
  chapter~7, pages 163--164.
\newblock Oxford University Press, Oxford, second edition, 2006.

\bibitem{Feng:2019wgq}
Wen-Fan Feng, Hai-Tian Wang, Xin-Chun Hu, Yi-Ming Hu, and Yan Wang.
\newblock {Preliminary study on parameter estimation accuracy of supermassive
  black hole binary inspirals for TianQin}.
\newblock {\em Phys. Rev. D}, 99(12):123002, 2019.

\bibitem{Ruan:2020smc}
Wen-Hong Ruan, Chang Liu, Zong-Kuan Guo, Yue-Liang Wu, and Rong-Gen Cai.
\newblock {The LISA-Taiji network}.
\newblock {\em Nature Astron.}, 4:108--109, 2020.

\bibitem{Ruan:2019tje}
Wen-Hong Ruan, Chang Liu, Zong-Kuan Guo, Yue-Liang Wu, and Rong-Gen Cai.
\newblock {The LISA-Taiji network: precision localization of massive black hole
  binaries}.
\newblock 10 2019.

\bibitem{OLeary:2005vqo}
Ryan~M. O'Leary, Frederic~A. Rasio, John~M. Fregeau, Natalia Ivanova, and
  Richard~W. O'Shaughnessy.
\newblock {Binary mergers and growth of black holes in dense star clusters}.
\newblock {\em Astrophys. J.}, 637:937--951, 2006.

\bibitem{Kremer:2018tzm}
Kyle Kremer, Sourav Chatterjee, Katelyn Breivik, Carl~L. Rodriguez, Shane~L.
  Larson, and Frederic~A. Rasio.
\newblock {LISA Sources in Milky Way Globular Clusters}.
\newblock {\em Phys. Rev. Lett.}, 120(19):191103, 2018.

\bibitem{Barack:2003fp}
Leor Barack and Curt Cutler.
\newblock {LISA capture sources: Approximate waveforms, signal-to-noise ratios,
  and parameter estimation accuracy}.
\newblock {\em Phys. Rev. D}, 69:082005, 2004.

\bibitem{Chen:2017gfm}
Xian Chen and Pau Amaro-Seoane.
\newblock {Revealing the formation of stellar-mass black hole binaries: The
  need for deci-Hertz gravitational wave observatories}.
\newblock {\em Astrophys. J. Lett.}, 842(1):L2, 2017.

\bibitem{Randall:2019znp}
Lisa Randall and Zhong-Zhi Xianyu.
\newblock {Eccentricity without Measuring Eccentricity: Discriminating among
  Stellar Mass Black Hole Binary Formation Channels}.
\newblock {\em Astrophys. J.}, 914(2):75, 2021.

\bibitem{Gayathri:2020coq}
V.~Gayathri, J.~Healy, J.~Lange, B.~O'Brien, M.~Szczepanczyk, I.~Bartos,
  M.~Campanelli, S.~Klimenko, C.~Lousto, and R.~O'Shaughnessy.
\newblock {GW190521 as a Highly Eccentric Black Hole Merger}.
\newblock 9 2020.

\bibitem{Gayathri:2020fbl}
V.~Gayathri, J.~Healy, J.~Lange, B.~O'Brien, M.~Szczepanczyk, I.~Bartos,
  M.~Campanelli, S.~Klimenko, C.~Lousto, and R.~O'Shaughnessy.
\newblock {Hubble Constant Measurement with GW190521 as an Eccentric Black Hole
  Merger}.
\newblock 9 2020.

\bibitem{Samsing:2018isx}
Johan Samsing and Daniel~J. D'Orazio.
\newblock {Black Hole Mergers From Globular Clusters Observable by LISA I:
  Eccentric Sources Originating From Relativistic $N$-body Dynamics}.
\newblock {\em Mon. Not. Roy. Astron. Soc.}, 481(4):5445--5450, 2018.

\bibitem{Abazajian:2008wr}
Kevork~N. Abazajian et~al.
\newblock {The Seventh Data Release of the Sloan Digital Sky Survey}.
\newblock {\em Astrophys. J. Suppl.}, 182:543--558, 2009.

\bibitem{DeVicente:2015kyp}
Juan De~Vicente, Eusebio S\'anchez, and Ignacio Sevilla-Noarbe.
\newblock {DNF \textendash{} Galaxy photometric redshift by Directional
  Neighbourhood Fitting}.
\newblock {\em Mon. Not. Roy. Astron. Soc.}, 459(3):3078--3088, 2016.

\bibitem{DeLuca:2020qqa}
V.~De~Luca, G.~Franciolini, P.~Pani, and A.~Riotto.
\newblock {Primordial Black Holes Confront LIGO/Virgo data: Current situation}.
\newblock {\em JCAP}, 06:044, 2020.

\bibitem{Clesse:2020ghq}
Sebastien Clesse and Juan Garcia-Bellido.
\newblock {GW190425, GW190521 and GW190814: Three candidate mergers of
  primordial black holes from the QCD epoch}.
\newblock 7 2020.

\bibitem{DeLuca:2020sae}
V.~De~Luca, V.~Desjacques, G.~Franciolini, P.~Pani, and A.~Riotto.
\newblock {GW190521 Mass Gap Event and the Primordial Black Hole Scenario}.
\newblock {\em Phys. Rev. Lett.}, 126(5):051101, 2021.

\bibitem{DeLuca:2021wjr}
V.~De~Luca, G.~Franciolini, P.~Pani, and A.~Riotto.
\newblock {Bayesian Evidence for Both Astrophysical and Primordial Black Holes:
  Mapping the GWTC-2 Catalog to Third-Generation Detectors}.
\newblock {\em JCAP}, 05:003, 2021.

\bibitem{Deng:2021ezy}
Heling Deng.
\newblock {A possible mass distribution of primordial black holes implied by
  LIGO-Virgo}.
\newblock {\em JCAP}, 04:058, 2021.

\bibitem{Visser:2003vq}
Matt Visser.
\newblock {Jerk, snap and the cosmological equation of state}.
\newblock {\em Class. Quant. Grav.}, 21:2603--2616, 2004.

\bibitem{Gong:2004sd}
Yun-Gui Gong.
\newblock {Model independent analysis of dark energy. 1. Supernova fitting
  result}.
\newblock {\em Class. Quant. Grav.}, 22:2121--2133, 2005.

\bibitem{Li:2019nux}
En-Kun Li, Minghui Du, Zhi-Huan Zhou, Hongchao Zhang, and Lixin Xu.
\newblock {Testing the effect of $H_0$ on $f\sigma_8$ tension using a Gaussian
  process method}.
\newblock {\em Mon. Not. Roy. Astron. Soc.}, 501(3):4452--4463, 2021.

\bibitem{Bartos:2014spa}
Imre Bartos, Arlin P.~S. Crotts, and Szabolcs M\'arka.
\newblock {Galaxy Survey On The Fly: Prospects of Rapid Galaxy Cataloging to
  Aid the Electromagnetic Follow-up of Gravitational-wave Observations}.
\newblock {\em Astrophys. J. Lett.}, 801(1):L1, 2015.

\bibitem{Chen:2015nlv}
Hsin-Yu Chen and Daniel~E. Holz.
\newblock {Facilitating follow-up of LIGO-Virgo events using rapid sky
  localization}.
\newblock {\em Astrophys. J.}, 840:88, 2017.

\bibitem{Klingler:2019fbl}
N.~J. Klingler et~al.
\newblock {$Swift$-XRT Follow-up of Gravitational Wave Triggers in the Second
  Advanced LIGO/Virgo Observing Run}.
\newblock {\em Astrophys. J. Suppl.}, 245(1):15, 2019.

\bibitem{DES:2005dhi}
T.~Abbott et~al.
\newblock {The dark energy survey}.
\newblock 10 2005.

\bibitem{Sevilla-Noarbe:2020jpu}
I.~Sevilla-Noarbe et~al.
\newblock {Dark Energy Survey Year 3 Results: Photometric Data Set for
  Cosmology}.
\newblock 11 2020.

\bibitem{Zhang:2019oet}
Chunyu Zhang, Qing Gao, Yungui Gong, Dicong Liang, Alan~J. Weinstein, and Chao
  Zhang.
\newblock {Frequency response of time-delay interferometry for space-based
  gravitational wave antenna}.
\newblock {\em Phys. Rev. D}, 100(6):064033, 2019.

\bibitem{Toubiana:2020cqv}
Alexandre Toubiana, Sylvain Marsat, Stanislav Babak, John Baker, and Tito
  Dal~Canton.
\newblock {Parameter estimation of stellar-mass black hole binaries with LISA}.
\newblock {\em Phys. Rev. D}, 102:124037, 2020.

\bibitem{Karki:2016pht}
S.~Karki et~al.
\newblock {The Advanced LIGO Photon Calibrators}.
\newblock {\em Rev. Sci. Instrum.}, 87(11):114503, 2016.

\bibitem{Sun:2020wke}
Ling Sun et~al.
\newblock {Characterization of systematic error in Advanced LIGO calibration}.
\newblock {\em Class. Quant. Grav.}, 37(22):225008, 2020.

\bibitem{Estevez:2020pvj}
D.~Estevez, P.~Lagabbe, A.~Masserot, L.~Rolland, M.~Seglar-Arroyo, and
  D.~Verkindt.
\newblock {The Advanced Virgo Photon Calibrators}.
\newblock {\em Class. Quant. Grav.}, 38(7):075007, 2021.

\bibitem{vanderWalt:2011bqk}
St\'efan van~der Walt, S.~Chris Colbert, and Ga\"el Varoquaux.
\newblock {The NumPy Array: A Structure for Efficient Numerical Computation}.
\newblock {\em Comput. Sci. Eng.}, 13(2):22--30, 2011.

\bibitem{Virtanen:2019joe}
Pauli Virtanen et~al.
\newblock {SciPy 1.0--Fundamental Algorithms for Scientific Computing in
  Python}.
\newblock {\em Nature Meth.}, 17:261, 2020.

\bibitem{lalsuite}
{LIGO Scientific Collaboration}.
\newblock {LIGO} {A}lgorithm {L}ibrary - {LALS}uite.
\newblock free software (GPL), 2018.

\bibitem{Hunter:2007ouj}
John~D. Hunter.
\newblock {Matplotlib: A 2D Graphics Environment}.
\newblock {\em Comput. Sci. Eng.}, 9(3):90--95, 2007.

\bibitem{corner}
Daniel Foreman-Mackey.
\newblock corner.py: Scatterplot matrices in python.
\newblock {\em The Journal of Open Source Software}, 1(2):24, jun 2016.

\end{thebibliography}

\end{document}